\documentclass[fleqn,usenatbib]{mnras}

\usepackage{float}
\usepackage{bm}
\usepackage{footnote}
\usepackage{booktabs}
\usepackage{mwe}
\usepackage{stfloats} 
\usepackage{multirow}

\usepackage[T1]{fontenc}
\DeclareRobustCommand{\VAN}[3]{#2}
\let\VANthebibliography\thebibliography
\def\thebibliography{\DeclareRobustCommand{\VAN}[3]{##3}\VANthebibliography}

\usepackage{graphicx}   
\usepackage{amsmath}    
\usepackage{amssymb}    
\usepackage{enumitem}
\usepackage[normalem]{ulem}

\newcommand\nside{N_{\rm side}}
\newcommand\healpix{\textsc{HEALPix}}
\newcommand\lenspix{\textsc{LensPix}}

\newcommand\um{\textsc{UniverseMachine}}

\newcommand\rockstar{\textsc{Rockstar}}
\newcommand\mdpl{\textsc{MDPL2}}
\newcommand\bp{\textsc{Bolshoi-Planck}}

\newcommand\graytrix{\textsc{GRayTrix}}
\newcommand\bahamas{\textsc{Bahamas}}
\newcommand\camb{\textsc{CAMB}}
\newcommand\trinity{\textsc{Trinity}}
\newcommand\msol{{\rm M}_{\odot}}
\newcommand\kcmb{\kappa_{\rm CMB}}

\newcommand\planck{{\it Planck}}
\newcommand\gpch{h^{-1}{\rm Gpc}}
\newcommand\mpch{h^{-1}{\rm Mpc}}
\newcommand\kpch{h^{-1}{\rm kpc}}
\newcommand\cnn{\delta_{\rm g}\delta_{\rm g}}
\newcommand\ckk{\kappa\kappa}
\newcommand\cggl{\delta_{\rm g}\gamma_{t}}
\newcommand\mum{\mu {\rm m}}
\newcommand\muk{\mu {\rm K}}

\usepackage{xcolor}
\definecolor{basecol}{HTML}{0D5661}
\usepackage{hyperref}
 \hypersetup{
     colorlinks=true,
     linkcolor=basecol,
     citecolor = basecol,      
     urlcolor=basecol,
     }
     
\usepackage{newtxtext,newtxmath}
\title[\textsc{Agora}]{\textsc{Agora}: Multi-Component Simulation for Cross-Survey Science}

\author[Y. Omori]{
Yuuki Omori$^{1,2,3}$\thanks{E-mail: yomori@uchicago.edu}
\\
$^{1}$ Department of Astronomy and Astrophysics, University of Chicago, Chicago, IL 60637, USA\\
$^{2}$Kavli Institute for Cosmological Physics, University of Chicago, Chicago, IL 60637\\
$^{3}$Kavli Institute for Particle Astrophysics and Cosmology and Department of
Physics, Stanford University, Stanford, CA, USA, 94305\\
}

\date{Accepted XXX. Received YYY; in original form ZZZ}

\pubyear{2023}

\begin{document}
\label{firstpage}
\pagerange{\pageref{firstpage}--\pageref{lastpage}}
\maketitle

\begin{abstract}
The tightest cosmological constraints currently available are obtained by combining complementary data sets. When combining correlated data sets, various astrophysical biases that affect the measurements must be identified and treated. There are numerous such biases, and they are  often intricately related with one another via complex astrophysical effects, making them difficult to characterize analytically. Consequently, a simulation with multiple components implemented coherently is required to investigate these biases simultaneously and as a whole. In this work, a suite of simulated extragalactic skies is presented, including maps and/or catalogues of cosmic microwave background (CMB) lensing, thermal and kinetic Sunyaev-Zel'dovich (tSZ/kSZ) effects, cosmic infrared background (CIB), radio sources, galaxy overdensity and galaxy weak lensing. Each of these probes is implemented in the lightcone using halo catalogues and/or particles from the \textsc{Multidark-Planck2}\ ($\mdpl$) $N$-body simulation, and the modelling is calibrated using hydrodynamic simulations and publicly available data. The auto- and cross-spectra of the individual probes, as well as the cross-spectra between the observables, are shown to be consistent with theoretical models and measurements from data. The simulation is shown to have a wide range of applications, including forecasting, pipeline testing, and evaluating astrophysical biases in cross-correlation studies. It is further demonstrated that the simulation products produced in this work have sufficient accuracy to recover the input cosmology when subjected to a full cosmological analysis and are ready for application in real-world analyses for ongoing and future surveys.

\end{abstract}


\clearpage 

\begingroup
\let\clearpage\relax
\tableofcontents
\endgroup

\section{Introduction} \label{sec:intro}

Ongoing surveys and experiments are amassing immense amounts of high-resolution and high-quality data for use in cosmological analyses. Using such data, measurements with unprecedented accuracy and precision will be made, necessitating the use of equally sophisticated tools and models for effective analysis.

Large-scale structure (LSS) studies have recently come to rely more heavily on cosmological $N$-body simulations, particularly in modelling the nonlinear regime of cosmological evolution, which is challenging to derive analytically. $N$-body simulations have been used in a variety of ways, including making precise predictions of given observables in various cosmologies \citep{lawrence2010,derose2019,nishimichi2019,villaescusanavarro2020,euclidemulator2}, estimating covariances of observables \citep{harnoisderaps2018,shirasaki2019,villaescusanavarro2020}, and investigating the growth of structure in a non-$\Lambda$CDM universe (e.g.,  \citealt{rocha2013,banerjee2016,kacprzak2022,harnoisderaps2022}). As measurements get pushed to smaller physical scales in efforts to extract more information from data, $N$-body simulations will become increasingly important as an analysis  tool.

Simulations are also extensively utilized in cosmic microwave background (CMB) studies; however, many of these are produced for a specific observable. For example, the thermal and kinetic Sunyaev Zel'dovich effects (tSZ/kSZ; \citealt{sunyaev1972,sunyaev1980}) have been studied extensively using hydrodynamical simulations due to their characteristic of involving gas and electron pressure \citep{battaglia2012,vogelsberger2014,dolag2016,mccarthy2017,emberson2019}. Other simulations have focused on the infrared sources \citep{lacey2016,bethermin2017,lovell2021,bisigello2021}, radio sources \citep{bonaldi2019,li2021}, as well as the effect of CMB lensing \citep{das2008,castorina2015}. While each of these simulations has been invaluable to the scientific community, simulations with multiple astrophysical effects \textit{implemented coherently in the same lightcone} are urgently needed since these observables are naturally correlated in our Universe, and neglecting such correlations in our model may lead to misinterpretation of our data.

There are two such multi-component simulations in the literature that are publicly available. The first is the work presented in \citet{sehgal2010}\footnote{Available at \url{https://lambda.gsfc.nasa.gov/simulation/full_sky_sims_ov.html}}, which used a $1000\ h^{-1}{\rm Mpc}$ box containing $1024^{3}$ particles to simulate various components of the millimetre sky. The suite includes CMB lensing, the tSZ/kSZ effects, the cosmic infrared background (CIB) and radio sources. The physical modelling was calibrated against available observational data at the time (see references therein). 
This simulation has been utilized in a number of studies, including Simon's Observatory (SO; \citealt{simons}) forecasts, and has proven to be a valuable tool for verifying analysis pipelines and performing accurate forecasts based on realistic maps. 
\cite{han2021} expanded on this work by using machine learning techniques to generate multiple realizations of the simulated skies.

More recently, \citet{stein2020} used the mass-Peak Patch approach \citep{bond1996a,bond1996b,stein2019} to evolve a large cosmological box onto which astrophysical effects were pasted. The mass-Peak Patch method operates on the initial density field, after which the halos are displaced to their final positions using Lagrangian perturbation theory \citep{bond1996a}. This approach has been validated against true $N$-body simulation as well as alternate simulation methods \citep{bond1996b,stein2020}. The advantage of this method is that it has a lower computational cost (because it does not numerically evolve particle positions at low redshifts, where time steps are shorter), allowing multiple and/or larger cosmological volumes to be simulated when compared with the standard $N$-body approach. Lightcones reaching higher redshifts could be generated without repeating structures, and the high-mass end of the halo mass function could be well characterized with larger simulation boxes. Furthermore, because of the low computational cost, it is feasible to repeat the simulation with different assumptions, such as the underlying cosmology.

Simulations that include both CMB and LSS observables are far less developed, but they will be required for future cross-correlation studies. Building such simulations is difficult because each observable has different simulation requirements (such as redshift and halo mass ranges), and it is computationally challenging to meet all the requirements simultaneously. However, it is clear that these simulations are becoming increasingly important, for example, in investigating biases from higher-order effects in cross-correlation studies (\citealt{omori2019a,omori2019b,fabbian2021,lembo2021,omori2022}).

\begin{figure*}
\begin{center}
\includegraphics[width=0.9\linewidth]{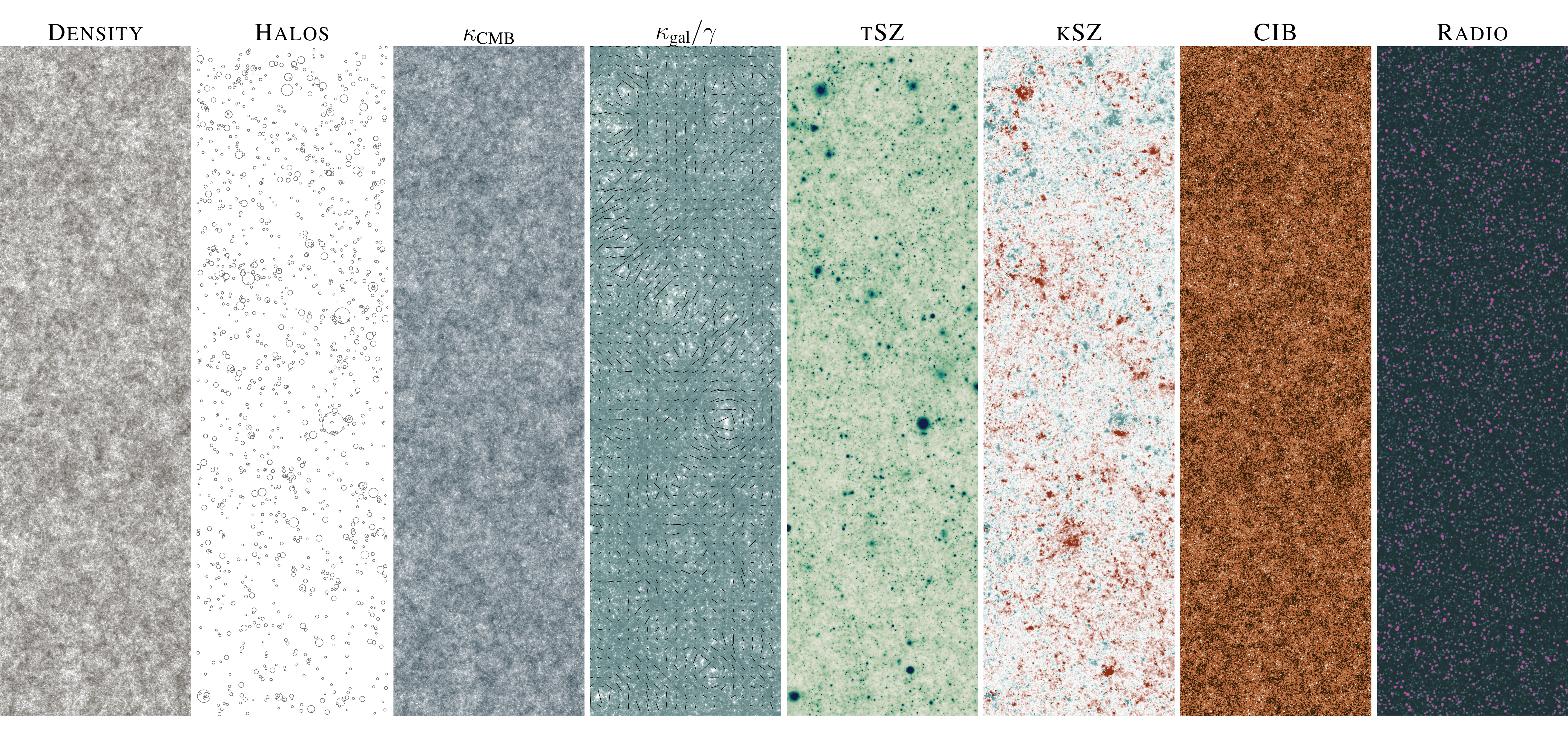}
\caption{Visualization of the components in the simulation. From left to right, dark matter density field, distribution of dark matter halos with $M_{\rm h}>10^{14}{\rm M}_{\sun}$, CMB convergence field $\kappa_{\rm CMB}$, galaxy shear plotted over galaxy convergence field, tSZ effect (as Compton-$y$ parameter), kSZ effect, CIB at 353 GHz, and radio sources. Note that some of these components have minimal redshift overlap, and therefore the correlation between such fields is expected to be small.}
\label{fig:agora_maps}
\end{center}
\end{figure*}

This paper presents a multi-component simulation with both CMB and LSS observables implemented coherently in a single lightcone. A similar approach to  \citet{sehgal2010} is used, in that observationally constrained analytical prescriptions are applied to halos and particles of a dark matter-only $N$-body simulation, from which maps and catalogues of extragalactic observables are created. All the CMB secondary effects are simulated to sub-arcminute resolution, and the various observables are validated in the multipole range $50\lesssim\ell\lesssim3000$, from where ongoing and future CMB experiments will primarily extract cosmological information. The output maps and catalogues are calibrated using data sets including ${\it Planck}$ \citep{planckoverview}, SPT-SZ \citep{carlstrom2011}, and  SPTpol \citep{bleem2012} such that the simulation products are ready to be used for real data analyses. For the LSS observables, maps of galaxy overdensity, galaxy weak lensing, and intrinsic alignment are produced and validated in the multipole range $30\lesssim\ell\lesssim3000$. Observational effects such as survey systematics, photo-$z$ uncertainties and shape noise can be incorporated depending on the specific dataset the user wishes to simulate. The output maps are shown in Figure \ref{fig:agora_maps}.

The simulation presented in this paper enables us to investigate several key aspects of modern CMB analysis, including foreground propagation in component separation, the impact of non-Gaussianity of extragalactic foregrounds, non-Gaussianity of the lensing potential, and the effect of lensing on the CMB secondaries. It can also be used to investigate the non-Gaussianity of the galaxy weak lensing field \citep{gatti2022b}, as well as higher-order effects like source clustering and lens-source clustering \citep{krause2021,prat2022}. The most powerful and novel aspect of this simulation, however, is its ability to facilitate cross-correlation studies between CMB and LSS observables (and others in the future). 

This paper is organized as follows. In Section \ref{sec:mdpl2}, a description of the underlying dark matter-only $N$-body simulation that the synthetic skies are based on is given. Section \ref{sec:seconadary_modeling} provides descriptions of how the astrophysical/cosmological observables are implemented. In Section \ref{sec:secondary_validation}, various measurements (such as auto- and cross-spectra) are made using the simulation outputs, and the results are compared with existing measurements from data to verify the accuracy of the simulation. In Section \ref{sec:applications}, the simulation products are used in a variety of ways to demonstrate their versatility. In Section \ref{sec:discussion}, caveats and prospective efforts relevant to this work are described, and finally, a summary of this work is given in Section \ref{sec:summary}. Appendix \ref{sec:simproducts} contains a list of all the simulation products that will be available upon publication.

\begin{table}
\centering

\begin{tabular}[t]{ccc}
\toprule
Parameters & Values  \\
\midrule
Box size & 1 $\gpch$\\
$N_{\rm part}$ & $3840^{3}$\\
Mass resolution &$1.51\times10^{9}\
 h^{-1}{\rm M}_{\sun}$\\
${\rm Force\ resolution}$ & $\sim13\ \kpch$ (at high $z$)\\
&$\sim 5\ \kpch$ (at low $z$)\\
${\rm Initial\ redshift}$ & $120$\\
$N_{\rm snap}$ & $130$\\
\midrule
$h_{0}$&0.6777\\
$\Omega_{\rm m}$&0.307\\
$\Omega_{\rm c}$&0.048\\
$\Omega_{\Lambda}$&0.693\\
$\sigma_{8}$&$0.818^{*}$\\
\bottomrule

\end{tabular}
\caption{Specification of the MDPL2 simulation. More details can be found at \url{https://www.cosmosim.org/metadata/mdpl2/}, and in \protect\cite{klypin2016}. Note that the original simulation uses $\sigma_{8}=0.8288$. However, it is found that the amplitude of the matter power spectrum matches better with the value given above, and hence this  value is assumed hereafter.}\label{table:cosmoparam_mdpl2}
\end{table}

\section{MultiDark Planck 2 simulation}\label{sec:mdpl2}
The \textsc{MultiDark\ Planck 2}  ($\mdpl$) simulation\footnote{Description of the simulation and some data products are queryable at \url{https://www.cosmosim.org/metadata/mdpl2/}.} is a dark matter-only $N$-body simulation that contains $3840^{3}$ particles in a $1\ h^{-1}{\rm Gpc}$ box \citep{klypin2016}. The specifications of the simulation and the adopted cosmology are listed in Table \ref{table:cosmoparam_mdpl2}.
In this work, both the dark matter particles\footnote{Requested through private communications.} and the $\rockstar$ \citep{behroozi2013} halo catalogues are used.\footnote{Available at  \url{http://halos.as.arizona.edu/simulations/MDPL2/hlists/}.} Numerous analyses have been carried out on the MDPL2 simulation, including studies of the halo concentration, halo profiles \citep{klypin2016}, and halo mass functions \citep{benson2017}. The foremost advantage of using MDPL2 over other publicly available simulations is the wide range of existing derived products, including galaxy catalogues from semi-analytical models \citep{benson2012,croton2016,cora2018}, studies of other tracers such as emission line galaxies \citep{alam2020}, and intensity mapping \citep{yang2020}, all of which could be implemented in practice in a single lightcone.

\begin{figure}
\begin{center}
\includegraphics[width=1.0\linewidth]{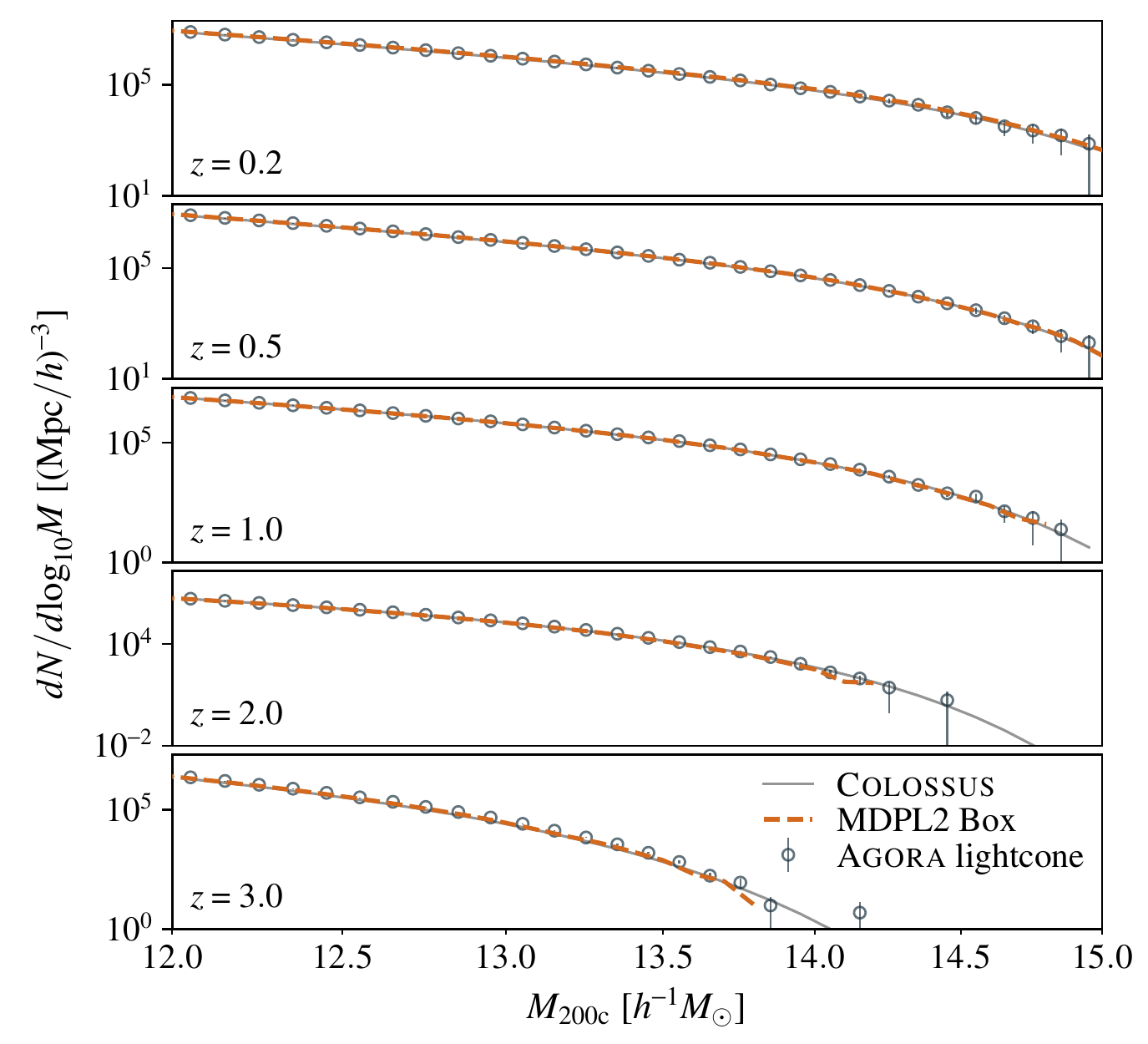}
\caption{The halo mass functions of the MDPL2 \rockstar{} halo catalogue, from the simulation boxes (orange lines), and lightcone (navy points) compared with a theoretical halo mass function assuming the \protect\cite{tinker2008} model, computed using the package \textsc{Colossus} \protect\citep{diemer2018} (solid grey lines).}
\label{fig:massfunction}
\end{center}
\end{figure}

To fill the cosmological volume required for this work, the original high-resolution simulation snapshots and their associated halo catalogues are tessellated making use of the periodic boundary condition (as opposed to stacking multi-resolution boxes to generate a non-repeating lightcone as done in \citealt{derose2019,fosalba2015a,fosalba2015b,takahashi2017}). The primary advantage of this approach is that it retains the halo mass resolution that is required to construct CIB maps at high redshifts. This can be confirmed by inspecting the halo mass functions at $z=0.2, 0.5, 1, 2, 3$, as shown in Figure \ref{fig:massfunction}. The high-resolution density maps also allow us to create high-resolution CMB and galaxy lensing maps, which are key observables of interest and are also required for applying deflection to the other components in the simulation.

\begin{figure}
\begin{center}
\includegraphics[width=0.7\linewidth]{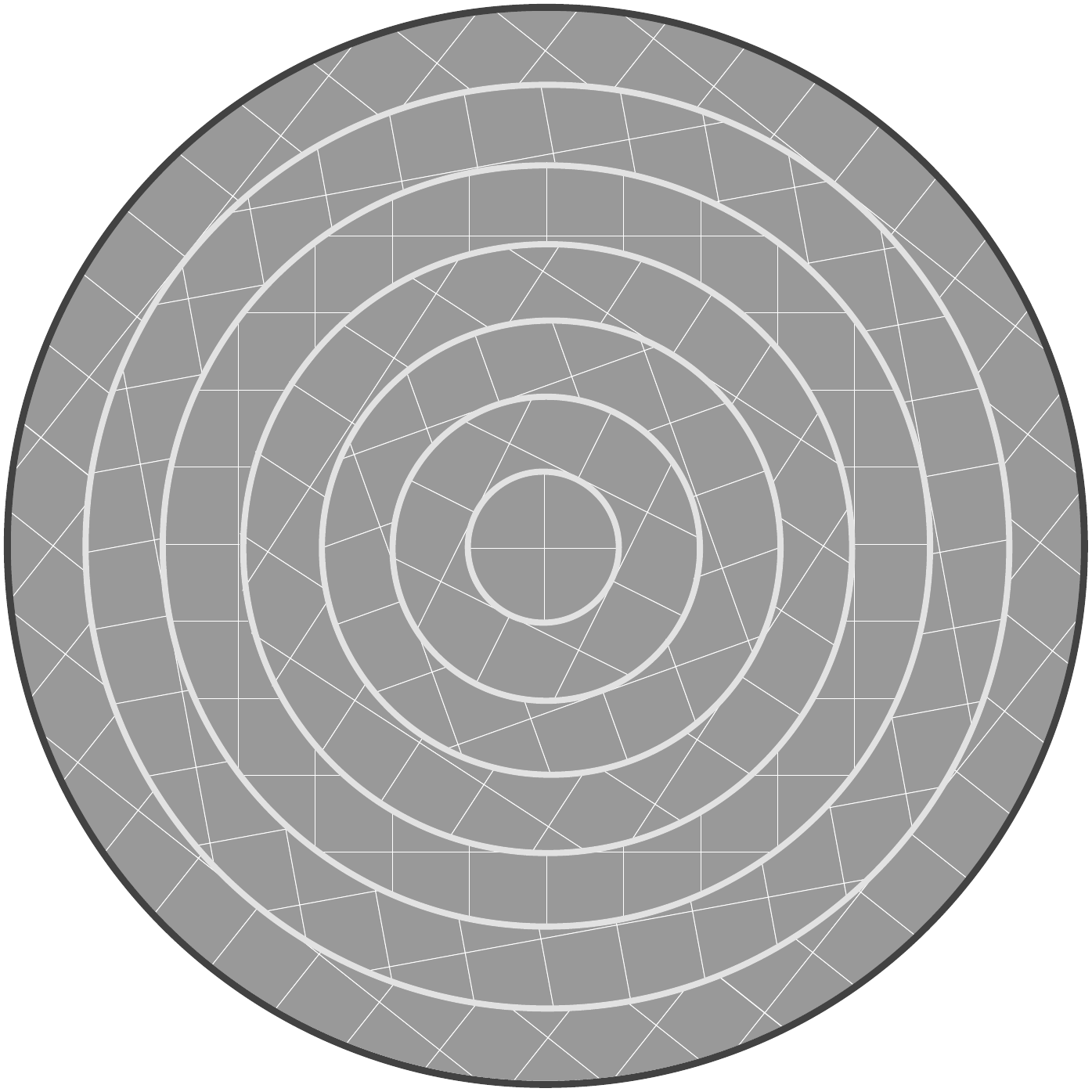}
\caption{The shell rotation scheme used to generate the lightcone from the MDPL2 simulation boxes. Each individual grid represents a 1 $\gpch$ box.}
\label{fig:boxrotation}
\end{center}
\end{figure}

Concentric spherical shells of thickness $25\ \mpch{}$ are extracted from the tiled volume, and the dark matter particles and their distance-weighted velocities $v_{\rm los}/d_{\rm A}^{2}$ are projected onto  $\healpix$ shells\ \citep{gorski2005} of $N_{\rm side}=8192$ (corresponding to 0.43 arcminutes in pixel size). To avoid repeating structures along the line of sight, the shells in the lightcone are rotated randomly every $ 1 \gpch$ (1 box length), as illustrated in Figure \ref{fig:boxrotation}. To keep the correlations between the observables, the same rotations are applied to the density/velocity shells and halo catalogues.

\section{Modelling of the secondary components}\label{sec:seconadary_modeling}

\begin{table}
\begin{center}
\begin{tabular}[t]{cc}
\toprule
Observable & Redshift range  \\
\midrule
CMB lensing  & \hspace{0.2cm}$0<z<1089$   \\
tSZ  & $0<z<3.0$  \\
kSZ  & $0<z<3.0$  \\
CIB     & $0<z<8.6$ \\
Radio sources   & $0<z<4.0$ \\
Density field  & $0<z<8.6$ \\
Galaxy lensing ($\kappa_{\rm g},\gamma_{1},\gamma_{2}$)  & $0<z<8.6$ \\
\bottomrule
\end{tabular}
\caption{The redshift ranges covered by the individual astrophysical components implemented in the simulation. For CMB lensing, ray tracing is carried out up to $z=8.6$, and a Gaussian realization of the contributions from $8.6<z<1089$ is added.}
\label{tab:zcoverage}
\end{center}
\end{table}

Table \ref{tab:zcoverage} summarizes the list of implemented observables, as well as the redshift range that each of the observables covers. CMB lensing is integrated up to $z=8.6$ and a Gaussian realization of the contribution from $8.6<z<1089$ is added. This redshift cut-off is set so that the underlying density field is sufficiently Gaussian beyond this redshift (which is verified by comparing the linear and nonlinear matter power spectra, and is found to differ by $<0.1$\% at $\ell=3000$). The CIB, dark matter density, and galaxy weak lensing fields are also integrated up to $z=8.6$. The tSZ and kSZ effects are integrated up to $z=3$, but the first 100 $\mpch$ of the lightcone are discarded because the most massive and nearby clusters introduce significant variance to the measured tSZ/kSZ power spectra (see e.g. \citealt{osato2020}). Radio sources up to $z=4$ are added, which is the redshift up to which the luminosity function adopted in this work is defined. The methodology and modelling used to implement each component are discussed in the sections that follow.

\subsection{CMB Weak Lensing} \label{sec:cmblensing}
The gravitational potential of the large-scale structure deflects a photon's trajectory, as it travels to us from the last scattering surface.  As a result, the observed CMB is a distorted image composed of photons coherently deflected by a few arcminutes over scales of approximately a degree. Since the primary CMB's statistical properties are well known, the deflection field can be estimated by studying the coupling of modes in the distorted image of the CMB (see \citealt{lewis2006} for a review). Recent observations from experiments such as \planck{}, ACT/ACTpol, POLARBEAR, and SPT-SZ/SPTpol have demonstrated that CMB lensing is a powerful probe for measuring the density and amplitude of fluctuations in the Universe \citep{vanengelen2012,planck2013xvii,story2015,planck2015xv,omori2017,sherwin2017,simard2018,planck2018viii,wu2019,faundez2020,bianchini2020,darwish2021}, and that it provides unique constraints on cosmological parameters such as  $H_{0}$ \citep{baxter2021}, and can be used for cross-correlation studies with large-scale structure to provide an independent measurement of structure growth \citep{giannantonio2016,kirk2016,omori2019a,omori2019b,desy1_6x2,darwish2021,krolewski2021,white2022,chang2022,desy3_6x2}.

In generating CMB lensing maps, two approaches are used. The first method employs the so-called Born approximation, which computes the effective total deflection along an undeflected ray path \citep{das2008}:
\begin{equation}
\kappa_{\rm CMB}(\hat{n})=\sum_{i}d\chi W_{\rm CMB}(\chi_{i})\delta^{i}_{\rm DM}(\chi_{i},\hat{n}),
\end{equation}
where
\begin{equation}\label{eq:W_kcmb}
W_{\rm CMB}(\chi)=\frac{3\Omega_{\rm m} H_{0}^{2}}{2c^{2}}\frac{\chi}{a(\chi)}\frac{\chi_{\rm CMB}-\chi}{\chi_{\rm CMB}},
\end{equation}
$\chi$ is the comoving radial distance, $\chi_{\rm CMB}$ is the comoving radial to the last scattering surface, $a(\chi)$ is the cosmological scale factor, and the $\healpix$ density shells are used in place of $\delta_{\rm DM}(\chi_{i},\hat{n})=(\rho(\chi_{i},\hat{n})-\bar{\rho})/\bar\rho$. Since the Born approximation only depends on the matter density field, it is less prone to resolution issues and computational challenges that ray tracing algorithms may have at small angular scales.

The second approach is multi-plane (sphere) ray tracing, which tracks the deflection of light rays as they propagate through the density fields \citep{becker2013,fabbian2018,gouin2019}. In contrast to the Born approximation, this method retains higher-order terms of the gravitational potential (i.e. Born corrections and lens-lens coupling, \citealt{hilbert2009}), and is thus capable of describing the nonlinear weak lensing field more accurately, and is thus critical for making measurements beyond the two-point statistics, such as skewness and kurtosis \citep{petri2017}.

Following \cite{das2008} and \cite{shirasaki2015}, the lensing field for the $j$-th shell is generated using
\begin{equation}
\kappa^{j}(\hat n)=\frac{4\pi G}{c^{2}}\frac{\Delta^{j}_{\Sigma}(\hat n)}{a^{j}d_{\rm A}^{j} },
\end{equation}
where $a^{j}$ and $d^{j}_{\rm A}$ are the scale factor and comoving angular diameter distance respectively, and the surface mass density is defined as:  
\begin{equation}
\Delta_{\Sigma}^{j}(\hat n)=\int_{\rm shell}  d\chi(\rho(\hat{n},\chi)-\bar\rho)d^{2}_{\rm A}(\chi).
\end{equation}
This can be recast in terms of simulation parameters:
\begin{equation}
\kappa^{j}(\hat n)=\frac{3\Omega_{\rm m}}{2a^{j}(\chi)d_{\rm A}^{j}}\left(\frac{H_{0}}{c}\right)^{2}\frac{V_{\rm sim}}{N_{\rm part}}\frac{N_{\rm pix}}{4\pi}(n_{\rm part}(\hat n)-\bar{n}_{\rm part}).
\end{equation}
where $V_{\rm sim}\equiv L_{\rm sim}^{3}$ is the simulation volume, $N_{\rm part}$ is the total number of particles in the simulation,  $N_{\rm pix}$  is the number of $\healpix$ pixels used to grid the particle density field, and $n_{\rm part}$ is the number of particles per \healpix{} pixel. These convergence shells are converted into shells of gravitational potential using:
\begin{equation}\label{eq:kap2phi}
\phi_{\ell m}=\frac{2}{\ell(\ell+1)}\kappa_{\ell m},
\end{equation}
and the first and second derivatives of the potential are computed on a \healpix{} grid of $N_{\rm side}=16384$. The light rays are then propagated through layers of deflection fields up to $z=8.6$ using the code $\graytrix$ \citep{hamana2015,shirasaki2015} to produce the $\kappa_{\rm CMB}$ map, and the output map is degraded back to $N_{\rm side}=8192$ to reduce the post-processing computational costs. 
Finally, the power spectrum of the lensing field between $8.6<z<1089$ is computed, from which a Gaussian realization is generated and added to produce the full $\kappa_{\rm CMB}$ map. A small Gaussian contribution is also added to correct the $\ell\lesssim10$ modes (due to the tiling strategy used).  

Throughout this paper, the convention of using $L$ to refer to multipoles of reconstructed CMB lensing maps and $\ell$ to refer to multipoles of all other observables is followed.

\subsection{Lensed Primary CMB}\label{sec:implementation_primarycmb}

To produce maps of lensed CMB, the unlensed CMB maps are first generated using the prescription given by \cite{giannantonio2008} (see also Appendix \ref{sec:appendix_corrgauss}):
\begin{align}
a_{\ell m}^{\kappa} &=\eta_{\kappa}\sqrt{C_{\ell}^{\kappa\kappa}}\label{eq:gauss_kappa}\\
a_{\ell m}^{T} &= \eta_{\kappa}\frac{C_{\ell}^{\kappa T}}{\sqrt{C_{\ell}^{\kappa\kappa}}}+\eta_{T}\sqrt{C_{\ell}^{TT}-(C_{\ell}^{\kappa T})^2/C_{\ell}^{\kappa\kappa} }\\
a_{\ell m}^{E} &= \eta_{\kappa}\frac{C_{\ell}^{\kappa E}}{\sqrt{C_{\ell}^{\kappa\kappa}}}+\eta_{T}\frac{\left(C_{\ell}^{TE}-C_{\ell}^{\kappa T}C_{\ell}^{\kappa E}/C_{\ell}^{\kappa\kappa}\right)}{\sqrt{C_{\ell}^{TT}-(C_{\ell}^{\kappa T})^2/C_{\ell}^{\kappa\kappa} }}\nonumber\\
&+\eta_{E}\sqrt{C_{\ell}^{EE}-(C_{\ell}^{\kappa E})^2/C_{\ell}^{\kappa\kappa}-\frac{(C_{\ell}^{TE}-C_{\ell}^{\kappa T}C_{\ell}^{\kappa E} /C_{\ell}^{\kappa\kappa})^{2}}{C_{\ell}^{TT}-(C_{\ell}^{\kappa T})^2/C_{\ell}^{\kappa\kappa}} }\\
a_{\ell m }^{B}&=0,
\end{align}
where $\eta_{\kappa},\eta_{T},\eta_{E}$ are complex random numbers with unit variance, and $C_{\ell}^{TT},C_{\ell}^{EE},C_{\ell}^{TE}$, $C_{\ell}^{T\kappa}$, $C_{\ell}^{E\kappa}$, $C_{\ell}^{\kappa\kappa}$ are spectra computed with $\camb$\footnote{\url{https://camb.info/}} \citep{lewis1999}, using the cosmological parameters listed in Table \ref{table:cosmoparam_mdpl2}, which are consistent with both \textsc{Planck+WP+highL+BAO} best-fit parameters from \cite{planck2013xvi} and the cosmology assumed for the underlying MDPL2 $N$-body simulation. 

$\lenspix$\footnote{\url{https://cosmologist.info/lenspix/}}  \citep{lenspix} is then used to deflect $T/Q/U$ maps (generated from $a_{\ell m}^{T}, a_{\ell m}^{E}, a_{\ell m}^{B}$) using the lensing field $\kappa$ (generated from $a_{\ell m}^{\kappa}$).  Two additional sets of lensed CMB maps are produced: the first is generated by replacing $a_{\ell m}^{\kappa}$ in Equation \eqref{eq:gauss_kappa} with that from the ray traced lensing map, and the second is obtained by replacing ``Gaussianized" lensing  maps (i.e., random Gaussian realizations generated from the measured power spectrum of the ray traced lensing map). In this paper, the maps in the first set are treated as the fiducial lensed CMB maps unless otherwise specified, whereas the maps generated using the second approach are used in scenarios that require multiple independent realizations with matching power spectra.

\subsection{Thermal Sunyaev-Zel'dovich effect}
The tSZ effect is the inverse-Compton scattering of CMB photons off high-energy electrons in hot gas in and around massive objects such as galaxy groups or clusters \citep{carlstrom2002,hill2015}. The effect has been used to detect galaxy clusters over a large sky area and at high redshifts (where detecting low-mass clusters using optical and X-ray observations is particularly challenging), and the number density and clustering signatures of these clusters have been used to extract cosmological information \citep{hasselfield2013,planck2015xxiv,dehaan2016,bocquet2019}.

Maps of the tSZ effect are produced by combining multiple frequency channels and exploiting the known frequency dependence of the tSZ effect. Such maps have been released by \planck{} \citep{planck2015xxii}, and numerous studies have characterized these maps by analysing their power spectra \citep{bolliet2018} and cross-correlations with galaxy density and galaxy weak lensing maps \citep{hojatti2017,makiya2018,pandey2019,tanimura2020,koukoufilippas2020,osato2020,makiya2020,sanchez2022}. More recently, higher-resolution tSZ maps produced by combining data from \planck{} with ground-based experiments such as ACT \citep{madhavacheril2020} and SPT \citep{bleem2021} have been released. While such maps contain a wealth of interesting astrophysical information, distinguishing the true tSZ signal from contamination from other sources such as Galactic dust, CIB, and radio sources remains a significant challenge \citep{madhavacheril2020,bleem2021}.

There are numerous hydrodynamical simulations with the tSZ effect implemented, ranging from zoom-ins of individual galaxy clusters to large-scale simulation boxes (e.g., \citealt{battaglia2010,dolag2016,mccarthy2017,cui2018,villaescusanavarro2022,pakmor2022}). Hydrodynamical simulations are essential for studying the tSZ effect since astrophysical effects such as Active Galactic Nuclei (AGN) and supernova feedback affect the distribution and temperature of gas, which govern the characteristics of the tSZ effect in and around halos. While these simulations are necessary for producing realistic tSZ maps, running them with sufficient volume to cover the high redshift clusters and enough resolution to study the gas distribution in detail across a wide range of cluster masses is computationally challenging.

In this work, tSZ maps are generated by pasting electron pressure profiles inferred  from an external hydrodynamical simulation onto halos with masses $M_{\rm h}>10^{12}  h^{-1} {\rm M}_{\sun}$ in the lightcone. For the complete description of the extraction and profile fitting methods, the readers are referred to \cite{mead2020}.  The method for pasting pressure profiles onto halos is described in greater detail below.

The first step is to compute the fractional gas mass bound to a halo: 
\begin{equation}
f_{\rm bnd}(M_{\rm vir})=\frac{\Omega_{\rm b}}{\Omega_{\rm m}}\frac{(M_{\rm vir}/M_{0})^{\beta}}{1+(M_{\rm vir}/M_{0})^{\beta}},
\end{equation}
where $M_{0}$ is a free parameter that is fitted by comparing with the BAryons and HAloes of MAssive Systems (\bahamas{}; \citealt{mccarthy2017}) simulation,\footnote{\url{https://www.astro.ljmu.ac.uk/~igm/BAHAMAS/}} and $\beta$ is a constant, which is fixed to 0.6. Gas is assumed to be gravitationally bound to halos with the profile:
\begin{equation}\label{eq:gasdensityprofile}
\rho_{\rm gas}^{\rm bnd}(M_{\rm vir},r)=\rho_{0}\left[\frac{{\rm ln}(1+r/r_{\rm s})}{r/r_{\rm s}}\right]^{1/(\Gamma-1)},
\end{equation}
which is a simplified form of the \cite{komatsu2001} profile adopted in \cite{martizzi2013}. Here $\Gamma$ is the effective polytropic index, and $r_{\rm s}$ is the scale radius defined as: 
\begin{equation}
r_{\rm s}=r_{\rm vir}/c(M_{\rm vir}),
\end{equation}
where $c(M_{\rm vir})$ is the concentration parameter.\footnote{While the mass-concentration relation from \cite{duffy2008}, $c(M)=7.85\left(M/( 2\times10^{12}h^{-1}\msol) \right)^{-0.081}(1+z)^{-0.71}$ is adopted in \cite{mead2020} to compute $r_{\rm s}$. In this work, $r_{\rm s}$ computed directly by $\rockstar$ are used, and are scaled using the modification term in Equation \eqref{eq:gasconcentration}.} 
In the presence of astrophysical feedback effects, the concentration parameter is modified such that:
\begin{equation}\label{eq:gasconcentration}
c(M_{\rm vir})\rightarrow c(M_{\rm vir})\left[1+\epsilon_{1}+(\epsilon_{2}-\epsilon_{1})\frac{f_{\rm bnd}}{\Omega_{\rm b}/\Omega_{\rm m}}\right],
\end{equation}
where $\epsilon_{2}$ is set to 0, and $\epsilon_{1}$ is a two-component function with two free parameters $\epsilon_{1}=\epsilon_{1,a}+\epsilon_{1,b}z$.
Next, the gas temperature profile is computed using:
\begin{equation}
T_{\rm gas}(M_{\rm vir},r)=T_{\rm vir}(M_{\rm vir})\frac{{\rm ln}(1+r/r_{\rm s})}{r/r_{\rm s}},
\end{equation}
where the virial temperature is obtained using:
\begin{equation}\label{eq:Tvir}
\frac{3}{2}k_{\rm B}T_{\rm vir}(M_{\rm vir})=\alpha\frac{GM_{\rm vir} m_{\rm p}\mu_{\rm p}}{ar_{\rm vir}},
\end{equation}
where $m_{\rm p}$ and $\mu_{\rm p}$ are the proton mass and the mean gas particle mass divided by the proton mass.\footnote{Following \citet{mead2020}, the values  $\mu_{\rm p}=0.61$ and $\mu_{\rm e}=1.17$ are adopted in this work in Equations \eqref{eq:Tvir} and \eqref{eq:gasdens2pe}.} 

The electron pressure profile can be separated into bound and unbound gas components. The bound component is given by: 
\begin{equation}\label{eq:gasdens2pe}
P_{\rm e}^{\rm bnd}(M_{\rm vir},r)=\frac{\rho^{\rm bnd}_{\rm gas}(M_{\rm vir},r)}{m_{\rm p}\mu_{\rm e}}k_{\rm B} T_{\rm gas}(M_{\rm vir},r),
\end{equation}
where $\mu_{\rm e}$ is the mean gas particle mass per electron divided by the proton mass. For the unbound component, the treatment from \cite{schneider2015} is adopted, and the gas profile is computed as:
 \begin{equation}\label{eq:rho_gas_unb}
 \rho^{\rm ejc}_{\rm gas}(M_{\rm vir}, r)=\frac{M_{\rm vir}}{(2\pi r_{\rm ejc}^{2})^{3/2}}\exp\left[-\frac{r^{2}}{2r_{\rm ejc}^{2}}\right],
 \end{equation}
 where the functional form is derived from the presumption that astrophysical feedback effects induce velocity kicks on particles that follow a Maxwell-Boltzmann distribution. $r_{\rm ejc}$ in Equation \eqref{eq:rho_gas_unb} is obtained by solving: 
 \begin{equation}
 1.0-{\rm erf}\left[\frac{\eta_{\rm b}r_{\rm esc}}{\sqrt{2}r_{\rm ejc}}\right]+ \sqrt{\frac{2}{\pi}}\frac{\eta_{\rm b}r_{\rm esc}}{r_{\rm ejc}}{\rm exp}\left[-\frac{\eta_{\rm b}^{2}r_{\rm esc}^{2}}{2r_{\rm ejc}^{2}}\right]\equiv\frac{\Omega_{\rm m}}{\Omega_{\rm b}}f_{\rm ejc}(M_{\rm vir}),
 \end{equation}
 where $\eta_{\rm b}=0.5$, $f_{\rm ejc}=1-f_{\rm bnd}$, and $r_{\rm esc}$ is defined as:
 \begin{equation}
 r_{\rm esc}=t_{0}v_{\rm esc}\sim t_{0}r_{200}\sqrt{\frac{8\pi}{3}G\Delta_{200}\rho_{\rm crit}}\sim 0.5r_{200}\sqrt{\Delta_{200}},
 \end{equation}
 with $\Delta_{200}=200$. To obtain the electron pressure profile for the unbound component, Equation \eqref{eq:gasdens2pe} is used, but the temperature is assumed to be $T_{\rm w}=T_{\rm w,0}\exp(T_{\rm w,1}z)$ as defined in the \cite{mead2020} model.\\

Three different tSZ maps are produced in this work, based on the three variants of the \cite{mead2020} model, which are based on AGN heating temperatures of $10^{7.6}$, $10^{7.8}$, and $10^{8.0}$ K in the \bahamas{} simulation. The best-fit model parameter values for the three models are summarized in Table  \ref{tab:meadprofile}, and will be referred to as the \bahamas{} 7.6, 7.8, and 8.0 models hereafter.
\begin{table}
\begin{center}
\begin{tabular}[t]{cccc}
\toprule
Parameter & $10^{7.6}$\ [{\rm K}] & $10^{7.8}$\ [{\rm K}] & $10^{8.0}$\ [{\rm K}]\\
\midrule
$\epsilon_{1,a}$ & -0.1002 & -0.1065 & -0.1253  \\
$\epsilon_{1,b}$ & -0.0456 & -0.1073 & -0.0111 \\
$\Gamma$  & 1.1647 & 1.1770 & 1.1966 \\
$\log_{10}(M_{0}/h^{-1}{\rm M}_{\sun})$   & 13.1949 &  13.5937 & 14.2480 \\
$\alpha$  & 0.7642 & 0.8471 & 1.0314\\ 
$\beta$   & 0.6     & 0.6 & 0.6\\
$\log_{10}(T_{{\rm w},0}/{\rm K})$ & 6.6762 & 6.6545& 6.6615 \\
$T_{{\rm w},1}$ & -0.5566 & -0.3652 & -0.0617\\
\bottomrule
\end{tabular}
\caption{Best-fit parameters for the tSZ halo model from \protect\cite{mead2020}.}
\label{tab:meadprofile}
\end{center}
\end{table}
Finally, both the bound and unbound pressure profiles are integrating along the line of sight to produce Compton-$y$ maps:
\begin{equation}
y^{\rm bnd/unb}(\hat n)=\frac{\sigma_{\rm T}}{m_{\rm e}c^2}\int_{\rm los} dl\ P_{\rm e}^{\rm bnd/unb} , 
\end{equation}
where $\sigma_{\rm T}$ is the Thomson cross-section and $m_{\rm e}$ is the electron mass, and the bound and unbound components are added to produce the total Compton-$y$ map:
\begin{equation}
y^{\rm tot}(\hat{n})= y^{\rm bnd}(\hat{n}) + y^{\rm unb}(\hat{n}).
\end{equation}
The Compton-$y$ maps are also converted into temperature units using Equation \eqref{eq:y2uk} depending on the application.

\subsection{Kinetic Sunyaev-Zel'dovich effect}
The late-time kSZ effect arises from the Doppler shifting of CMB photons induced by the bulk motion of the free electrons in galaxy groups and clusters \citep{sunyaev1980}. While the amplitude of the tSZ effect is determined by the integrated electron pressure along the line of sight, the strength of the kSZ effect is determined by the electron number density and the bulk velocity of the gas with respect to the CMB. The effect is also less mass-dependent, making it a promising method for mapping gas distribution, particularly in low-mass systems \citep{schaan2020}.

Several studies have measured the kSZ effect, most notably by examining the characteristic signature imprinted on the background CMB by pairs of in-falling galaxy clusters, also known as the pair-wise kSZ effect \citep{hand2012,soergel2016,li2018}. Other approaches, such as the projected-field \citep{hill2016,ferraro2016,kusiak2021} and the velocity-field reconstruction \citep{schaan2016} methods can be used to measure the kSZ effect. Although detections using these methods have only been made at the level of a few sigmas thus far, the signal-to-noise ratio is expected to improve significantly with upcoming galaxy surveys such as the Dark Energy Spectroscopic Instrument (DESI; \citealt{desi}) and {\it SPHEREx} \citep{spherex}.

The temperature fluctuation caused by the kSZ effect can be  computed using:
\begin{equation}
\left(\frac{\Delta T}{T}\right)_{\rm kSZ}=-\frac{\sigma_{\rm T}}{c}\int_{\rm los}\hspace{0.1em} dl \hspace{0.2em} n_{\rm e}\hspace{0.2em} v_{\rm los},
\end{equation}
where $n_{\rm e}$ is the number density of electrons and $v_{\rm los}$ is the line-of- sight velocity of the electrons.

To produce maps of the kSZ effect, the gas profiles obtained from the tSZ modelling in Equations \eqref{eq:gasdensityprofile} and \eqref{eq:rho_gas_unb}
are used to estimate the number density of electrons $n_{\rm e}=\frac{\rho_{\rm gas}}{m_{\rm p}\mu_{\rm e}}$.  This is then multiplied with the pixelized particle velocity maps to create kSZ maps.\footnote{In practice, computing the exact number density of gas around particles with velocity requires simultaneous processing of halos and particles and is hence computationally intensive. Here, it is assumed that shells of 25 $\mpch{}$ are sufficiently thin such that the approximation $\sum \int{n_{\rm e} v_{\rm los}}dl \approx  \sum v_{\rm los} \int{n_{\rm e} }dl$ is valid.}  The differential temperature maps are computed for each shell, and are integrated up to $z=3$.

\subsection{Cosmic Infrared Background}\label{sec:model_cib}
The CIB is composed of diffuse infrared emission from dust surrounding star-forming galaxies at $z\sim2$. Astrophysical quantities such as their emission properties,  cosmic star formation rate and the environment that these star-forming galaxies live in can be studied by characterizing the CIB.

The CIB has been studied using infrared satellites such as {\it IRIS} \citep{iris}, {\it Herschel} \citep{griffin2010,pilbratt2010} and 353/545/857 GHz channels of {\it Planck}. However, while not dominant, it is also detected at lower frequency channels, including those used in high-resolution ground-based CMB experiments \citep{dunkley2013,addison2013,george2015, reichardt2021}. As a result, accurate CIB modelling is critical for investigating contamination in, for example, Compton-$y$ maps. However, predicting the amplitude of the CIB at lower frequencies is known to be difficult since its amplitude at those frequencies is subdominant relative to the CMB and other CMB secondary effects, and is therefore poorly constrained by data.

The first step in implementing the CIB is to use the code $\um$\footnote{\url{https://bitbucket.org/pbehroozi/universemachine}} \citep{behroozi2019} to assign a star formation rate (SFR) and $M_{*}$ to each individual halo in the simulation. These two quantities (and many more) are assigned by $\um$ based on halos' properties such as redshift, potential well depth, and mass assembly history, and are constrained by observational measurements such as stellar mass functions, SFRs, and quenched fractions \citep{behroozi2019}.

\begin{figure}
\begin{center}
\includegraphics[width=1.0\linewidth]{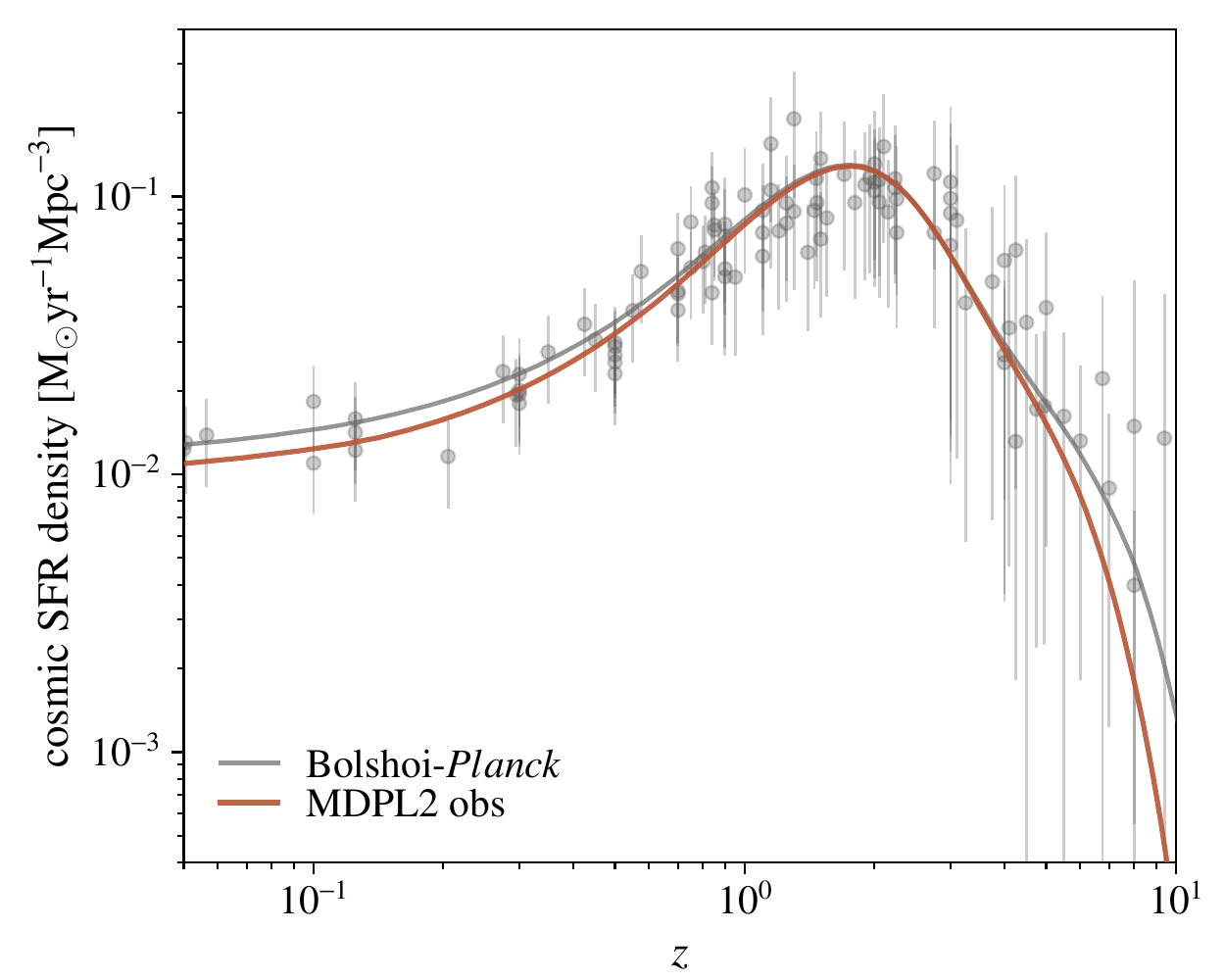}
\caption{Cosmic SFR density as a function of redshift for MDPL2 (orange line), \bp{} (grey line), and the observational data points used to constrain the \um{} model (grey points). The data points are a compilation of \citet{salim2007,bauer2013,whitaker2014,zwart2014,karim2011,kajisawa2010,schreiber2015,tomczak2016,salmon2015,smit2014,labbe2013,mclure2011}.}
\label{fig:um_cosmicsfr}
\end{center}
\end{figure}

In this work, the \um{} catalogues created by applying the \um{} model and model parameters provided with data release 1\footnote{These can be found at: \url{http://halos.as.arizona.edu/UniverseMachine/DR1/umachine-dr1-code.tar.gz}} to the MDPL2 simulation are used.\footnote{The star formation catalogues can be accessed at \url{http://halos.as.arizona.edu/UniverseMachine/DR1/MDPL2_SFR/}}  A comparison of the cosmic SFR density from \bp{} (the simulation which the calibration procedure was performed on), and  MDPL2 simulation (the results of applying the model)  as well as observational data that were used to constrain the model are shown in Figure \ref{fig:um_cosmicsfr}. The resulting total cosmic SFR density\footnote{Here, the {\it observed} star formation rate which includes scatter is used instead of the {\it true} star formation rate from the output of $\um$. } for MDPL2 agrees well with the results from that of $\bp$, despite having slightly worse mass resolution ($1.38\times 10^{8}\ h^{-1}{\rm M}_{\sun}$ versus  $1.51\times10^{9}\ h^{-1}{\rm M}_{\sun}$). Following \citet{bethermin2017}, only galaxies with an observed SFR of less than $1000\ \msol {\rm yr}^{-1}$ are considered in this work, because observations at 870 $\mum{}$ and in radio wavelengths indicate a rapid decline in number density of galaxies with SFR beyond that rate \citep{dacunha2015,barger2014,barger2017} , and galaxies exceeding SFR=$1000\ \msol {\rm yr}^{-1}$ are rare.\footnote{More recent studies from ALMA (e.g. \cite{casey2021}) have detected galaxies with SFRs exceeding $1000\ \msol{\rm yr}^{-1}$, and hence this assumption will be validated with future observations.}

The bolometric infrared ($8-1000$\  $\mu {\rm m})$ luminosity of galaxies is known to be proportional to their SFR \citep{kennicutt1998}. However, a simple linear relationship between $L_{\rm IR}$ and SFR is known to break down for low-mass galaxies with low dust content \citep{heinis2014,wu2017}. An improved model takes this into account and can be written as:
\begin{equation}\label{eq:sfr_LIR}
L_{\rm IR}=\frac{\rm SFR}{K_{\rm IR}+K_{\rm UV}10^{-{\rm log}_{10}{\rm IRX}(M_{*})}},
\end{equation}
where the values\footnote{A Kroupa-to-Chabrier IMF conversion factor of 0.92 extrapolated from the conversion factor given in \cite{madau2014} is applied.} $K_{\rm UV}=1.53\times10^{-10},K_{\rm IR}=1.38\times10^{-10}$ from  \citet{kennicutt2012} are adopted and IRX is the infrared excess, for which the prescription from \citet{bouwens2020},  derived from ALMA observations of dust-enshrouded star-formation galaxies is adopted. In particular, their prescription:  
\begin{equation}\label{eq:IRX}
{\rm IRX}=(M_{*}/M_{\rm s})^{\alpha_{\rm IRX}},
\end{equation}
with $M_{\rm s}=10^{9.63} {\rm M}_{\sun}$ and $\alpha_{\rm IRX}=1.37$ derived using the \textsc{Prospector} code is used in this work.

\begin{figure}
\begin{center}
\includegraphics[width=1.0\linewidth]{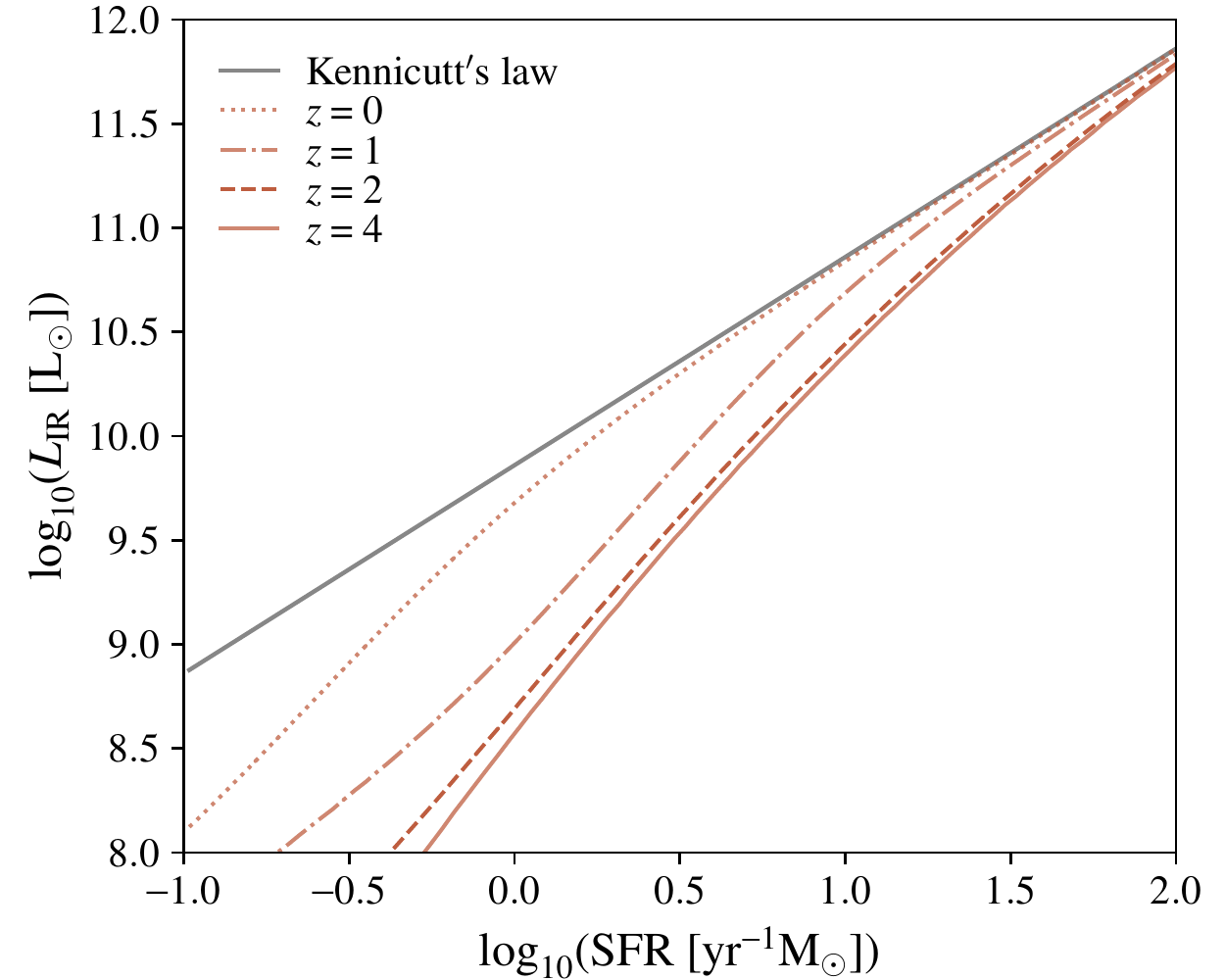}
\caption{The relationship between SFR and $L_{\rm IR}$ in the range $8-1000\ \mu {\rm m}$, at redshifts $z=0, 1, 2, 4$ calculated using Equations \eqref{eq:sfr_LIR} and \eqref{eq:IRX}. As a reference, the solid grey line corresponds to the Kennicutt's law \citep{kennicutt1998,kennicutt2012}.}
\label{fig:um_lir}
\end{center}
\end{figure}

The spectral energy distribution (SED) of infrared galaxies is assumed to be a modified blackbody with a power-law transition \citep{planck2015xv}: 
\begin{equation}\label{eq:sed_greybody}
\Phi(\nu,T_{\rm d})=
\begin{cases}
\left[\exp(\frac{h\nu}{kT_{\rm d} })-1\right]^{-1}\nu^{\beta_{\rm d}+3}, & (\nu \leq \nu') \\
\left[\exp(\frac{h\nu'}{kT_{\rm d} })-1\right]^{-1}\nu'^{\beta_{\rm d}+3}\left(\frac{\nu}{\nu'}\right)^{-\alpha_{\rm d}}, & (\nu>\nu')
\end{cases}
\end{equation}
where $\nu'$ is the frequency at which the SED transitions from a modified blackbody to a power-law,\footnote{The exact value of $\nu'$ is chosen such that the derivatives on both ends match at $\nu'$. } and $T_{\rm d}$ is the SED dust temperature. $\beta_{\rm d}$ is assumed to be correlated with $T_{\rm d}$ such that:
\begin{equation}
\beta_{\rm d}=\frac{\zeta_{\rm d}}{0.4+0.008 T_{\rm d}},
\end{equation}
which is a relationship derived by  \citet{dupac2003}, but with a free amplitude scaling parameter $\zeta_{\rm d}$ to take into account for the difference between local and high-redshift infrared sources. Dust temperature is modelled using the relation:
\begin{equation}\label{eq:Tdust}
T_{\rm d}=A_{\rm d}\left( \frac{L_{\rm IR}}{M_{\rm dust}}\right)^{1/(4+\beta_{\rm d})},
\end{equation}
where $A_{\rm d}$ is a free parameter that is marginalized over, and $M_{\rm dust}$ is the estimated dust mass of galaxies based on the relation \citep{donevski2020}:\footnote{As noted in \cite{donevski2020}, this relation uses the assumption that $\log(M_{\rm dust}/M_{\rm gas})\propto \log(Z/{\rm Z}_{\odot})$ for massive galaxies.}
\begin{equation}\label{eq:ratio_dts}
\frac{M_{\rm dust}}{M_{*}} \propto \frac{M_{\rm gas }}{M_{*}}\times Z_{\rm gas}.
\end{equation}
To obtain the gas-to-stellar mass ratio, the empirical relation from \cite{tacconi2020} is used:
\begin{align}\label{eq:tacconi2020}
&\log_{10}\left(\frac{M_{\rm gas}}{M_{*}} \right)=\nonumber\\
&A+B({\rm log}_{10}(1+z)-F)^{2}+C{\rm log}_{10}(\Delta_{\rm MS})+D(\log_{10}M_{*}-10.7),
\end{align}
with $\Delta_{\rm MS}={\rm sSFR}/{\rm sSFR}_{\rm MS}$, where sSFR is the specific star formation rate (i.e., ${\rm SFR}/M_{*}$) and  ${\rm sSFR}_{\rm MS}$ is the main sequence specific star formation rate (for which, the model from \citealt{tacconi2018} is used) and the best-fit values $A, B, F, C, D= (0.06, -3.33, 0.65, 0.51, -0.41)$ from \cite{tacconi2020} are adopted. The gas metallicity $Z_{\rm gas}$ in Equation \eqref{eq:ratio_dts} is estimated using an empirical relation from \cite{sanders2021}:
\begin{equation}
Z_{\rm gas}\equiv12+\log_{10}({\rm O/H})=8.80+0.188y-0.220y^2-0.0531y^3
\end{equation}
where $y=\mu_{0.60}-10$ with
\begin{equation}
\mu_{0.60}=\log_{10}(M_{*}/{\rm M}_{\sun})-0.60\log_{10}({\rm SFR}/{\rm M}_{\sun}{\rm yr}^{-1}).
\end{equation}
In addition to the fiducial model described above, two modifications are made:
\begin{enumerate}[leftmargin=\parindent,align=left,labelwidth=\parindent,labelsep=0pt]
\item The prescription above results in a $M_{\rm dust}/M_{*}$ relation that underestimates the dust mass fraction at low redshifts, resulting in high dust temperatures relative to observations. Therefore, one of the parameters in Equation \eqref{eq:tacconi2020} (namely, $B$) is allowed to vary, and the data is allowed to self-constrain this parameter.
\item Results from \cite{donevski2020} show hints of a mildly decreasing trend in the $M_{\rm dust}/M_{*}$ ratio at $z>2$ (see Figures 4 and 5 of that work). While this trend has yet to be confirmed, a redshift-dependent suppression factor $f(z)$ is introduced to account for the possibility of this effect:
\begin{equation}
f(z) = \begin{cases}
  1  & z<2 \\
  (1-0.05z)^\alpha & z>2.
\end{cases}
\end{equation}
\end{enumerate}
Once the bolometric luminosity and the SED of an infrared source are determined, the luminosity of the source observed in a specific frequency band is calculated as: 
\begin{equation}
L_{\nu}=L_{\rm IR,bolo}(M,z)\frac{\int d\nu \hspace{0.05cm}\Phi(\nu,T_{\rm d}) \tau(\nu)}{\int d\nu\hspace{0.05cm} \Phi(\nu,T_{\rm d})},
\end{equation}
where $\tau(\nu)$ is the spectral transmission function of a particular channel for a given experiment. The flux is then calculated using the source's luminosity, comoving radial distance, and redshift:
\begin{equation}\label{eq:cibflux}
S_{\nu}=\frac{L_{\nu(1+z)}}{4\pi \chi^{2}(1+z)},
\end{equation}
and a Gaussian scatter of $\sigma=0.25$ is applied to the flux values, which is necessary to reconcile the observed number counts. This procedure is repeated for all the sources in the lightcone, and a source catalogue is produced.

A Markov Chain Monte Carlo (MCMC) is used to determine the best-fit values for the model parameters  $A_{\rm d}$, $\zeta_{\rm d}$, $B$, and $\alpha$ that produce maps with auto- and cross-spectra that match with observational data. This is accomplished as follows:
\begin{enumerate}[leftmargin=\parindent,align=left,labelwidth=\parindent,labelsep=0pt]
\item Starting with the \um{} catalogue, Equations (\ref{eq:sfr_LIR}-\ref{eq:cibflux}) are used to generate CIB maps at 75 parameter points within the ranges tabulated in Table \ref{tab:cibemul}, and the 353/545/857 GHz auto-/cross-power spectra are computed at each point.
\item A power spectrum emulator is built using these training points, allowing auto- and cross-power spectra to be computed quickly at any given point within the parameter space spanned by the training points.
\item The auto- and cross-spectra measured from the \cite{lenz2019} 353/545/857\ GHz CIB maps\footnote{Maps with HI column density threshold of $1.5\times 10^{20} \ {\rm cm}^{-1}$, which are the most conservative maps provided, are used here. While the auto-spectra of 353 and 545 GHz channels shift minimally between the $1.5\times 10^{20} \ {\rm cm}^{-1}$  and $2.5\times 10^{20} \ {\rm cm}^{-1}$ maps, we find that the 857 GHz channel shifts by a non-negligible amount} are compared with emulated power spectra in the multipole range of $200<\ell<3000$. To explore the parameter space, a simple Gaussian likelihood is used, with:
\begin{eqnarray}
\label{eq:likelihood}
\ln \mathcal{L}(\vec{d}| \vec{\theta}) = -\frac{1}{2} \left[  \vec{d} - \vec{d}_{\rm e}(\vec{\theta})\right]^T \Sigma^{-1} \left[ \vec{d} - \vec{d}_{\rm e}(\vec{\theta}) \right],
\end{eqnarray}
where $\vec{d}$ and $\vec{d}_{\rm e}$ are the measured band powers from data and those predicted by the emulator at a given parameter point $\vec{\theta}$, respectively. The covariance matrix $\Sigma$ is estimated using a simple Gaussian approach: 
 \begin{equation}
 \Sigma^{XYWZ}=\frac{1}{(2\ell+1)\Delta{\ell} f_{\rm sky}}[C_{\ell}^{XY}C_{\ell}^{WZ}+C_{\ell}^{XW}C_{\ell}^{YZ}]
 \end{equation}
where $W,X,Y,Z\in\{353/545/857\}$ GHz, and the $C_{\ell}$ are the auto- and cross-spectra measured from \citet{lenz2019} CIB maps. While this calculation is a simplification of the true covariance, it does account for systematic effects such as galactic dust because power spectra from observational data are used.
\end{enumerate}

\begin{figure}
\begin{center}
\includegraphics[width=1.0\linewidth]{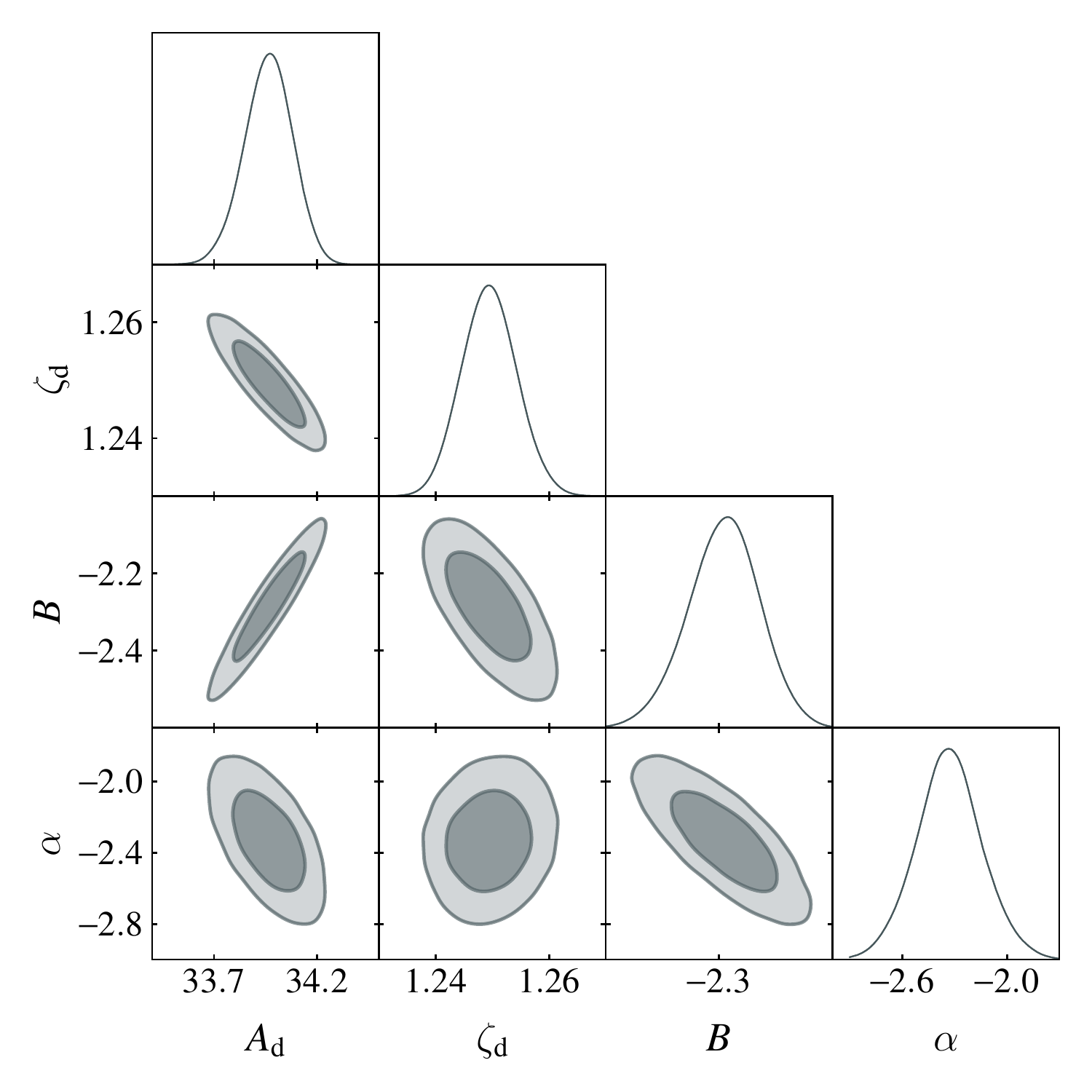}\llap{\makebox[\wd1][l]{\raisebox{0.25cm}{\includegraphics[bb=155 503 600 -600]{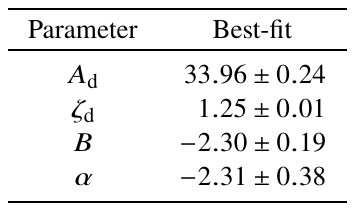}}}}
\caption{Constraints obtained for the CIB model parameters $A_{\rm d}$, $\zeta_{\rm d}$, $B$ and $\alpha$ using the measured 353/545/857 GHz auto- and cross-spectra from \citet{lenz2019} CIB maps as the data vector. }
\label{fig:cib_mcmc}
\end{center}
\end{figure}

\begin{figure}
\begin{center}
\includegraphics[width=1.0\linewidth]{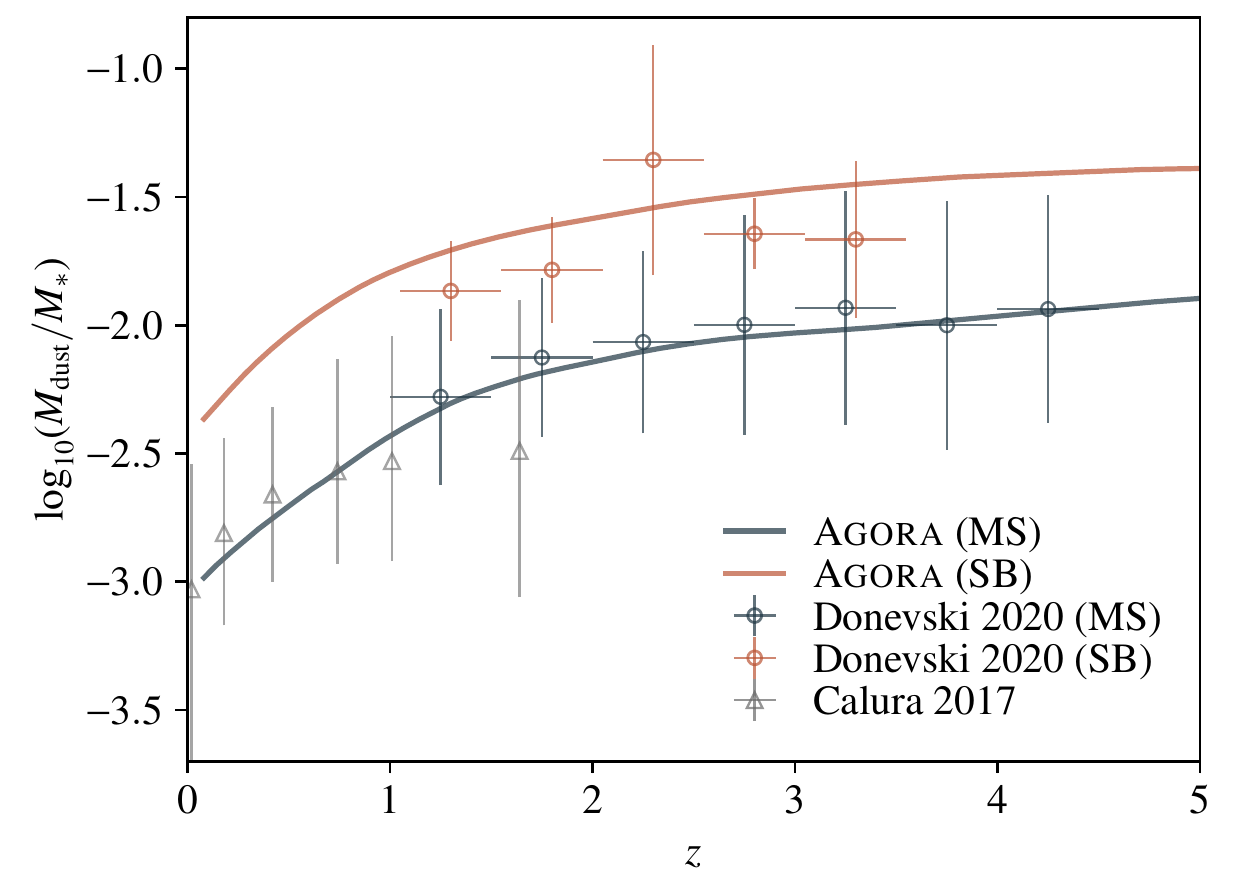}
\caption{The derived ${\rm log}_{10}(M_{\rm dust}/M_{*})$ - $z$ relation compared against observational data from \protect\cite{donevski2020} and \protect\cite{calura2017}. The amplitudes of the derived relation are scaled arbitrarily, since the $A_{\rm d}$ parameter in Equation \eqref{eq:Tdust} encapsulates both the normalization of the ${\rm log}_{10}(M_{\rm dust}/M_{*})$ ratio and the relationship between $T_{\rm d}$ and $L_{\rm IR}/M_{\rm dust}$. The samples are split into main sequence (MS) and starburst (SB) galaxies, defined as ${\rm sSFR}<4\times{\rm sSFR}_{\rm MS}$ and ${\rm sSFR}>4\times{\rm sSFR}_{\rm MS}$ respectively, where ${\rm sSFR}_{\rm MS}$ is the main sequence specific star formation rate from \protect\cite{tacconi2018}. }
\label{fig:Mdust_z}
\end{center}
\end{figure}

Figure \ref{fig:cib_mcmc} depicts the constraints obtained on the parameters $A_{\rm d}$, $\zeta_{\rm d}$, $B$, and $\alpha$. Using the best-fit values, maps and catalogues of the CIB for various experiments and frequency channels are generated. Furthermore, astrophysical quantities such as $M_{\rm dust}/M_{*}$ are computed and compared with data to ensure that the model is physically plausible (shown in Figure \ref{fig:Mdust_z}).

\begin{table}
\begin{center}
\begin{tabular}[t]{cc}
\toprule
Parameters & Ranges \\
\midrule
$A_{\rm d}$     & [33.5,34.5]  \\
$\zeta_{\rm d}$ & [1.07,1.29]  \\
$B$             & [-2.70,-2.00]\\
$\alpha$        & [-2.9, -1.7] \\
\bottomrule
\end{tabular}
\caption{Parameter ranges used to construct the CIB power spectrum emulator.}
\label{tab:cibemul}
\end{center}
\end{table}

\subsection{Radio galaxies}
While the majority of point sources detected by CMB experiments at frequencies $\nu\gtrsim150$ GHz are dusty star-forming galaxies, radio galaxies dominate in number at lower frequencies. A significant number of these radio sources are thought to be blazars. These sources are known to be spatially correlated with CIB/tSZ to some extent (see e.g., \citealt{delvecchio2021}), and are polarized \citep{datta2018,gupta2019}. While the population of radio sources detectable at a few GHz is well studied, the physical properties of radio galaxies detectable at frequencies used in CMB experiments ($\nu\gtrsim90\ {\rm GHz}$) are less well understood due to their low number density.

Radio galaxies are implemented in the lightcone as follows:
\begin{enumerate}[leftmargin=\parindent,align=left,labelwidth=\parindent,labelsep=1pt]
\item Starting from the $M_{*}$ values in the $\um$ catalogue, bulge masses are estimated using the empirical fitting function from \cite{zhang2021}:
\begin{equation}
M_{\rm bulge} =\frac{f_{z}(z)M_{*}}{1+\exp(-1.13(\log_{10}M_{*}-10.2))},
\end{equation}
where
\begin{equation}
f_{z}(z)=\frac{z+2}{2z+2},
\end{equation}
which is derived from the measured $M_{\rm bulge}/M_{*}$ ratios of the SDSS catalogue in the redshift range $0<z<2$ \citep{lang2014,mendel2014}.
\vspace{0.2cm}

\item From the estimated bulge masses, the results of the $M_{\rm bulge}$-$M_{\rm BH}$ relation computed by \textsc{Trinity}\footnote{ \url{https://github.com/HaowenZhang/TRINITY}} \citep{zhang2021} are used to assign $M_{\rm BH}$ to all the halos in the lightcone.
\vspace{0.2cm}
\item Using the active black hole mass function computed by \trinity{}, a flag that represents the presence of an active black hole is assigned to a subset of halos, such that the overall number density of active black holes matches with the number density predicted by \textsc{Trinity} (see Figures 7 and 8 for comparison of their model with observational data from \citealt{schulze2010,schulze2015,kelly2013}).

\vspace{0.2cm}
\item The 5 GHz luminosity function from \cite{tucci2021} is used to assign radio luminosities to the halos. The luminosity function is statistically modelled on the basis of physical and phenomenological relationships that connect the physical properties of the supermassive black hole at their centre via the fundamental plane of black hole activity. The luminosity function is divided into flat/steep spectrum AGNs, as well as low and high-kinetic-mode (LK/HK) AGNs (classified by their Eddington ratios $\lambda=L_{\rm bolo}/L_{\rm Edd} <0.01$ or $\lambda >0.01$ respectively). In each redshift shell, the total number of halos (i.e., the sum of flat low-kinetic, steep low-kinetic, flat high-kinetic, and steep high-kinetic sources) is matched with the total number of active black holes predicted by \trinity{} obtained in the previous step. In this work, only the flat-spectrum sources are kept in the simulation, since steep-spectrum sources are expected to be dim at frequencies  above $90$ GHz.

\vspace{0.2cm}
\item Since the correlation between supermassive black hole mass and radio luminosity is known to be small \citep{woo2002}, each halo with an active black hole is assigned a luminosity drawn at random from the luminosity function, and their fluxes are computed:
\begin{equation}
S_{\nu}=\frac{L_{\nu}}{4\pi D_{L}^2(z)}(1+z)^{1+\alpha_{5}},
\end{equation}
where $\alpha_{5}=0$ is adopted for the flat-spectrum sources that are considered in the simulation. The recovered 5 GHz differential number counts are shown in Figure \ref{fig:counts_5ghz}.

\begin{figure}
\begin{center}
\includegraphics[width=1.0\linewidth]{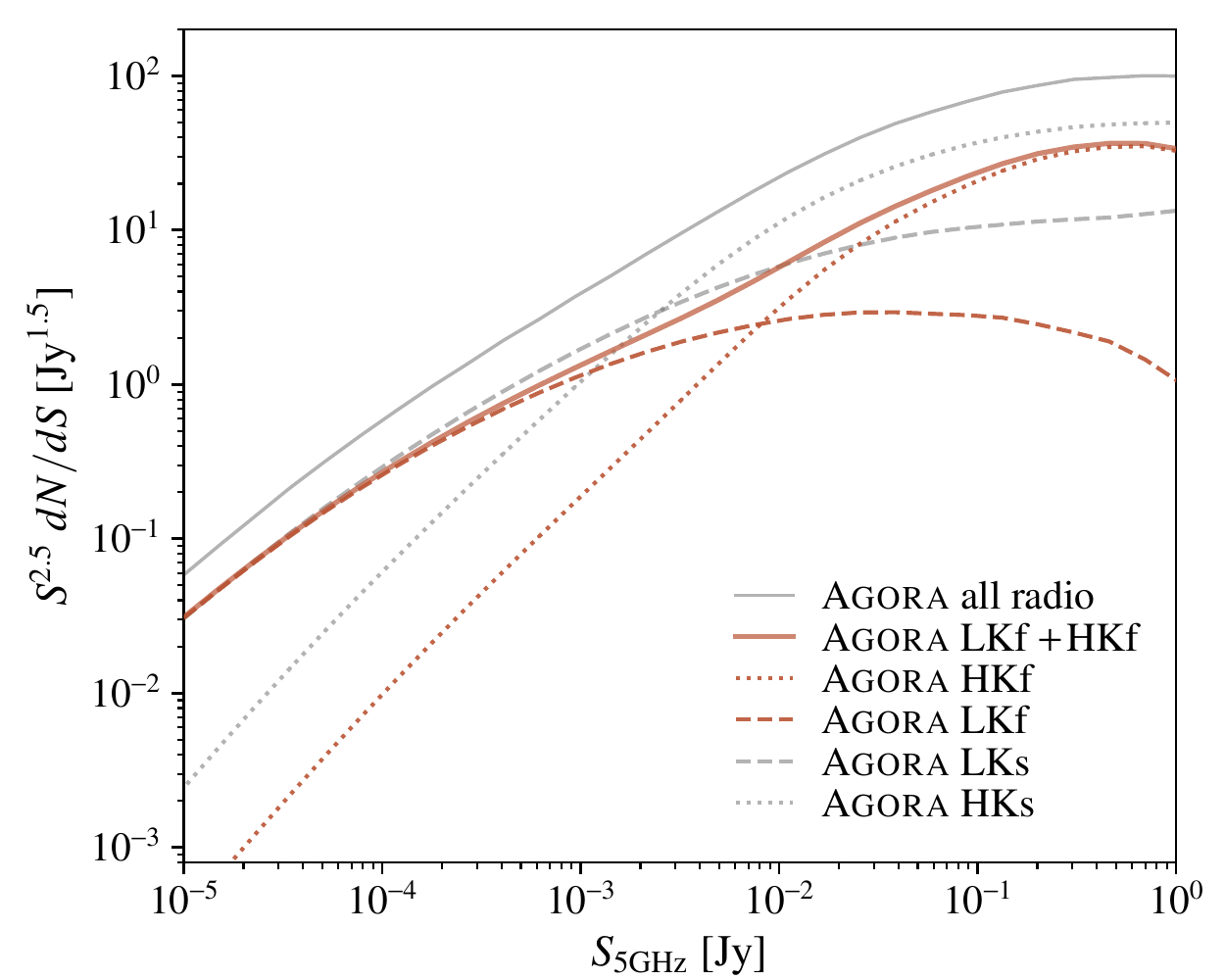}
\caption{The simulation's differential number counts of radio sources as a function of 5 GHz flux. The solid orange line depicts the sum of the LK and HK flat spectrum sources, which are used to model the radio sources at higher frequencies. Contributions from HK and LK are indicated by dotted/dashed orange lines. The number counts for steep-spectrum  HK and LK sources are shown as dotted/dashed grey lines for reference, and the total of all source types is shown as the solid grey line.}
\label{fig:counts_5ghz}
\end{center}
\end{figure}
\vspace{0.2cm}
\item The 5 GHz fluxes are translated into 150 GHz fluxes using the scaling relation:
\begin{equation}\label{eq:radio_freqscaling}
S_{\rm 150GHz}=S_{\rm 5GHz}\left(\frac{\rm 150\ [GHz]}{\rm 5\ [GHz]}\right)^{\alpha_{5}^{150}}.
\end{equation}
Here, the LK and HK samples are considered separately, and the spectral index $\alpha_{5}^{150}$ is modelled to have a Gaussian distribution with mean $\mu_{\alpha}$ and spread $\sigma_{\alpha}$. The best-fit values for $\mu_{\alpha}^{\rm LK}$, $\sigma_{\alpha}^{\rm LK}$, $\mu_{\alpha}^{\rm HK}$, and $\sigma_{\alpha}^{\rm HK}$ are determined by comparing the model\footnote{In practice, $P(S_{\rm 150GHz}|S_{\rm 5GHz})$ is estimated using a lognormal distribution with ${\rm loc}=\sigma_{\alpha}\log(\nu/5{\rm GHz})$ and ${\rm scale}=S_{\rm 5GHz}\exp(\mu_{\alpha}\log(\nu/5 {\rm GHz}))$, and sum the probability density function to produce a deterministic estimate of the differential number counts $dN/dS_{\rm 150GHz}$.} with the number counts from \cite{planck2018liv}\footnote{The catalogue is available at \url{http://pla.esac.esa.int/pla/aio/product-action?SOURCE_LIST.NAME=COM_PCCS_PCNT_R2.00.fits}.} and \cite{everett2020}\footnote{The catalogue is available at  \url{https://pole.uchicago.edu/public/data/everett20/}.} at 143/150\footnote{Given the large uncertainties in the measured differential counts for \planck{} due to the small sample size of radio sources in the high flux regime, the small bandpass mismatch between the two experiments is disregarded.} GHz in the flux range $6 < S_{\rm 143/150GHz} < 1000$ mJy. This is accomplished by employing a method similar to that described in Section \ref{sec:model_cib}, but comparing the observed and model differential number counts instead of $C_{\ell}$.
Due to the flux limit of \citet{everett2020}, only a weak constraint on $\sigma_{\alpha}^{\rm LK}$ is obtained, and is found to be highly degenerate with $\mu_{\alpha}^{\rm LK}$. Therefore, a fixed value of $\sigma_{\alpha}^{\rm LK}=0.1$ is adopted, and the fitting procedure is allowed to constrain $\mu_{\alpha}^{\rm LK}$. While the choice of this value is somewhat arbitrary, it has been verified that the results are not impacted by small changes to this value.

Figure \ref{fig:radio_mcmc} shows the constraints obtained for $\mu_{\alpha}^{\rm LK}$, $\mu_{\alpha}^{\rm HK}$ and  $\sigma_{\alpha}^{\rm HK}$ and the best-fit values obtained. These best-fit values are then used to generate a probability distribution $P(S_{\rm 150GHz}|S_{\rm 5GHz})$ from which 150 GHz fluxes are drawn at random. \vspace{0.2cm}

\begin{figure}
\begin{center}
\includegraphics[width=1.0\linewidth]{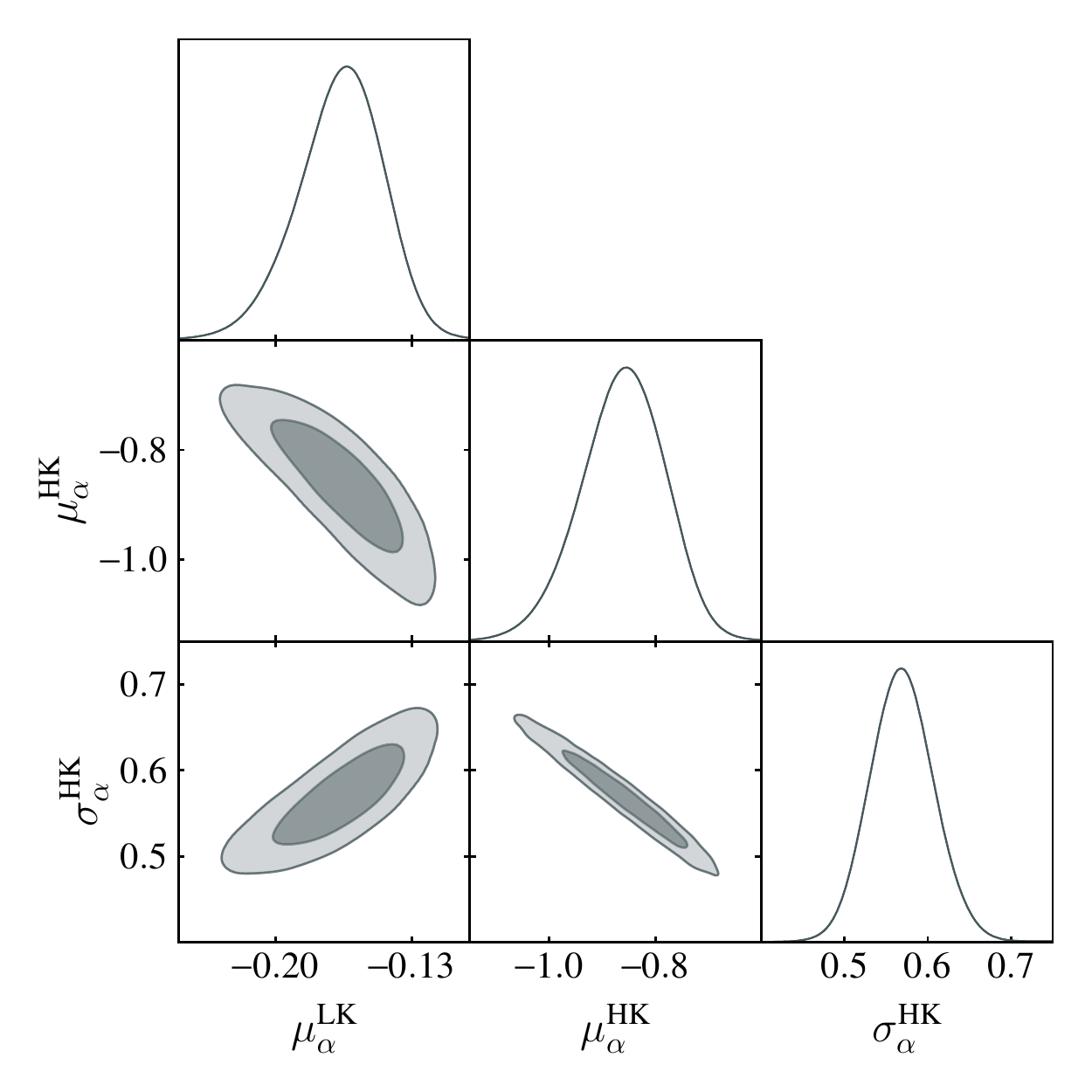}\llap{\makebox[\wd1][l]{\raisebox{0.25cm}{\includegraphics[bb=158 505 600 -600]{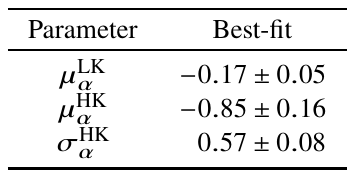}}}}
\caption{The parameter constraints obtained for  $\mu_{\alpha}^{\rm LK}, \sigma_{\alpha}^{\rm HK}$, $\mu_{\alpha}^{\rm HK}$. $\sigma_{\alpha}^{\rm LK}$ has been fixed to 0.1 as described in the text.}
\label{fig:radio_mcmc}
\end{center}
\end{figure}

\item Once 150 GHz fluxes are assigned to each halo, 95 and 220 GHz fluxes are also calculated by scaling the 150 GHz fluxes using the spectral slopes $\alpha^{95}_{150}$ and $\alpha^{220}_{150}$ (defined in the same way as Equation \ref{eq:radio_freqscaling}), measured from \citet{everett2020}. The distributions of $\alpha_{150}^{95}$ and $\alpha_{150}^{220}$ for the LK sources are estimated using sources in the $6<S_{\rm 150GHz}<20$ mJy range, while the distributions for the HK sources are estimated using the sources in the $100<S_{\rm 150GHz}<1000$ mJy range. A simple multivariate Gaussian distribution is used to jointly model these distributions with $\mu=-0.9,-0.8,-0.65,-0.75$ for 
$\alpha_{150}^{95,{\rm LK}}$, $\alpha_{150}^{220,{\rm LK}}$, $\alpha_{150}^{95,{\rm HK}}$ and $\alpha_{150}^{220,{\rm LK}}$ respectively, and the  covariances between $\alpha_{150}^{95}$, $\alpha_{150}^{220}$ are computed directly from data:
\begin{align}
\Sigma_{95-150,150-220}^{\rm LK}&=\begin{bmatrix}
0.20 & -0.0086 \\
-0.0086 & 0.75 
\end{bmatrix},\\
\Sigma_{95-150,150-220}^{\rm HK}&=\begin{bmatrix}
0.062 & 0.026  \\
0.026 & 0.077
\end{bmatrix}.
\end{align}
Figure \ref{fig:radio_alpha_95_220} shows that such a simplistic model can provide a reasonable fit (with an exception of $\alpha_{150}^{220,{\rm LK}}$, for which the fit is mildly offset due to the skewness of its distribution). The 95 and 220 GHz fluxes are randomly drawn from this model and are assigned to each halo.
\item The polarization properties of radio sources are determined by their polarization angles $\psi$ (which are drawn randomly from a uniform distribution in the range [$0$, $2\pi$]), and polarization fractions $p$ (drawn from a truncated Gaussian distribution with $\mu_{\rm truncG}=0.027$ and $\sigma_{\rm truncG}=0.572$ from \citealt{datta2018}). The stokes parameters $Q$ and $U$ are then assign by taking \citep{lagache2019}:
\begin{align}
Q=\hspace{1em}&I\times p \cos(2\psi),\\
U=-&I\times p \sin(2\psi).
\end{align}
\end{enumerate}

\begin{figure}
\begin{center}
\includegraphics[width=1.0\linewidth]{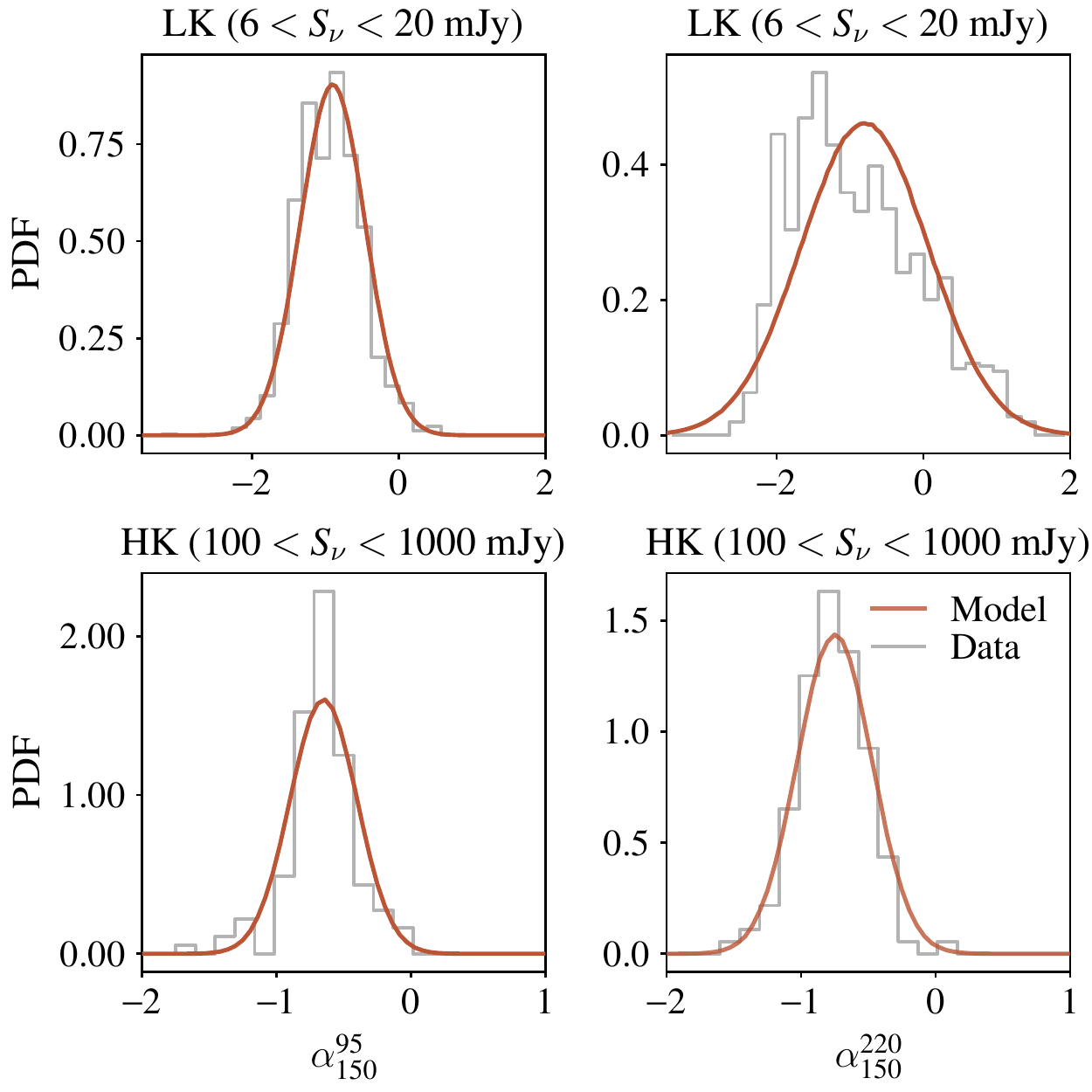}
\caption{Distributions of the spectral slopes $\alpha^{95}_{150}$ and $\alpha^{220}_{150}$ measured from the SPT-SZ point-source catalogue \citep{everett2020} in two flux regimes, $6<S_{\nu}<20$ mJy and $100<S_{\nu}<1000$ mJy, dominated by LK and HK sources, respectively (grey lines). The distributions obtained from the adopted model are also shown (orange lines). }
\label{fig:radio_alpha_95_220}
\end{center}
\end{figure}

\subsection{Galaxies}\label{sec:method_gal}
Implementing realistic galaxies in a dark matter-only simulation necessitates the pasting of observables such as magnitude and colour \citep{derose2019,wechsler2021}. These properties are difficult to assign precisely because they are heavily influenced by the galaxy evolution model used and the environment in which the galaxies reside. The characteristics of a given galaxy sample are heavily influenced by the selection cuts based on these properties. Because each survey has its own set of selection criteria (along with associated systematics), it is beyond the scope of this work to create realistic mock catalogues for each galaxy survey. However, it is noted that because $\um{}$ catalogues are employed in this work, astrophysical quantities such as $M_{*}$ and SFR are available in the lightcone. These quantities allow for a more realistic implementation of galaxies than a simple mass-based halo occupation distribution (HOD). Such work will be considered in the future.

While realistic galaxies with astrophysical properties are not implemented, maps of galaxy overdensity $\delta_{\rm g}$  are generated by multiplying the dark matter density shells with linear galaxy bias values and summing the shells, weighted by the redshift distributions of the clustering galaxy sample i.e.,
\begin{equation}\label{eq:lensgal}
\delta_{\rm g}^{j} (\hat{n}) = \sum_{i} d\chi W_{\delta}^{j}(\chi_{i})\delta_{\rm DM}^{i}(\chi_{i},\hat {n}),
\end{equation}
where $\delta_{\rm DM}^{i}$ is the $i$-th dark matter density shell and $W^j_{\delta}$  
is the galaxy density kernel given by:
\begin{align}\label{eq:W_gal}
W^j_{\delta}(\chi)=b_{\rm g}^{j}\frac{dn_{l}^{j}}{dz}\frac{dz}{d\chi}.
\end{align}

\subsection{Galaxy weak lensing}\label{sec:method_gwl}

As with CMB weak lensing, the intervening mass between distant galaxies and us causes background galaxies to appear slightly distorted. The distribution of matter in the Universe can be mapped by studying these distortions and drawing statistical inferences from a large number of galaxies.

To date, many optical surveys have used the galaxy weak lensing effect to place constraints on our understanding of cosmology. The Canada-France-Hawaii-Telescope Lensing Survey (CFHTLenS; \citealt{heymans2012}) was among the first to conduct a wide-field weak lensing analysis, and other experiments such as the Red Cluster Sequence Lensing Survey (RCSLenS; \citealt{hildebrandt2016}), the Kilo-Degree Survey (KiDS; \citealt{kids}), the Dark Energy Survey (DES; \citealt{desoverview}), and the Subaru Telescope Hyper Suprime-Cam Subaru Strategic Program (HSC-SSP; \citealt{aihara2018}) have since provided improvements. In the coming years, surveys such as the Rubin Observatory Legacy Survey of Space and Time  (LSST;  \citealt{lsstsciencebook}), {\it Euclid} \citep{euclid} and the Nancy Grace Roman Space Telescope ({\it Roman}; \citealt{roman1,roman2}) will measure the shapes of billions of galaxies with exquisite precision, allowing us to place tighter constraints on cosmological parameters and models.

\subsubsection{Mock catalogue generation}\label{sec:mock_shearcatalog}

The signal component of the galaxy weak lensing effect is generated by stacking the convergence and shear maps obtained from ray tracing in Section \ref{sec:cmblensing}. To produce noiseless full-sky galaxy weak lensing maps, the individual lensing shells are weighted by the redshift distributions of the background galaxies and are stacked:
\begin{equation}\label{eq:shearmap}
\gamma^{j}_{\rm g_{1},g_{2},\kappa_{\rm g}}(\hat{n})=\sum_{i}d\chi \frac{dn_{\rm s}^{j}}{dz}\frac{dz}{d\chi}\gamma_{g_{1},g_{2},\kappa_{\rm g}}^{i}(\chi_{i},\hat{n}).
\end{equation}
Realistic shape catalogues are generated by adding shape noise to the signal, which is achieved in two ways. One method is to add randomly rotated ellipticities measured from data to the simulation's signal component:
 \begin{align}
 e_{1}&=\gamma_{1}/(1-\kappa_{\rm g})+e_{1}^{\rm rot},\\
 e_{2}&=\gamma_{2}/(1-\kappa_{\rm g})+e_{2}^{\rm rot},
 \end{align}
 where the first term is the reduced shear and the second is the noise term:
\begin{align}
 e^{\rm rot}_{1}&=\hspace{0.5em}e_{1}'\cos(2\varphi)+e'_{2}\sin(2\varphi),\\
 e^{\rm rot}_{2}&=-e_{1}'\sin(2\varphi)+e'_{2}\cos(2\varphi),
 \end{align}
 where $e'_{1},e'_{2}$ are the ellipticities in the data catalogue, and $\varphi$ is a random angle drawn from a uniform distribution in the range $0$ and $2\pi$.
 
 The second approach is to draw random $e_{1},e_{2}$ values from a normal distribution with variance $\sigma_{\rm e}^{2}$, such that $\boldsymbol{e}^{\rm rand}=\mathcal{N}(0,\sigma_{\rm e})+i\mathcal{N}(0,\sigma_{\rm e})$ and 
 \begin{equation}
 e_{1}+ie_{2}=\boldsymbol{e} = \frac{\boldsymbol\gamma+\boldsymbol{e}^{\rm rand}}{1+\boldsymbol\gamma^{*}\boldsymbol{e}^{\rm rand}},
 \end{equation}
 where $\boldsymbol{\gamma}=\gamma_{1}+i\gamma_{2}$. The first approach is used to generate mock galaxy catalogues for publicly data sets (e.g., DES-Y1 and DES-Y3), while the second approach is used to generate mock catalogues for surveys whose data are not yet available (e.g., LSST).

\subsubsection{Intrinsic alignment}
A key systematic in galaxy weak lensing measurements is the effect of intrinsic alignment, in which intrinsic galaxy shapes and orientations become correlated as a result of the stretching or compression of initially spherically collapsing mass in a local gravitational gradient or galaxies acquiring angular momentum through tidal torquing \citep{troxel2015}.

In this paper, the effect of intrinsic alignment is implemented using the nonlinear alignment model (NLA; \citealt{hirata2004,bridle2007}), which is based on the redshift-dependent intrinsic alignment kernel and the matter density shells \citep{fluri2019}:
\begin{equation}\label{eq:kappa_IA}
\kappa^{j}_{\rm IA}(\hat{n})=\sum_{i} d\chi\ W^{j}_{\rm IA}(\chi_{i})\delta^{i}_{\rm DM}(\chi_{i},\hat{n}),
\end{equation}
where $\delta_{\rm DM}^{i}$ is the $i$-th dark matter density shell\footnote{For the dark matter density shells the {\it lensed} density shells are used.} and $W^{j}_{\rm IA}(\chi)$ is the intrinsic alignment effect's redshift kernel:
\begin{equation}
W^{j}_{\rm IA}(\chi)= -A_{\rm IA}C_{1}\rho_{\rm crit,0}\frac{\Omega_{\rm m,0 }}{D(z)}\left(\frac{1+z}{1+z_{0}}\right)^{\eta}\frac{dn^{j}_{\rm s}}{dz}\frac{dz}{d\chi},
\end{equation}
with $C_{1}=5\times10^{-14} (h^{2} {\rm M}_{\sun}{\rm Mpc}^{3})^{-2}$, $z_{0}=0.62$ and the best-fit parameter values $A_{\rm IA}=0.44$, $\eta=-0.7 $ from \citet{desy1_3x2} are used. $D(z)$ is the growth factor normalized to unity at the present day.
The negative sign implies that an amplification of this effect leads to a reduction in the measured cosmic shear correlation amplitude. It is noted that the amplitude, $A_{\rm IA}$, is known to be sensitive to the type of source galaxies used \citep{samuroff2019}, and therefore IA maps must be generated for each galaxy sample of interest separately.

Finally, the IA convergence maps are converted to shear maps in harmonic-space using the relation \citep{chang2018}:
\begin{align}
\gamma^{j}_{{\rm IA},\ell m}=-\sqrt{\frac{(\ell+2)(\ell-1)}{\ell(\ell+1)}}\kappa^{j}_{{\rm IA},\ell m},
\end{align}
and are converted into shear maps using:
\begin{equation}
\gamma^{j}_{{\rm IA},1}+i\gamma^{j}_{{\rm IA},2}=\sum_{\ell m}\gamma^{j}_{{\rm IA}, \ell m}\ _{2}Y_{\ell m},
\end{equation}
from which the values of  $\gamma_{{\rm IA},1}$ and $\gamma_{{\rm IA},2}$ are interpolated from, at the locations of galaxies, and are added to the shear catalogues.

\subsection{Lensing of observables}\label{sec:secondary_lensing}
All the observables in the lightcone are lensed, 
shell by shell, using the integrated lensing field up to the shell of the source. Two different lensing procedures are used, depending on the characteristics of the background source. For continuous sources such as tSZ and kSZ, \lenspix{} is used to deflect the background maps. For point sources, it is more efficient to apply the deflection and magnification operation per object, since the number density of sources is significantly less than the number of pixels.  For these sources, deflection and magnification are applied following the implementation described in \cite{lewis2005,diegopalazuelos2020}, which is described next.

Photons observed to be travelling from the direction $\vec{n}_{l}=(\theta_{l},\varphi_{l})$, were originally travelling from   $\vec{n}_{\rm u}=(\theta_{\rm u},\varphi_{\rm u})=(\theta',\varphi_{l}+\delta\varphi). $\footnote{As noted in \cite{diegopalazuelos2020}, Equation \eqref{eq:base_deflection} is assuming the Born approximation, which is calculated through potential gradients along the unperturbed path, which starts to fail above $\ell>10000$, which is in the regime beyond the scope of these simulations.} The mapping can be written as:
\begin{equation}\label{eq:base_deflection}
\tilde{X}(\vec{n}_{l})=X(\vec{n}_{\rm u})=X(\vec{n}_{l}+\vec{\alpha}(\vec{n}_{l})),
\end{equation}
where $X\in T/Q/U/\delta/\gamma_{1}/\gamma_{2}$ and $\vec{\alpha}(\vec{n})$ is the deflection angle.
The unlensed position can be computed using:
\begin{align}
\cos(\theta')&=\cos(\alpha)\cos(\theta_{l})-\sin(\alpha)\sin(\theta_{l})\cos(\beta),\\
\sin(\delta \varphi)&=\frac{\sin(\beta) \sin(\alpha)}{\sin(\theta')},
\end{align}
where $\alpha=|\vec{\alpha}(\vec{n})|$ is the magnitude of the deflection along the direction of $\vec{\alpha}(\vec{n})$, and $\beta$ is the direction between $\vec{\alpha}$ and $\vec{e}_{\theta}$ at the lensed position $\vec{n}_{l}$. For spin-2 fields (such as polarization of radio sources or galaxy shear), an additional  rotation is applied to the tensor $\tilde{P}(\vec{n}_{l})=e^{2i \eta} P(\vec{n}_{\rm u})$ where:
\begin{equation}
e^{2i \eta} =\frac{2(\alpha_{\theta}+\alpha_{\varphi}A)^{2}}{\alpha^{2}(1+A^{2})}-1+\frac{2i(\alpha_{\theta}+\alpha_{\varphi}A)(\alpha_{\varphi}-\alpha_{\theta}A)}{\alpha^{2}(1+A^2)}
\end{equation}
and 
\begin{equation}
A=\frac{\alpha_{\varphi}}{\alpha\sin\alpha\cot\theta_{l}+\alpha_{\theta}\cos\alpha}.
\end{equation}
In addition, the effect of magnification:
\begin{equation}
\mu=1/((1-\kappa)^{2}-|\gamma|^{2}),
\end{equation}
is also applied to all point sources, where $\kappa$ and $\gamma$ are taken from the outputs of ray tracing.  In practice, when the denominator is close to zero, the magnification value of a given pixel can take on large values. To avoid such scenarios, smoothing is applied (by taking the median of the neighbouring 8 pixels) to pixels with $\mu>100$, and their absolute values are taken to avoid flipping of signs. Finally, the unlensed flux values of sources are multiplied by the values of magnification inferred at the undeflected locations of the sources $S_{\nu}^{\rm len}=|\mu|S_{\nu}^{\rm unl}$.

\subsection{Masks}\label{sec:mask}
\begin{figure}
\begin{center}
\includegraphics[width=1.0\linewidth]{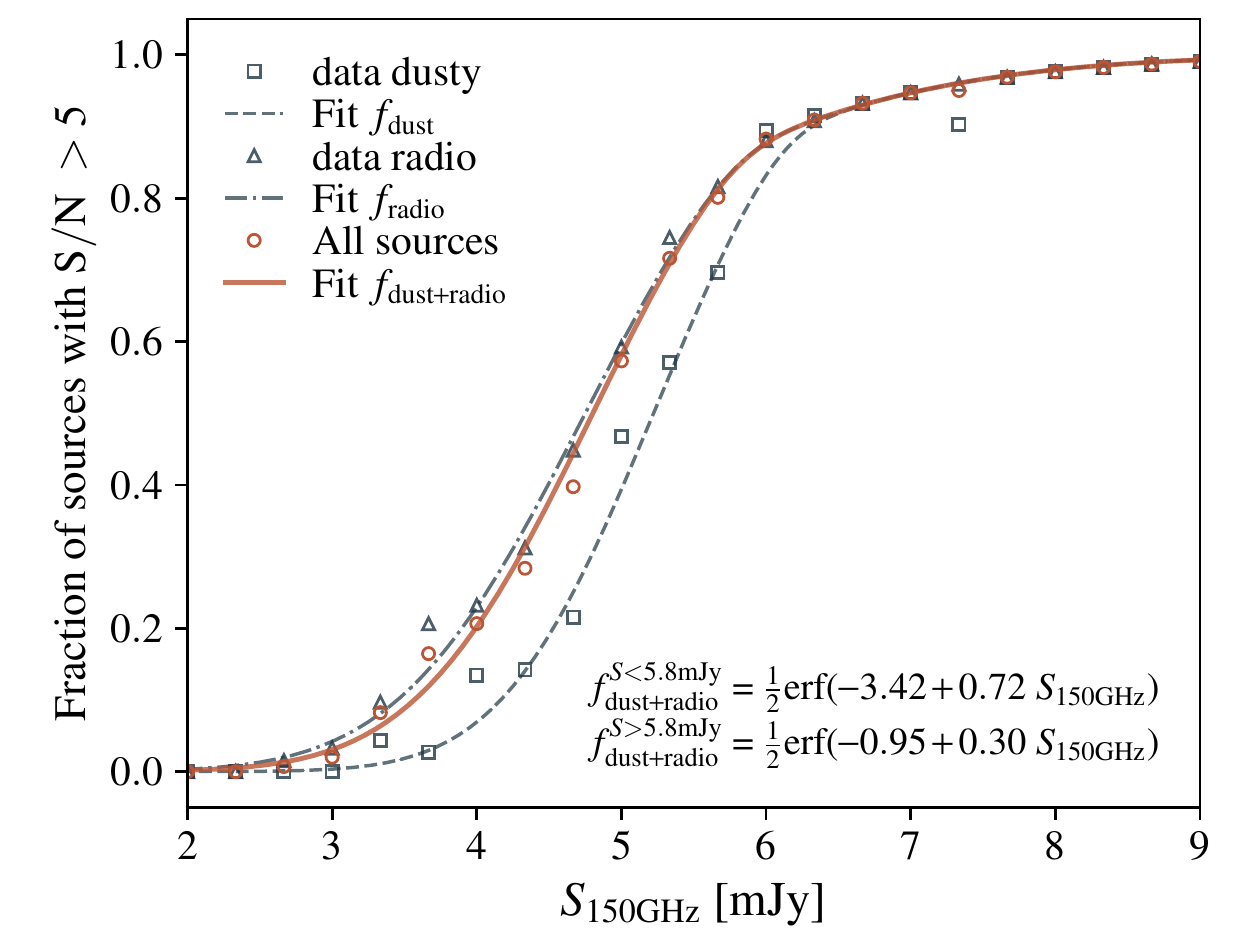}
\caption{ Fraction of sources with ${\rm S/N}>5$ as a function of 150 GHz flux calculated using the point-source catalogue from \protect\cite{everett2020}, split into dusty and radio sources, as well as the total population (navy square, navy triangle and orange circles points respectively). The empirical fitting functions given in Equation \eqref{eq:masking_thres}, for dusty, radio and the total population are overlaid (dashed navy, dot dashed navy and solid orange lines, respectively). }
\label{fig:masking_prob}
\end{center}
\end{figure}

To generate a point source mask based on a detection significance threshold, the threshold must be converted into an equivalent flux value. However, the relationship between detection threshold and flux is not always one-to-one, and an empirical relationship describing the fraction of sources with detection significance greater than a given threshold as a function of flux must be calculated using existing catalogues. This function can be parameterized as: 
\begin{equation}\label{eq:masking_thres}
f=\frac{1}{2}{\rm erf}(a+b\ S_{\nu}),
\end{equation}
where
\begin{equation}
(a,b) = \begin{cases}
  (-3.42,0.72) & S<5.8 {\rm mJy} \\
  (-0.95,0.30) & S>5.8 {\rm mJy}
\end{cases}
\end{equation}
for SPT-SZ at 150 GHz using a 5$\sigma$ threshold, as shown in Figure \ref{fig:masking_prob}. Although the masking probability functions for radio and dusty sources are found to be slightly different, the values obtained for $a,b$ using the total population are used. The same method is used to generate \planck{} point-source mask, with $(a,b)=(0.013,3.5),(0.02,3.5),(0.016,2.0),(0.006,0.46)$ for the $100-353$ GHz channels, and a flat flux cut of 353, 513 mJy are used for the 545 and 857 GHz channels, respectively.

To generate masks for clusters, local maxima in the Compton-$y$ map are found using the \texttt{hotspot} utility provided with the \healpix{} package, and the locations are ranked according to their peak Compton-$y$ values. The top $n$ locations of this list are used as proxies for galaxy cluster locations, where $n$ is the number of galaxy clusters above a specific detection threshold for a given experiment. The masking radius for each individual cluster is determined by increasing the masking radius until the fraction of pixels within the aperture with Compton-$y$ values greater than 5\% of the peak value falls below 0.95.

\section{Validation}\label{sec:secondary_validation}
In this section, the maps and catalogues generated in the previous sections are validated by comparing the measured auto-/cross-spectra and number counts with analytical predictions and measurements from data.

\subsection{$\kappa_{\rm CMB}$ auto-spectra}\label{sec:validation_kcmb}
\begin{figure}
\begin{center}
\includegraphics[width=1.0\linewidth]{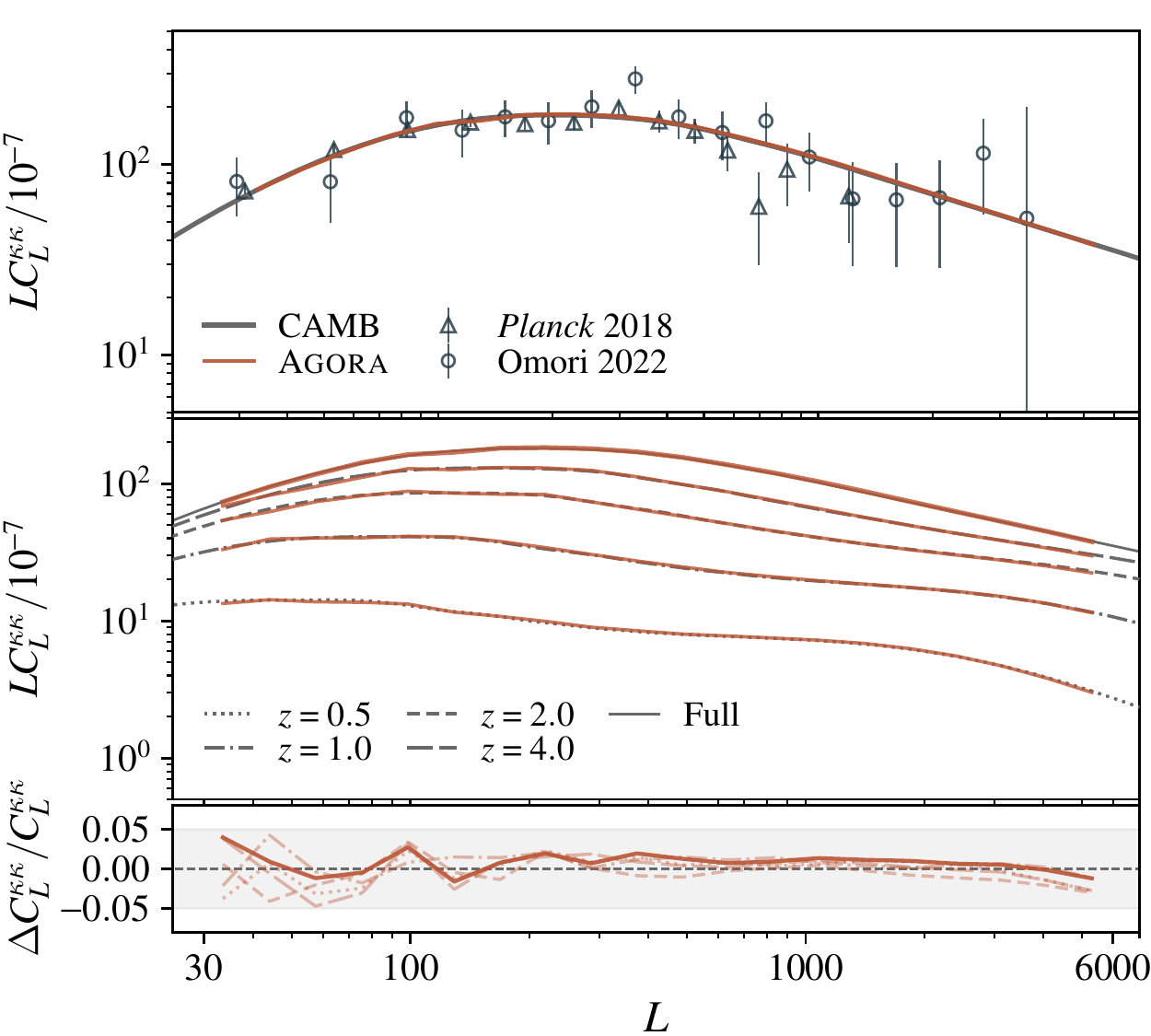}
\caption{
{\bf Upper:} The simulation's measured total CMB lensing power spectrum (orange solid line) compared to measurements from \citet{planck2018viii} (grey triangles) and \citet{omori2022} (grey circles), as well as the analytic calculation from CAMB (black solid line). The amplitude of the low-$L$ correction described in Section \ref{sec:cmblensing} is smaller than the visible range of this plot.  {\bf Center:} The measured CMB lensing spectrum at $z=0.5, 1.0, 2.0$, and $z=4.0$ (solid orange lines)  compared with calculations from CAMB (black dotted, dot-dashed, dashed, and long-dashed lines). {\bf Lower:} Same as above, but shown as fractional differences.
}
\label{fig:cls_mdpl2_clkk}
\end{center}
\end{figure}

In Figure \ref{fig:cls_mdpl2_clkk}, the total CMB lensing spectrum of the simulation is compared with an analytical CMB lensing spectrum computed using CAMB\footnote{While there exists updated nonlinear parametrization such as \textsc{HMCode} which takes into account for baryonic feedback effects, the Takahashi \citep{takahashi2012} prescription is used  
here, since this the lensing maps are constructed from a dark matter-only simulation.}, and an agreement of better than 5\% is found in the multipole range $30<L<5000$. The simulation CMB lensing spectrum is additionally compared with measured spectra from \citet{planck2018viii} and \citet{omori2022}, and is also found to be consistent with those measurements. Furthermore, CMB lensing spectra computed at various redshifts are found to be consistent with the analytically computed spectra to within 5\%, which is within the quoted uncertainties of \textsc{HaloFit} (see \citealt{takahashi2012}). 

\subsection{Lensed Primary CMB}
\begin{figure*}
\begin{center}
\includegraphics[width=1.0\linewidth]{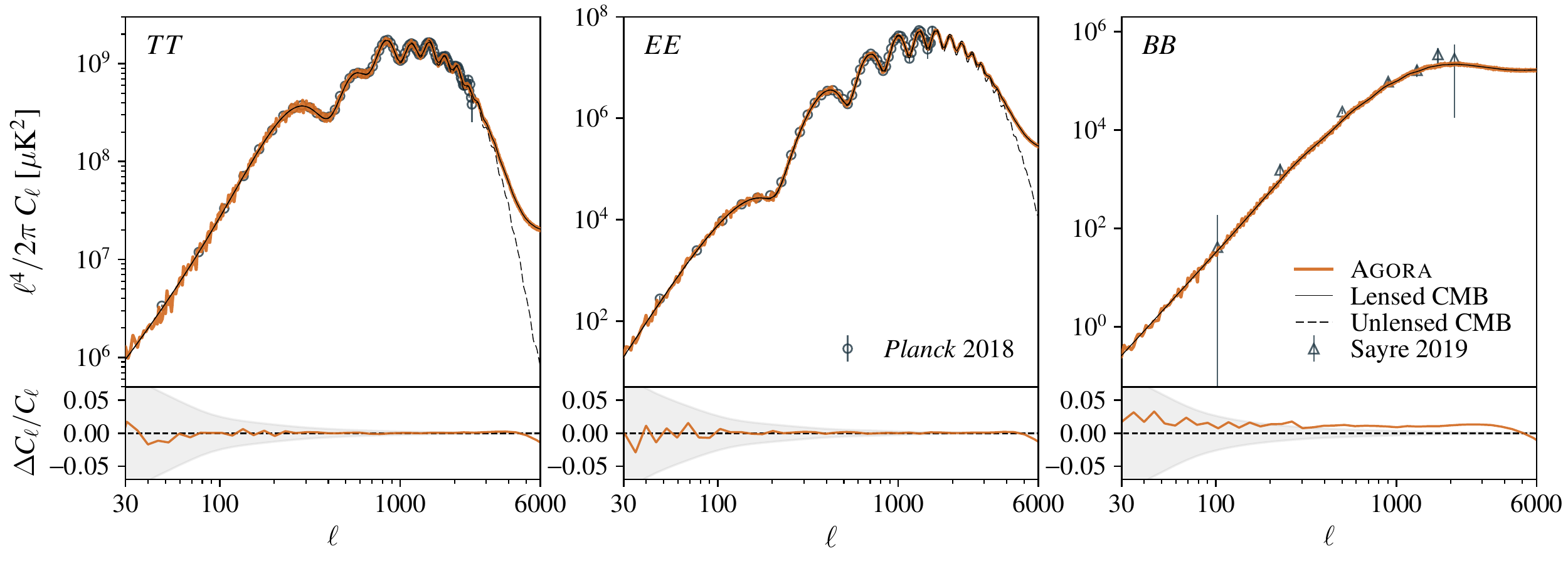}
\caption{{\bf Upper:} The measured lensed CMB spectra for temperature and polarization averaged over 20 CMB realizations (orange lines), compared with analytical calculations of lensed and unlensed spectra from CAMB (solid and dashed grey lines).  {\bf Lower:} Fractional difference between the average spectrum in the simulation and the theoretical lensed spectrum from CAMB (orange lines). The grey shaded region represents the variance of the 20 realizations.}
\label{fig:cls_mdpl2_teb}
\end{center}
\end{figure*}

The measured lensed $C_{\ell}^{TT}$, $C_{\ell}^{EE}$, and $C_{\ell}^{BB}$ power spectra averaged over 20 realizations\footnote{Since there is only one CMB lensing field available, 20 independent realizations of the primary CMB, lensed by a single lensing potential obtained from ray tracing are used here.} and their fractional differences compared with analytical predictions from CAMB are shown in Figure \ref{fig:cls_mdpl2_teb}. The agreement for $C_{\ell}^{TT}$ and $C_{\ell}^{EE}$ are better than 1\% in the multipole range $30<\ell<5000$. For $C_{\ell}^{BB}$, a bias of $\sim1\%$ relative to the theoretical prediction is observed, which can be attributed to the marginally high convergence spectrum amplitude in the range $200<L<2000$ as seen in Figure \ref{fig:cls_mdpl2_clkk}.

\begin{figure}
\begin{center}
\includegraphics[width=1.0\linewidth]{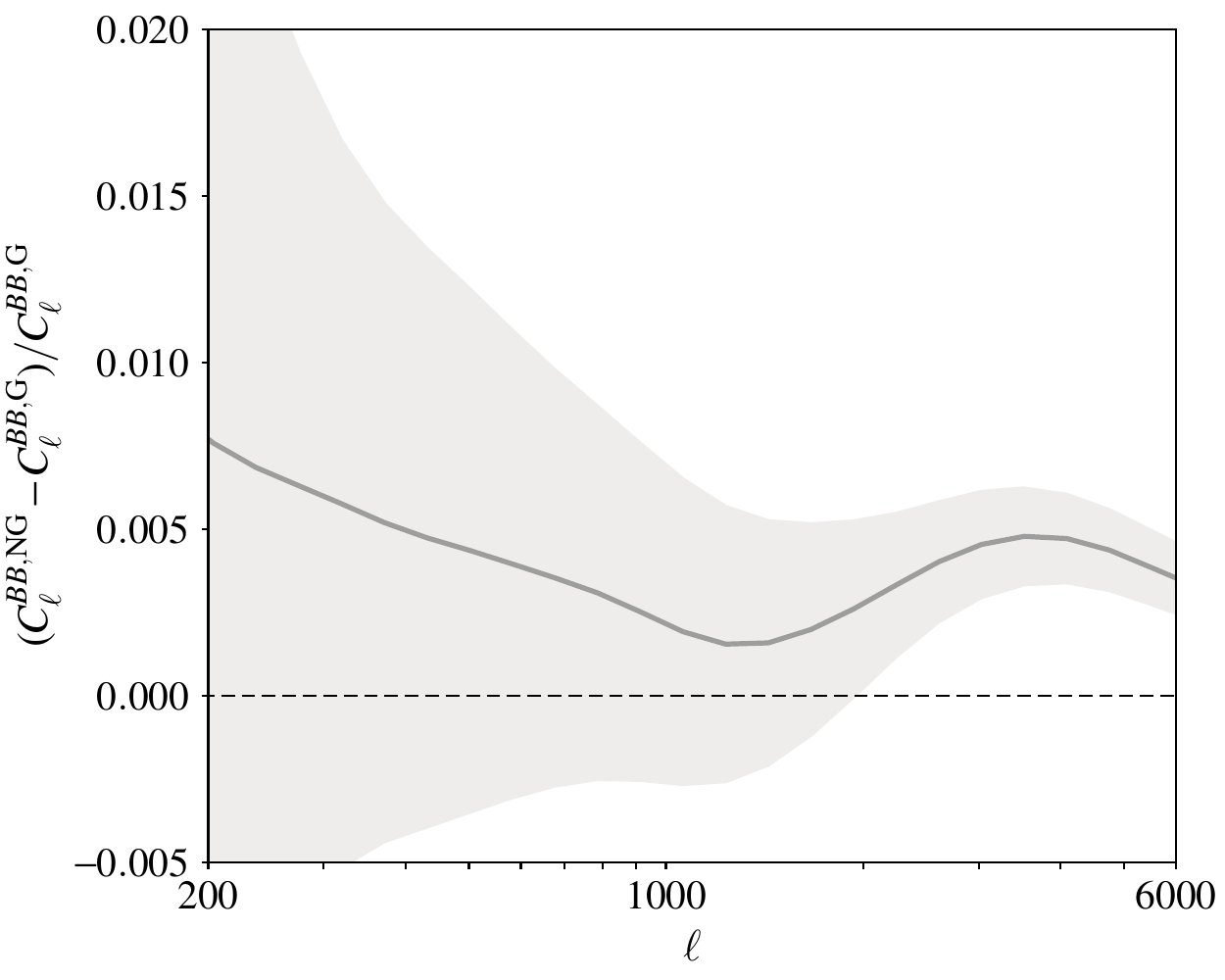}
\caption{The fractional difference in $C_{\ell}^{BB}$ produced by a non-Gaussian and Gaussian lensing fields, with statistical uncertainties from a CMB-S4-like experiment shown (assuming a noise level of $1\mu {\rm K}$-${\rm arcmin}$, beam FWHM $\theta_{\rm FWHM}=1'$, and $f_{\rm sky}=0.4$). }
\label{fig:cls_mdpl2_kcmbNG_kcmbG_ratio}
\end{center}
\end{figure}

In Figure \ref{fig:cls_mdpl2_kcmbNG_kcmbG_ratio}, the average $B$-mode spectrum measured from CMB maps lensed using Gaussian realizations of $\kcmb$ is compared with the average $B$-mode spectrum measured from the fiducial lensed CMB maps (i.e.,  $T/Q/U$ maps deflected using the non-Gaussian $\kcmb$ map from ray tracing). An excess of $\sim0.5\%$ is observed at $\ell>1500$, which is consistent with the results by \citet{takahashi2017}. Since this effect is only present above $\ell=1500$, and the statistical uncertainties for $C_{\ell}^{BB}$ are large for multipoles below that, the assumption of Gaussian lensing fields when measuring $C_{\ell}^{BB}$ is likely to be valid even for futuristic surveys such as CMB-S4\footnote{The comparison is made against the diagonal elements of a simple Gaussian covariance matrix assuming a noise level of $1\mu {\rm K}$-${\rm arcmin}, \theta_{\rm FHWM}=1', f_{\rm sky}=0.4$}.

\subsection{tSZ auto-spectra}
\begin{figure}
\begin{center}
\includegraphics[width=1.0\linewidth]{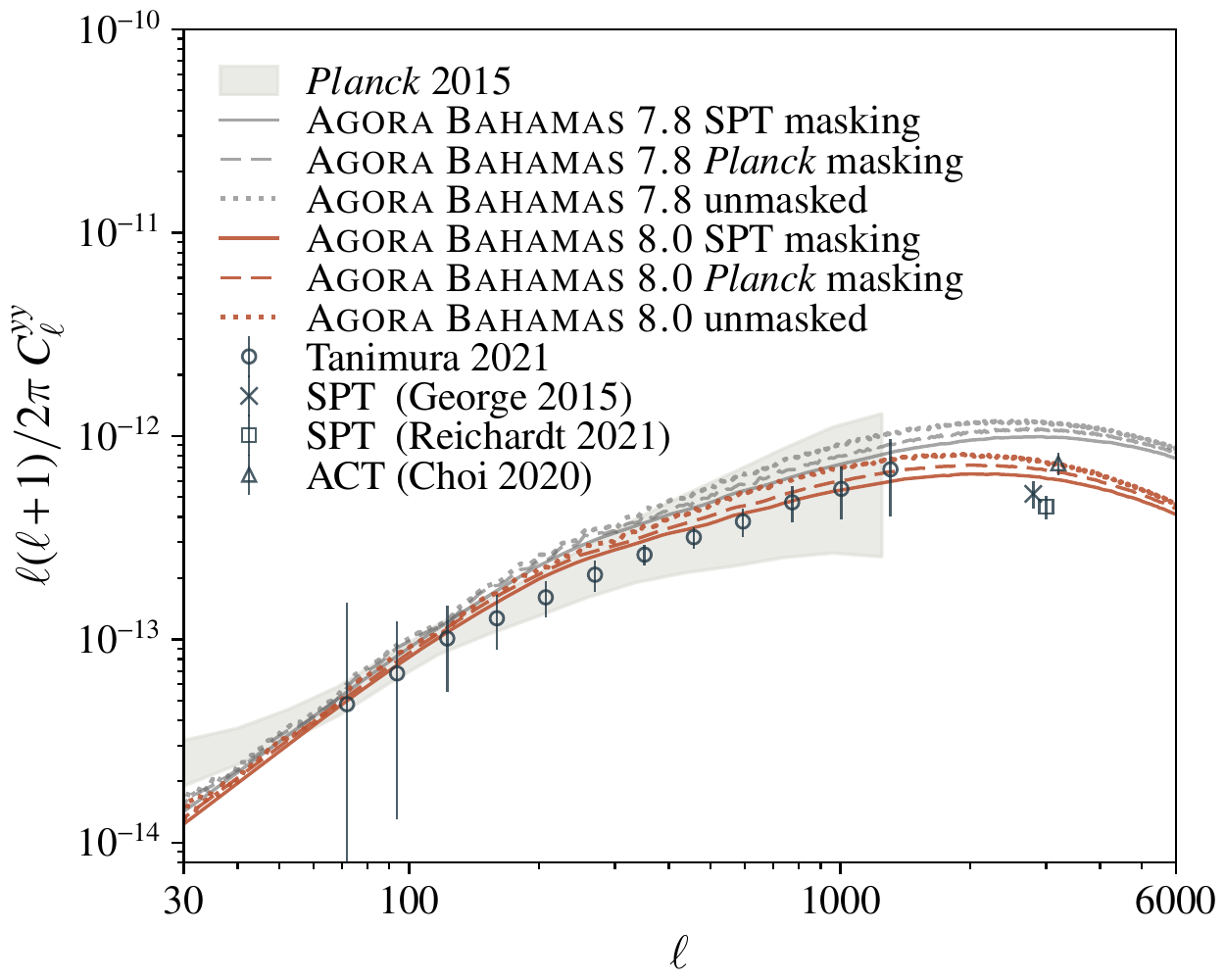}
\caption{
The measured auto-spectra of the Compton-$y$ maps based on the \bahamas{} 7.8 (grey) and 8.0 (orange) models, before (dotted) and after applying  the SPT-SZ (solid) or the \planck{} mask (dashed). 
The \bahamas{} 7.6 model is omitted from this plot since it significantly overpredicts the amplitude.
For reference, the data measurements from \planck{}  (\citealt{planck2015xxii,tanimura2021}; grey shade/navy circles), ACT (\citealt{choi2020}; navy triangle) and  SPT-SZ/SPTpol (\citealt{george2015,reichardt2021}; navy cross/square) are also shown.
}
\label{fig:cls_mdpl2_tsz}
\end{center}
\end{figure}

Figure \ref{fig:cls_mdpl2_tsz} shows the measured Compton-$y$ power spectra for both \bahamas{} 7.8 and 8.0 models in the simulation. With no mask applied, the power spectra for both models are consistently higher than the measurements from \planck{} and SPT-SZ, but are more or less in agreement with the amplitude inferred by ACTpol \citep{choi2020}. Since tSZ and IR/radio sources are known to be correlated, the Compton-$y$ power spectra are also measured after applying masks that remove sources down to 5$\sigma$  (described in Section \ref{sec:mask}). The amplitudes of $C_{\ell}^{yy}$ are reduced by as much as 16\% and 8\% when the SPT-SZ and the \planck{} masks are applied, respectively. The masking effect brings down the simulation spectra in between the $\ell=3000$ measurements from ACTpol and SPT-SZ. It is noted, however, that the masking threshold adopted by \citet{choi2020} is higher ($15$ mJy at 150 GHz) than that of SPT-SZ and SPTpol ($5\sigma\sim6.4$ mJy at 150 GHz) and therefore a higher amplitude is expected.

Between the three models, the \bahamas{} 8.0 model is found to be the most consistent with observations. This conclusion is also in agreement with the best-fit AGN heating temperature value of $T_{\rm AGN}=10^{7.96}$ K obtained from the joint constraint of tSZ-shear and cosmic shear  measurements by \cite{troester2021}, although with the caveat that these two probes are sensitive to different halo mass regimes as measured by the tSZ auto-spectrum (see e.g., \citealt{osato2020,pandey2022}). Throughout the rest of this paper, the \bahamas{} 8.0 is adopted as the fiducial tSZ model.
 
\subsection{kSZ auto-spectra}

\begin{figure}
\begin{center}
\includegraphics[width=1.0\linewidth]{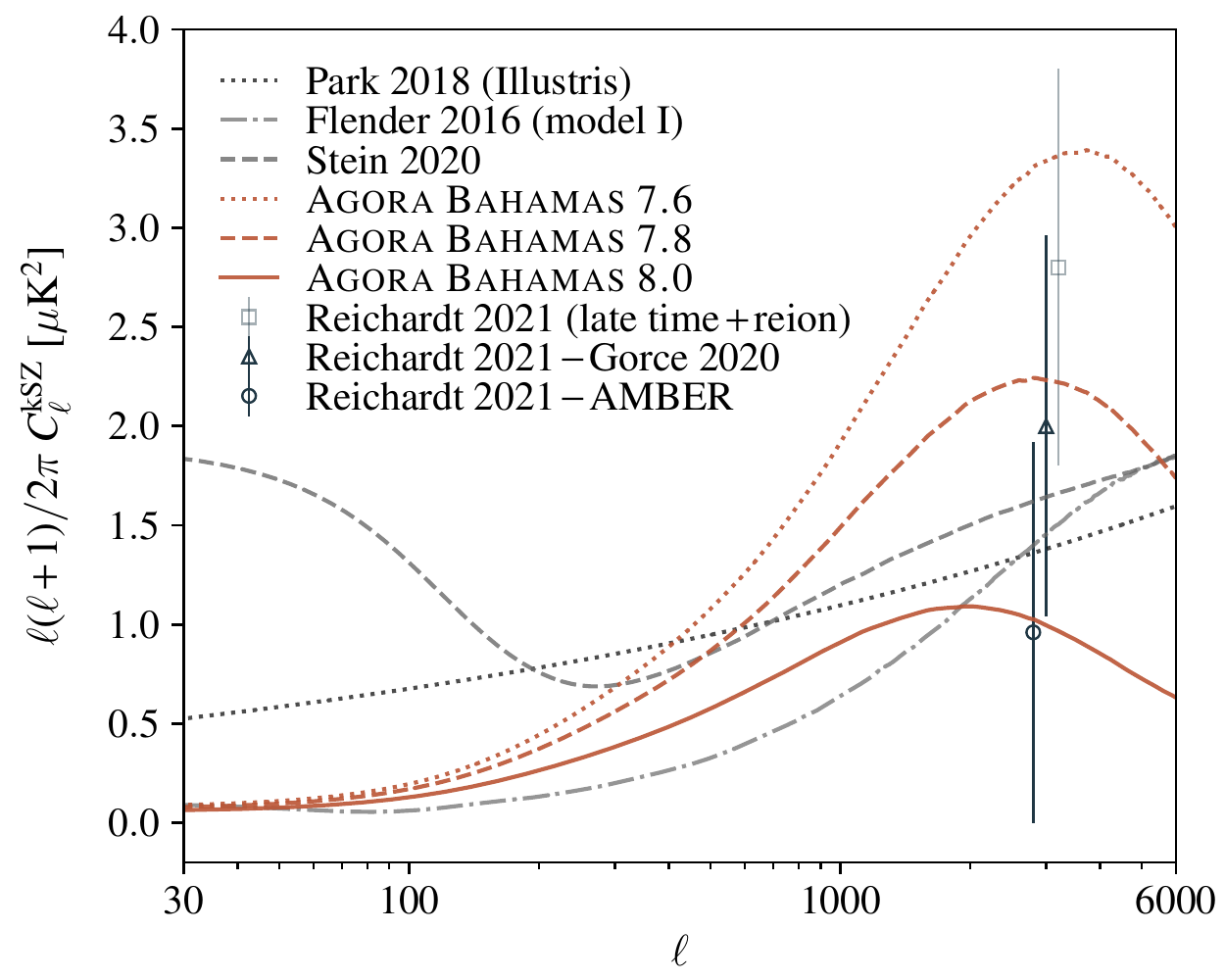}
\caption{The measured auto-spectra of the kSZ maps in the simulation using the \bahamas{} 7.6, 7.8 and 8.0 models (dotted, dashed, and solid orange lines). Also shown are measurements from other simulations  \citep{flender2016,park2018,stein2020}, and the best-fit total kSZ amplitude from \citet{reichardt2021} (light navy square), as well as the inferred late-time kSZ contribution obtained by subtracting the reionisation kSZ model from \citet{gorce2020} (navy triangle) and AMBER (navy circle). }
\label{fig:cls_mdpl2_ksz}
\end{center}
\end{figure}

Figure \ref{fig:cls_mdpl2_ksz} shows the measured power spectra from the kSZ maps built assuming the three \bahamas{} models, and are compared with results from the \textsc{Illustris} hydrodynamical simulation \citep{park2018}, dark matter-only simulations \citep{flender2016,stein2020}, and the observational constraint from \citet{reichardt2021}. Since the constraint\footnote{Here, the constraints obtained using the tSZ/kSZ templates from \cite{shaw2010,shaw2012} with a bispectrum prior $D_{3000}^{\rm kSZ}=2.8\pm0.9 \mu{\rm K}^2$ is used.} presented in \citet{reichardt2021} is for the \textit{total} kSZ amplitude (i.e., late-time + reionisation kSZ), a model for the reionisation component must be subtracted to isolate the late-time kSZ component. Two models are considered in this work: the first is the model by \cite{gorce2020}, which is an analytical prescription calibrated using radiative hydrodynamical simulations, and the second is computed using a semi-analytic model using the code \textsc{AMBER}\footnote{The simulation was run using the default reionisation parameters given in \url{https://github.com/hytrac/amber/blob/main/examples/input_cmbmap.txt}, and adopting the MDPL2 cosmology.} \citep{trac2022}. These two models predict the reionisation kSZ amplitude to be, $D_{\ell=3000}^{\rm kSZ,reion}=0.8,1.8$ $\muk{}^{2}$ respectively. The \bahamas{} 80 model is found to be in good agreement with the inferred late-time kSZ amplitude using AMBER as the model for the reionisation component, while the amplitude is $\sim\!1\sigma$ (ignoring the uncertainties on the patchy kSZ amplitude) lower compared with the amplitude inferred using the \citet{gorce2020} model. Since the amplitude of the kSZ spectrum is currently not tightly constrained by observational data, none of the \bahamas{} models can be ruled out strictly. Therefore, the \bahamas{} 8.0 model is used as the fiducial kSZ model for the remainder of this work to keep the modelling consistent with the modelling of the tSZ effect. It is expected that ongoing and future experiments will be able to better characterize the amplitude and shape of the kSZ power spectrum.

\subsection{CIB auto/cross-spectrum and number counts}
Using the CIB maps and catalogues generated with the best-fit values for $A_{\rm d}$, $\zeta_{\rm d}$, $B$ and $\alpha$ obtained in Section \ref{sec:model_cib}, the auto- and cross- spectra, source number counts, frequency decorrelation, as well as correlation with other observables, such as CMB lensing and tSZ, are measured in this section.

\subsubsection{Comparison of CIB power spectra}\label{sec:CIB_powerspectra}
\noindent {\it Planck \& Herschel frequencies}:\\
\noindent Since the CIB maps in the simulation are calibrated against CIB maps from \citet{lenz2019}, which are based on data products from the \planck{} Public Release 3 (PR3), differences in calibration between PR1 (2013) and PR3 (2018) must be taken into account when comparing the power spectra from the simulation with the results from \citet{planck2014xxx}. Calibration factors of 1.021, 0.98, and 0.96 (squared for the auto-spectra) are applied to the 353, 545, and 857 GHz measurements from \citet{planck2014xxx} to make comparisons, which are the calibration factors between PR1 and PR2 quoted in \citet{mak2017}, and the calibration factors between PR2 and PR3 are taken to be unity. For comparisons with measurements from \citet{viero2019}, colour correction factors of $0.976$ and $0.854$\footnote{These values are computed by taking the product of the conversion factors quoted in the caption of Figure 1 of \citet{viero2019} and the conversion factors between \planck{}  PR1 and PR2. } are used to convert {\it Herschel}'s 857 GHz and 600 GHz channels to {\it Planck}'s 857 and 545 GHz channels. 

\begin{figure*}
\begin{center}
\includegraphics[width=1.0\linewidth]{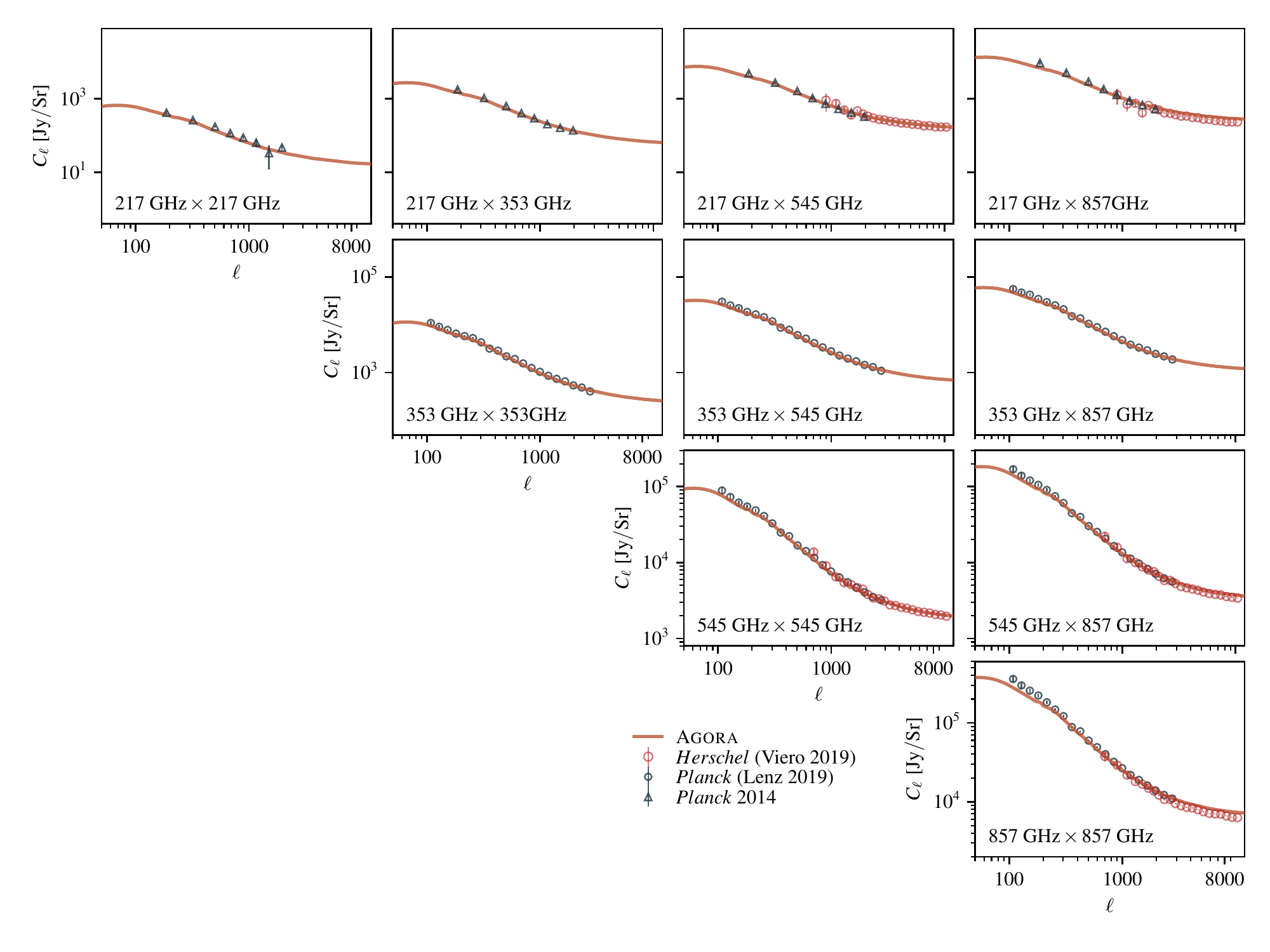}\llap{\makebox[\wd1][l]{\hspace{-17cm}\raisebox{0.25cm}{\includegraphics[bb=60 555 600 -600]{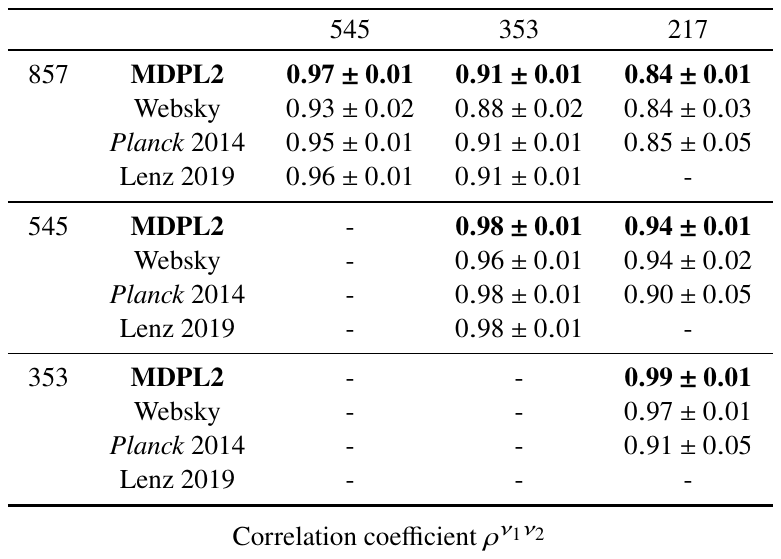}}}}
\caption{Auto- and cross-spectra of 217/353/545/857 GHz CIB maps. The inset table summarizes the correlation coefficients between the frequency channels computed using Equation \protect\eqref{eq:cib_corrcoeff}. }
\label{fig:cibautocross095150220}
\end{center}
\end{figure*}

Figure \ref{fig:cibautocross095150220} shows the measured auto- and cross-spectra of the CIB maps in the simulation at \planck{} frequencies (217/353/545/857 GHz) compared with measurements from data with the aforementioned calibration parameters applied. By construction, the CIB 353/545/857 GHz auto-/cross-spectra are consistent with the measurements from \citet{lenz2019} in the angular range $200<\ell<3000$. The simulation amplitudes for 217 GHz auto-spectrum is slightly lower than the measurements from \citet{planck2014xxx}. This could be due to leakages of other astrophysical components (such as the CMB) into the \planck{} 217 GHz CIB map, resulting in an overall excess amplitude. For comparisons with \citealt{viero2019}, an overall agreement is found at 545/600 GHz, but the measurements that use the 857 GHz channel are $\sim 10\%$ higher at $\ell>3000$.\\[0.15cm]

\noindent {\it SPT frequencies}:\\
\vspace{0.35cm}
\noindent The CIB amplitudes reported by \citet{george2015} and \citet{reichardt2021} at $\ell=3000$ are quoted at their reference frequencies $\nu^{\rm ref}$ = $90/150/220$ GHz. These are converted to the equivalent amplitudes at their effective frequencies by applying \citep{reichardt2021}:
\begin{equation}\label{eq:convert_nueff_nuref}
C_{\ell, \nu^{\rm ref}}=C_{\ell,\nu^{\rm eff}}\epsilon_{\nu^{\rm ref}}\left(\frac{\Phi(\nu^{\rm ref},T_{\rm d})}{ \Phi(\nu^{\rm eff},T_{\rm d})  }\right)^2,
\end{equation}
where $\Phi(\nu,T_{\rm d})$ is the modified blackbody SED from Equation \eqref{eq:sed_greybody}, and 
\begin{equation}
\epsilon_{\nu^{\rm ref}}\equiv 
\left(\frac{dB/dT\rvert_{\nu^{\rm eff}} }{ dB/dT\rvert_{\nu^{\rm ref}}  }\right)^{2},
\end{equation}
where $dB/dT$ is the derivative of the black body spectrum. The effective frequencies are taken to be $\nu^{\rm eff}=97.9/154.1/219.6$ GHz for SPT-SZ and $\nu^{\rm eff}=96.9/153.4/221.6$ GHz for SPT-SZ+SPTpol \citep{george2015,reichardt2021}.

\begin{figure}
\begin{center}
\includegraphics[width=1.0\linewidth]{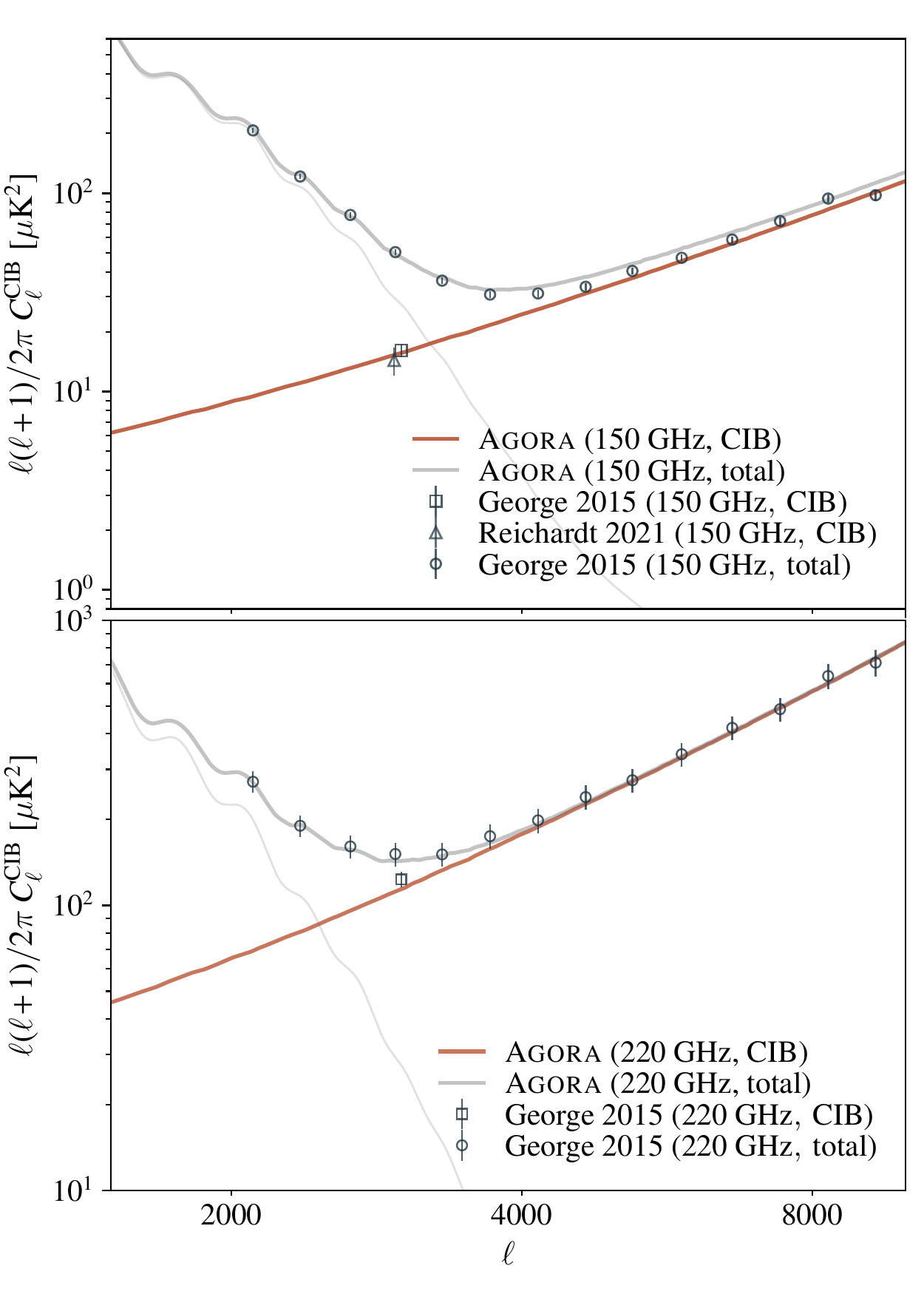}\\
\caption{Auto-spectra of the CIB maps at 150 and 220 GHz (orange lines), compared with the best-fit CIB amplitudes (quoted as $D_{\ell}\equiv\ell(\ell+1)/2\pi\hspace{0.05cm}C_{\ell}$ at $\ell=3000$) from \citealt{george2015} (navy square) and \citealt{reichardt2021} (navy triangle). The CIB points are scaled to $\nu_{\rm eff}$ using Equation \eqref{eq:convert_nueff_nuref}. The total power spectra measured from the simulated total 150 and 220 GHz frequency maps are shown as grey lines, and are compared with the measurements from \protect\citet{george2015} (navy points). }
\label{fig:cibauto150220}
\end{center}
\end{figure}

The measured CIB power spectra at 150 GHz and 220 GHz, after applying the mask generated in Section \ref{sec:mask}, are shown in Figure \ref{fig:cibauto150220}. The power spectra amplitudes at $\ell=3000$ are found to be 
 $D^{\rm CIB}_{\ell=3000}=15.5,113.4\ \mu {\rm K}^{2}$ for 150 and 220 GHz channels, respectively. These results are within $1.5\sigma$ of the measurements from \citet{george2015} and \citet{reichardt2021}.

\subsubsection{Frequency decorrelation}
The correlation coefficient between two frequency channels can be computed using the measured CIB auto- and cross-power spectra:
\begin{equation}\label{eq:cib_corrcoeff}
\rho^{\nu_{1}\nu_{2}}= \left\langle C_{\ell}^{\nu_{1}\nu_{2}} /\sqrt{ C_{\ell}^{\nu_{1}\nu_{1}}C_{\ell}^{\nu_{2}\nu_{2}}} \right\rangle,
\end{equation}
where the average is taken across the multipole range $150<\ell<1000$ and the contribution from shot noise is disregarded, as done in \cite{planck2013xxx} and \cite{stein2020}. The correlation coefficients are expected to be less than one because each frequency channel is sensitive to emission from dusty star-forming galaxies at slightly different redshifts \citep{planck2014xxx}.

The table shown within Figure \ref{fig:cibautocross095150220} summarizes the measured correlation coefficients between the 217/353/545/857 GHz frequency channels. The correlation coefficients are higher for frequency pairings that are closer together and lower for frequency pairings that are farther apart, as expected. The simulation results are generally marginally higher than those found by \citealt{planck2014xxx}, and the difference in correlation coefficient can be as large as 8\% (for the correlation coefficient between $217\times353$ GHz).
The source of this discrepancy could be in either the data or the simulation, however, further investigation on this point is left for future work.

\subsubsection{Redshift distribution}\label{sec:cib_dIdz}
The inferred redshift distributions of the infrared galaxies that make up the 353/545/857 GHz CIB maps in the simulation are compared with the observational results from \citet{schmidt2015} and \citet{zavala2017}.

\citet{schmidt2015} cross-correlated \planck{} 353/545/857 GHz maps with quasars from the Sloan Digital Sky Survey (SDSS) DR7 to estimate the redshift distribution of the CIB in the redshift range $0<z<5$. On the other hand, \citet{zavala2017} used deep 450 and 850 $\mum{}$ observations from the SCUBA-2 Cosmology Legacy Survey and  photometric redshifts from the 24$\mum{}$-selected sources to infer the redshift distributions of the sub-millimetre galaxies.

\begin{figure}
\begin{center}
\includegraphics[width=1.0\linewidth]{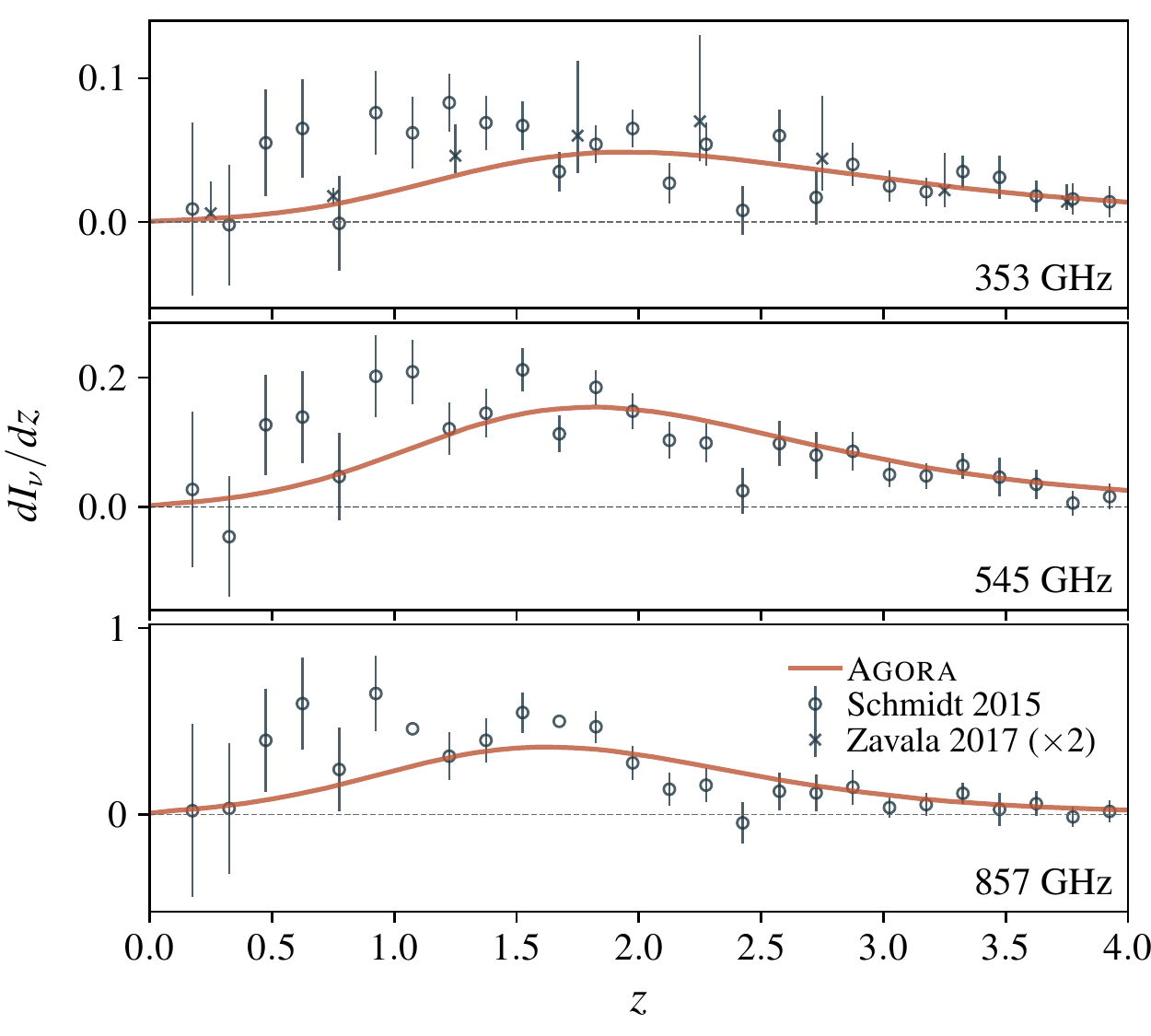}
\caption{Redshift distributions of the CIB, inferred from the 353, 545 and 857 GHz CIB maps (orange lines). In comparison, the measurements from \planck{} \protect\citep{schmidt2015}\protect\footnotemark and SCUBA-2 \protect\citep{zavala2017} are shown (navy circles and navy triangles). Since the raw result from \protect\cite{zavala2017} only accounts for approximately 50\% of the total CIB power, the points are multiplied by a factor of 2 in this plot.}
\label{fig:cib_dIdz}
\end{center}
\end{figure}

The inferred redshift distributions of infrared galaxies from the simulated 353/545/857 GHz CIB maps are presented in Figure \ref{fig:cib_dIdz}, and are shown to be consistent with the measurements from \citet{schmidt2015} at $z>1.5$.  However, a deficit in intensity at $z<1.5$ is found, which could be due to low-redshift infrared sources such as ultra luminous infrared galaxies (ULIRGs) or low-redshift dusty interlopers, which are not implemented in the simulation.

\footnotetext{The seventh and eleventh bins in the results for the 857 GHz channel from \citet{schmidt2015} have unrealistically large uncertainties. For visualization purposes, we have omitted the  uncertainties on these two points in the figure.}

The measured differential intensity in the simulation is higher at all redshifts when compared with the raw measurements from \citet{zavala2017}. This is expected, given that these raw measurements are based on detected sources at 450 and 850 $\mum{}$ and stacks of detected sources at 24 $\mum{}$, and it is well known that undetected sources contribute a significant portion of total CIB intensity. \citet{zavala2017} estimates that the sum of resolved sources and stacks of 24 $\mum{}$ sources account for 60\% and 50\% of total CIB power at 450 and 850 $\mum{}$, respectively. To account for the contributions from undetected sources, the $ dI_{\nu}/dz$ from \citet{zavala2017} is naively scaled by a factor of two. The simulation's differential intensity is consistent with the scaled results from \citet{zavala2017}.

The differential intensity function is also used to verify whether the redshift cut of $z<8.6$ used in this work sufficiently captures all the CIB information. This is checked by comparing the $dI_{\nu}/dz$ at $z=8.6$ and the redshift at which the intensity peaks ($z\sim1.8$). It is verified  that the contribution from sources at $z\sim8.6$ amounts to less than 0.1\% of the contributions from $z\sim1.8$, and therefore, it is concluded that contributions from sources above $z=8.6$ will have negligible impact on the CIB maps. 

\subsubsection{Number counts}\label{sec:cib_number_counts}
\begin{figure}
\begin{center}
\includegraphics[width=1.0\linewidth]{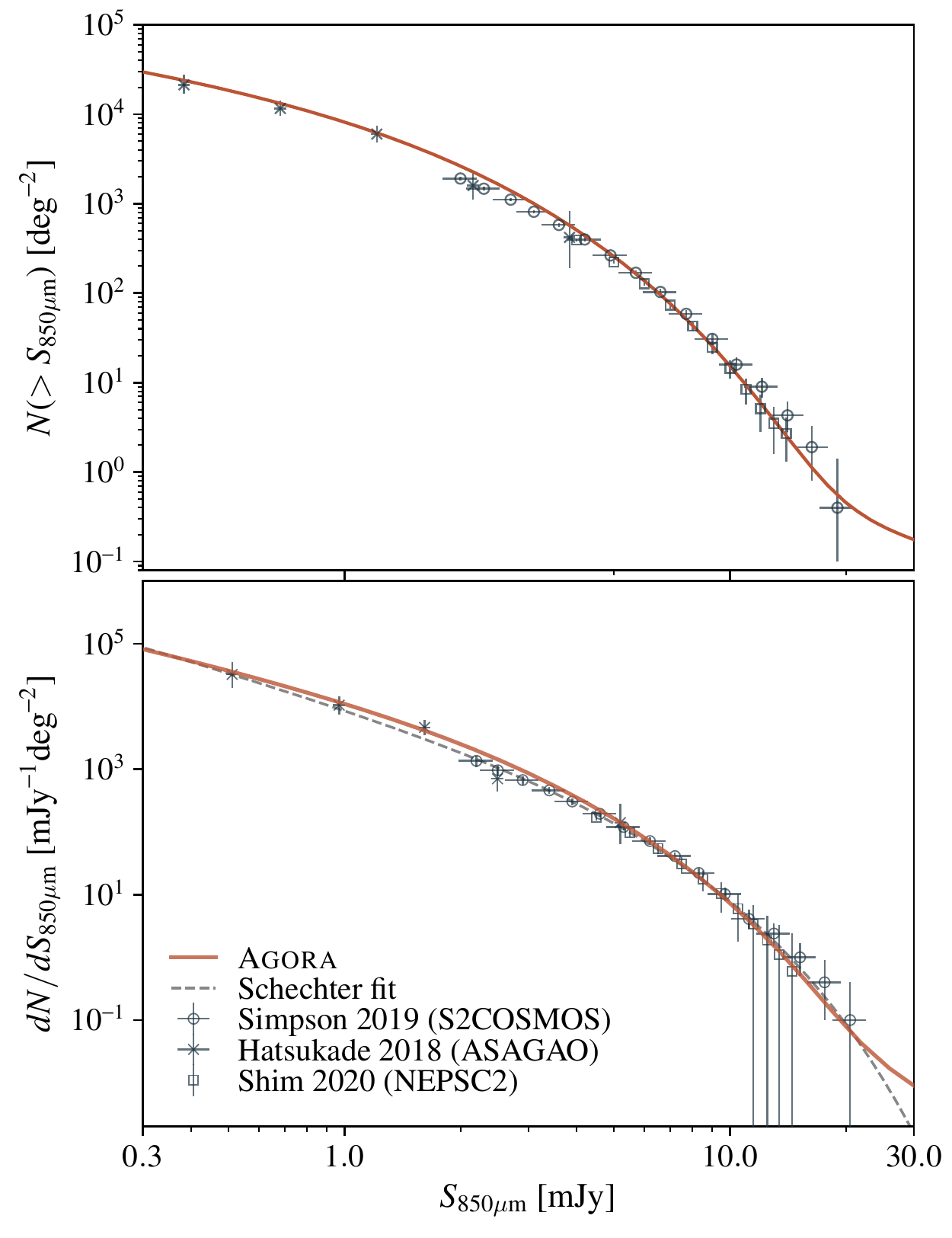}
\caption{{\bf Upper:} The cumulative number count of dusty sources at $353\ {\rm GHz}/850\ \mu{\rm m}$ in the simulation (orange line). As a comparison, measurements from the S2COSMOS survey (\protect\citealt{simpson2019}; navy circles), ASAGAO survey (\protect\citep{hatsukade2018}; navy crosses) and NEPSC2 survey  (\protect\citealt{shim2020}; navy squares) are also shown.
{\bf Lower:} The same as above, but for differential number counts. The dashed grey line represents the Schechter function fit to the compilation of data from \protect\citealt{geach2017,hatsukade2018,stach2018,simpson2019,simpson2020,shim2020}. }
\label{fig:cib_counts353}
\end{center}
\end{figure}

In this section, the differential and cumulative number counts of the 353 GHz (850\ $\mum{}$) sources in the simulation are compared with other studies. Many such analyses have been conducted using SCUBA-2 850 $\mum{}$ data and ALMA band 7 data (e.g., \citealt{oteo2016, geach2017,stach2018, simpson2019}). The comparisons of differential counts ${d}N/{d}S_{850 \mum{}}$ and cumulative counts $N(>S_{850\mum{}})$ are shown in Figure \ref{fig:cib_counts353}.
Since different surveys observe at slightly different wavelengths/frequencies, colour correction factors are applied to standardize the measurements to $850\ \mum{}$ $(353\ {\rm GHz})$. Following \citet{simpson2020}, a scaling factor of $S_{850\mu{\rm m}}/S_{870\mu{\rm m}}=0.94$ is used to convert ALMA's 870 $\mum{}$ (band 7) to 850 $\mum{}$. To convert the $1.2$ mm channel of the ASAGAO survey \citep{hatsukade2018} to 850 $\mum{}$, the product of the conversion factors  $S_{870\mu {\rm m}}/S_{1200\mu {\rm m}}=2.7$ from \citet{dudzeviciute2020} and $S_{850\mu{\rm m}}/S_{870\mu{\rm m}}=0.94$ is applied.

Additionally, a function of the form  \citep{schechter1976}:
\begin{equation}
\frac{{\rm d}N}{{\rm d}S}=\frac{N_{0}}{S_{0}}\left(\frac{S}{S_{0}}\right)^{-\gamma}{\rm exp}\left(-\frac{S}{S_{0}}\right)
\end{equation}
that is fitted to a combination of recent measurements from 
\citealt{geach2017,hatsukade2018,stach2018,simpson2019,simpson2020,shim2020} is also shown.\footnote{The best-fit parameters for the Schechter function are found to be  $N_{0}=4147.28\ {\rm deg}^{-2}$, $S_{0}=3.3\ {\rm mJy}$, and $\gamma=1.94$.} Overall, both the differential and cumulative number counts from the simulation are found to be in good agreement with measurements from observational data in all flux regimes, however with a slight excess in number counts at $\sim$2 mJy, although this feature is not identified when compared with the 1.2 mm measurements from \citealt{hatsukade2018}.

\subsection{tSZ $\times$ CIB}

The tSZ-CIB correlation factor is a crucial quantity in component separation, and is defined as:\footnote{ \citet{george2015} and \citet{reichardt2021} both use a frequency dependent definition, which differs from this definition by a factor of two.}
\begin{equation}\label{eq:tszcibcorr}
\rho_{\ell}^{{\rm tSZ}\times{\rm CIB}}=\frac{C_{\ell}^{\rm tSZ\times CIB}} {\sqrt{C_{\ell}^{\rm tSZ\times tSZ} C_{\ell}^{\rm CIB\times CIB}}}.
\end{equation}
To compare with existing constraints from SPT-SZ, power spectra of the simulated tSZ and CIB maps for the SPT-SZ 150 GHz channel are computed after applying the SPT-SZ 5$\sigma$ point-source mask.

\begin{figure}
\begin{center}
\includegraphics[width=1.0\linewidth]{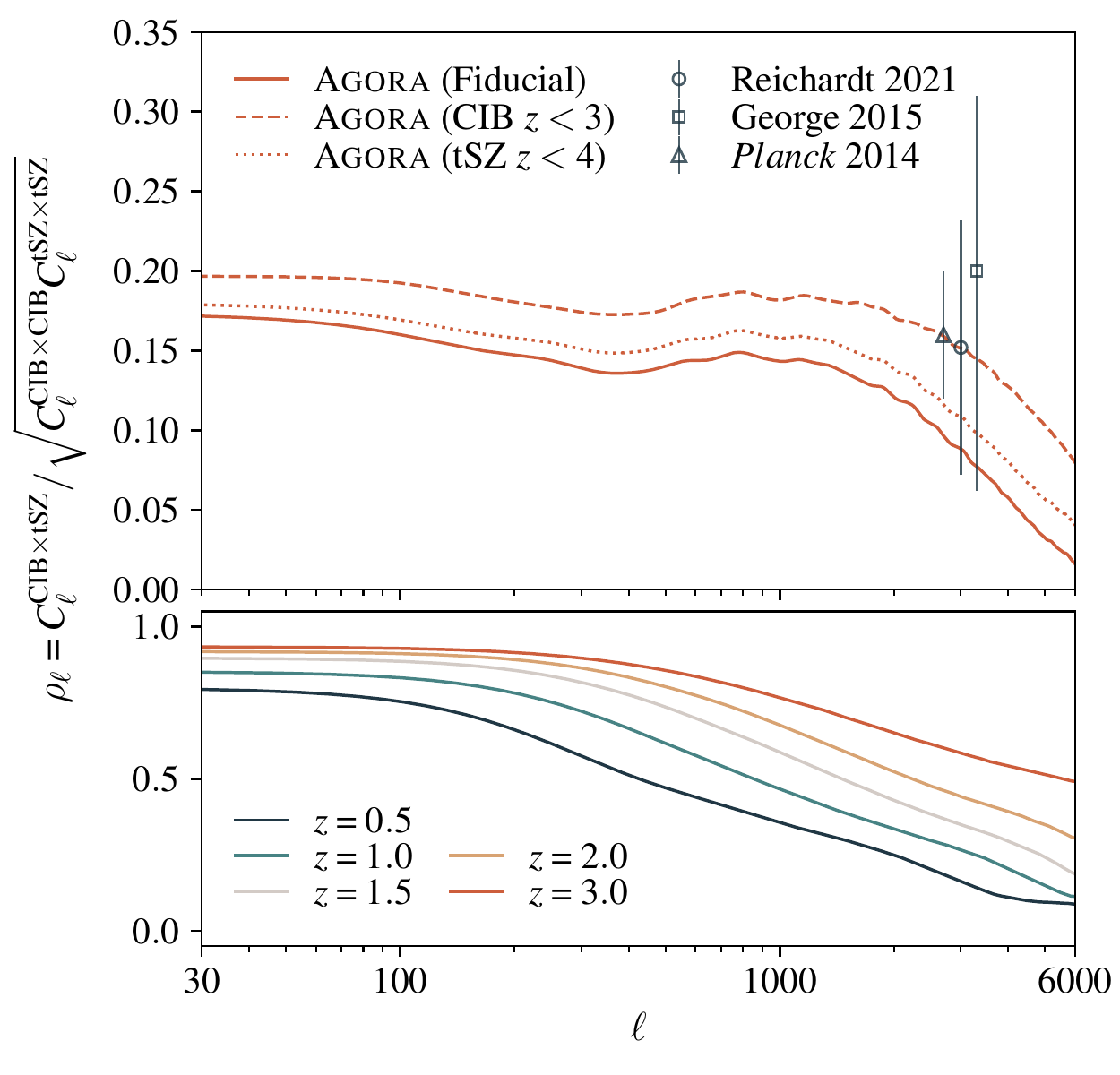}
\caption{ {\bf Upper:} The measured correlation factors between tSZ and CIB in the simulation (orange line) compared with observations from \protect\cite{planck2015xxiii,george2015,reichardt2021}. The solid orange line represents the measured correlation factor using the fiducial maps, for which both observables have been integrated out to their maximum redshifts ($z=8.6$ for CIB and $z=3$ for tSZ).  The dashed orange line represents the correlation factor computed using a CIB map truncated at $z=3$, and the dotted orange line represents the correlation factor computed using a tSZ map extended to $z=4$. {\bf Lower:} The correlation factor computed for thin slices at $z=0.5,1.0,1.5,2.0$ and $3.0$. }
\label{fig:tszxcib}
\end{center}
\end{figure}

Figure \ref{fig:tszxcib} shows the measured correlation factors.
Using the fiducial maps in the simulation (i.e., CIB map integrated up to $z=8.6$ and tSZ map integrated up to $z=3$), $\rho^{{\rm tSZ} \times {\rm CIB}}_{\ell}$ is found to be $\sim40-55\%$ lower than the constraints from \cite{planck2015xxiii}, \cite{george2015} and \cite{reichardt2021}. $\rho_{\ell}$ is also computed using a CIB map with an upper redshift limit of $z=3$, to match the cut-off applied to the tSZ map. In this case, the amplitude is in good agreement with observational constraints. This implies that the tSZ-CIB correlation coefficient is sensitive to the tSZ amplitude beyond $z=3$, which can also be extrapolated from the redshift evolution of the correlation factor measured at $z<3$ (shown in the lower panel of Figure \ref{fig:tszxcib}). To verify this claim, the correlation coefficient is also computed using a tSZ map that is integrated up to $z=4$. In this case, an increase of $\sim20$\%  in the correlation factor is seen, and therefore, the fiducial results shown in Figure \ref{fig:tszxcib} should be regarded as a lower limit. Since clusters below $z=3$ dominate the auto-spectrum signal (extending the upper redshift of the tSZ map to $z=4$ has negligible impact on the  auto-spectrum amplitude) and cross-correlations are often taken with tracers below $z=3$ \citep{battaglia2012,osato2020}, for most practical purposes, this feature could be disregarded.  

\subsection{Radio galaxy counts and spectra}

\subsubsection{Radio galaxy counts}
\begin{figure}
\begin{center}
\includegraphics[width=1.0\linewidth]{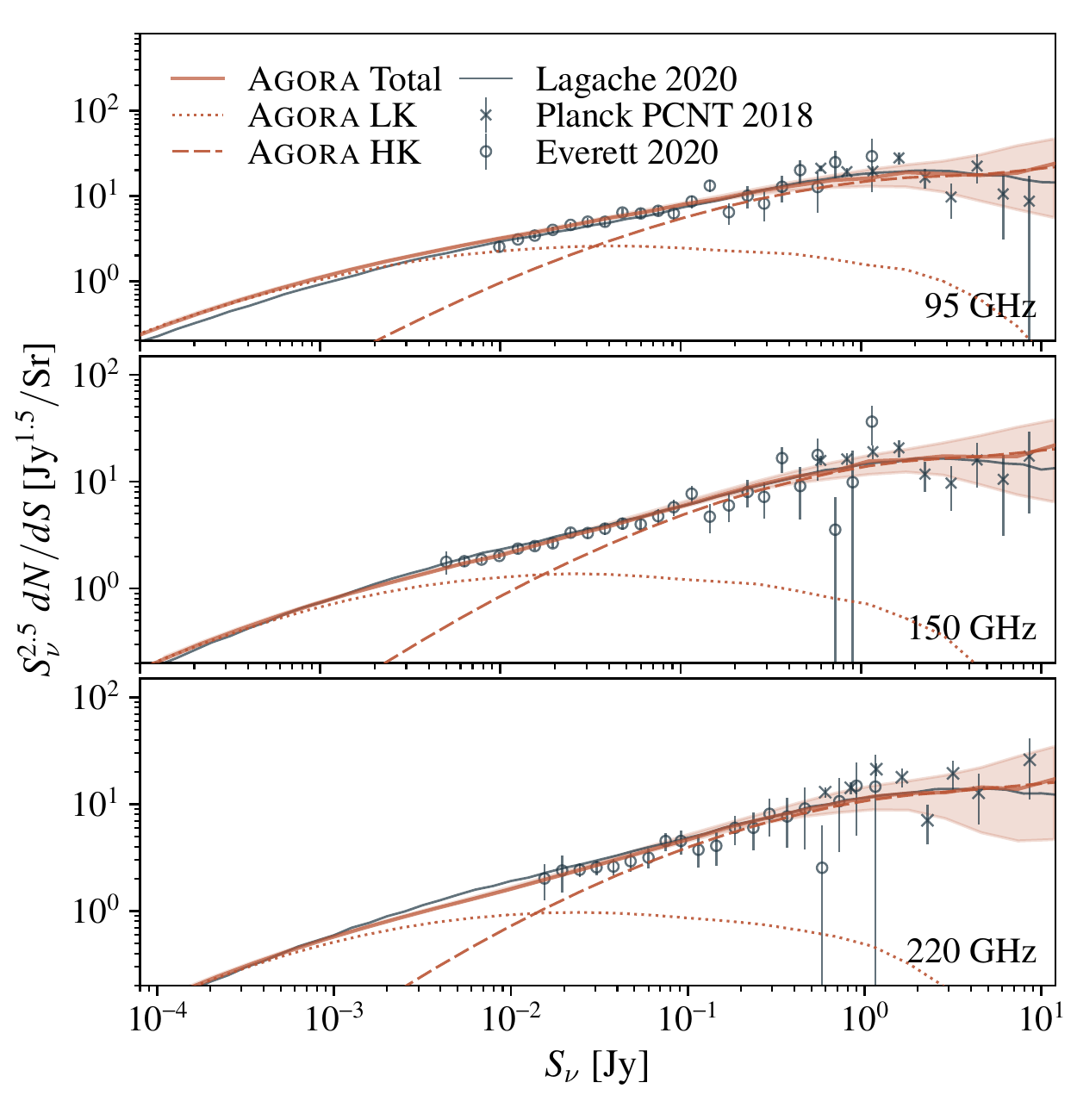}
\caption{Measured differential source counts $dN/dS$ at 95/150/220 GHz in the simulation. The total counts are shown in solid orange, while the LK and HK components are shown as dashed and dotted orange lines, respectively. Also shown are the measurements from \citet{everett2020} (navy circles), the measurements from \citet{planck2018liv} (navy crosses), and the number count models from \citet{lagache2019} (grey lines). }
\label{fig:radio_counts}
\end{center}
\end{figure}

Figure \ref{fig:radio_counts} shows the radio source number counts at 95, 150, and 220 GHz. In comparison to the counts from \citet{lagache2019}, small differences in the low ($S_{\nu}<10\ {\rm mJy}$) and high flux ($S_{\nu}>1000\ {\rm mJy}$) regimes are found, where the number counts in the simulation are slightly higher ($\sim10\%$). The differences in the high flux regime are within statistical uncertainties and are thus insignificant. The discrepancy in the low flux regime is primarily due to the low-kinetic AGN modelling: because the radio source component of the simulation is calibrated against the SPT-SZ point source catalogue, which has limited sensitivity below 10 and 6 mJy at 95 and 150 GHz respectively, the number counts in this flux regime are subject to large uncertainties, and data from ongoing experiments such as advACT \citep{advact} and SPT-3G are needed to improve the modelling.

\subsubsection{Power spectra}
Radio sources are projected onto a \healpix{} grid of $\nside{}=8192$ and the power spectra for 95/150/220 GHz maps are computed. The 5$\sigma$ threshold SPT-SZ mask described in Section \ref{sec:mask} is applied for measuring $C_{\ell}^{TT}$, and the resulting power spectra are compared with constraints from \citet{reichardt2021}. For $C_{\ell}^{BB}$, masking thresholds of 1.12, 1.68, and 4.45 mJy are used, and the resulting power spectra are compared with models from \cite{lagache2019}.

\begin{figure}
\begin{center}
\includegraphics[width=1.0\linewidth]{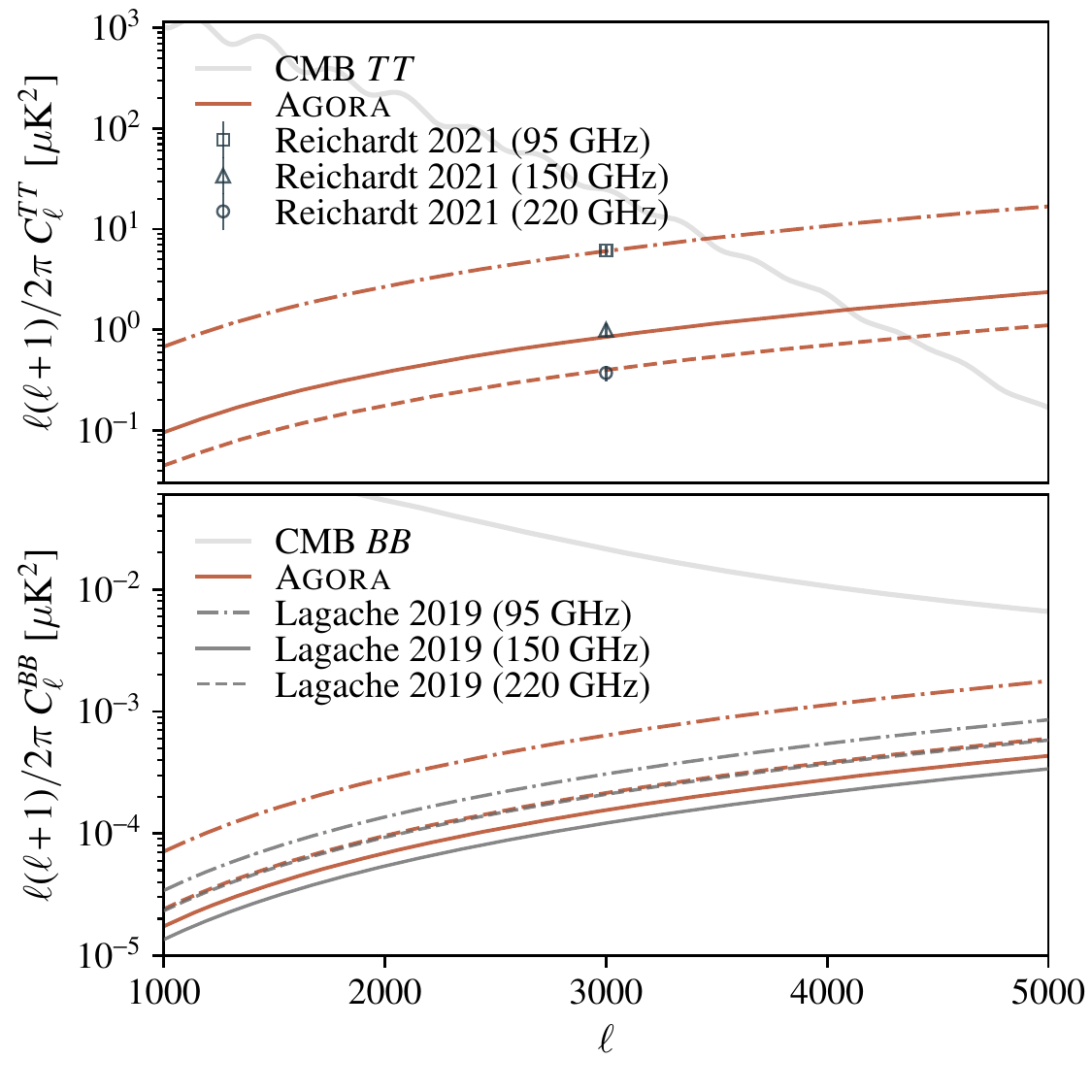}
\caption{{\bf Upper:} The measured $C_{\ell}^{TT}$ spectra at 95/150/220 GHz in the simulation (orange lines) and the best-fit radio amplitudes at $\ell=3000$ from \protect\cite{reichardt2021} (navy points). {\bf Lower:} The measured  $C_{\ell}^{BB}$ spectra in the simulation (orange lines) compared with the models from \protect\cite{lagache2019}  (dashed, solid and dot-dashed dark grey lines).}
\label{fig:radio_cltt_clbb}
\end{center}
\end{figure}

The measured radio source power spectra are shown in Figure \ref{fig:radio_cltt_clbb}. For $C_{\ell}^{TT}$, the 95 GHz power spectrum is consistent with the best-fit amplitude from \cite{reichardt2021}, but the amplitudes are approximately $-15$\% and $+8$\% offset for 150 GHz and 220 GHz, respectively.
For $C_{\ell}^{BB}$, the amplitude of the power spectrum at 220 GHz is found to be in good agreement with the model from \cite{lagache2019}, but the amplitudes at 95 and 150 GHz are discrepant by factors of 2 and 1.3, respectively.\footnote{While the forecast in their analysis was intended for SPT-3G, SPT-SZ band passes were used in the analysis, and therefore simulated SPT-SZ catalogues are used here as well.} These findings are not surprising, given that (a) the flux thresholds used are significantly lower than the limits of current surveys, and thus the source population is poorly characterized in this flux regime, and (b) different distribution of polarization fraction are assumed between the modelling used in this work and that used in \cite{lagache2019}. Evidently, more data is needed to determine which of these models is more realistic.

\subsection{$k_{\rm CMB}$-CIB cross-spectra}
Cross-correlations between CMB lensing and CIB allow us to calculate the masses of halos that star-forming galaxies inhabit \citep{planck2013xviii}. Several studies have measured such cross-correlations using data from {\it Planck}, ACT, SPT and {\it Herschel}  \citep{holder2013,planck2013xviii,vanengelen2015,omori2017,maniyar2018,cao2020}.

\begin{figure}
\begin{center}
\includegraphics[width=1.0\linewidth]{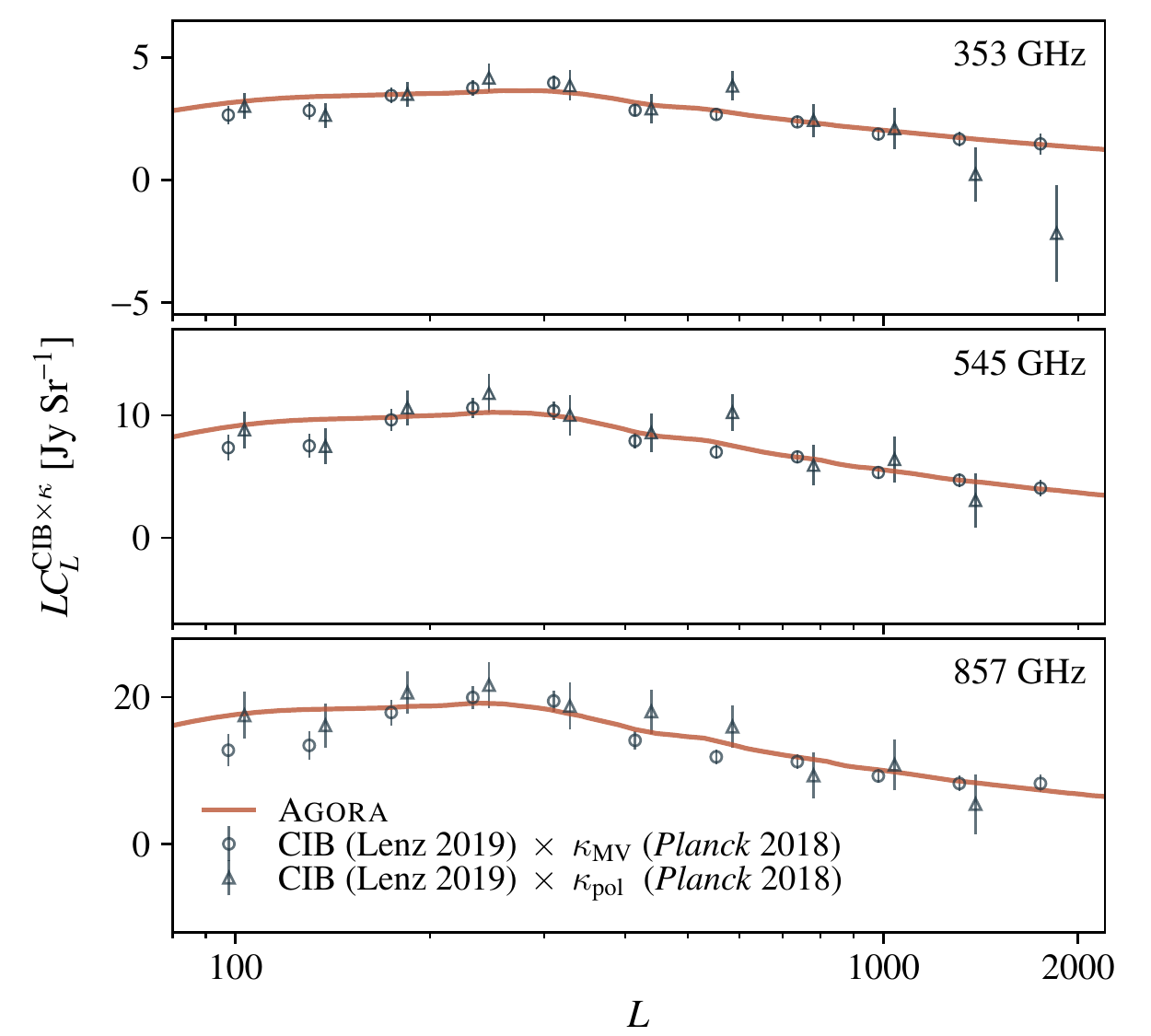}
\caption{Cross-correlation measurements between CMB lensing and CIB maps at 353/545/857 GHz from the simulation (orange lines), and between \protect\citet{lenz2019} CIB maps and minimum variance/polarization-only lensing maps from \protect\citet{planck2018viii} (navy circle/triangle points). }
\label{fig:cls_mdpl2_kcmb-cib}
\end{center}
\end{figure}

The cross-correlation measured in the simulation is compared with the measurements using CIB maps from \cite{lenz2019} and lensing maps from \cite{planck2015xv}. Both the minimum variance and the polarization-only CMB lensing maps are used for this comparison, since the minimum variance lensing map from \planck{} is dominated by information from the temperature channels, and could potentially be contaminated by Galactic dust. For this reason, the most conservative CIB map derived using a HI column density threshold of  $N_{\rm HI}=1.5\times 10^{20}\ {\rm cm}^{-2}$ ($f_{\rm sky}=0.11$) from \cite{lenz2019} is cross-correlated with the minimum variance lensing map. The impact of dust contamination, on the other hand, is expected to be significantly lower for the lensing map derived from polarization channels only, and thus the less conservative CIB map with a HI column density threshold of $N_{\rm HI}=2.5\times 10^{20}\ {\rm cm}^{-2}$ ($f_{\rm sky}=0.19$) is used. The results are shown in Figure \ref{fig:cls_mdpl2_kcmb-cib}, and the simulation cross-spectra are shown to be consistent with both the minimum variance and polarization-only cross-correlation measurements from the data.

\subsection{Auto/cross-spectra of frequency maps}\label{sec:frequency_auto_cross}

\begin{figure*}
\begin{center}
\includegraphics[width=1.0\linewidth]{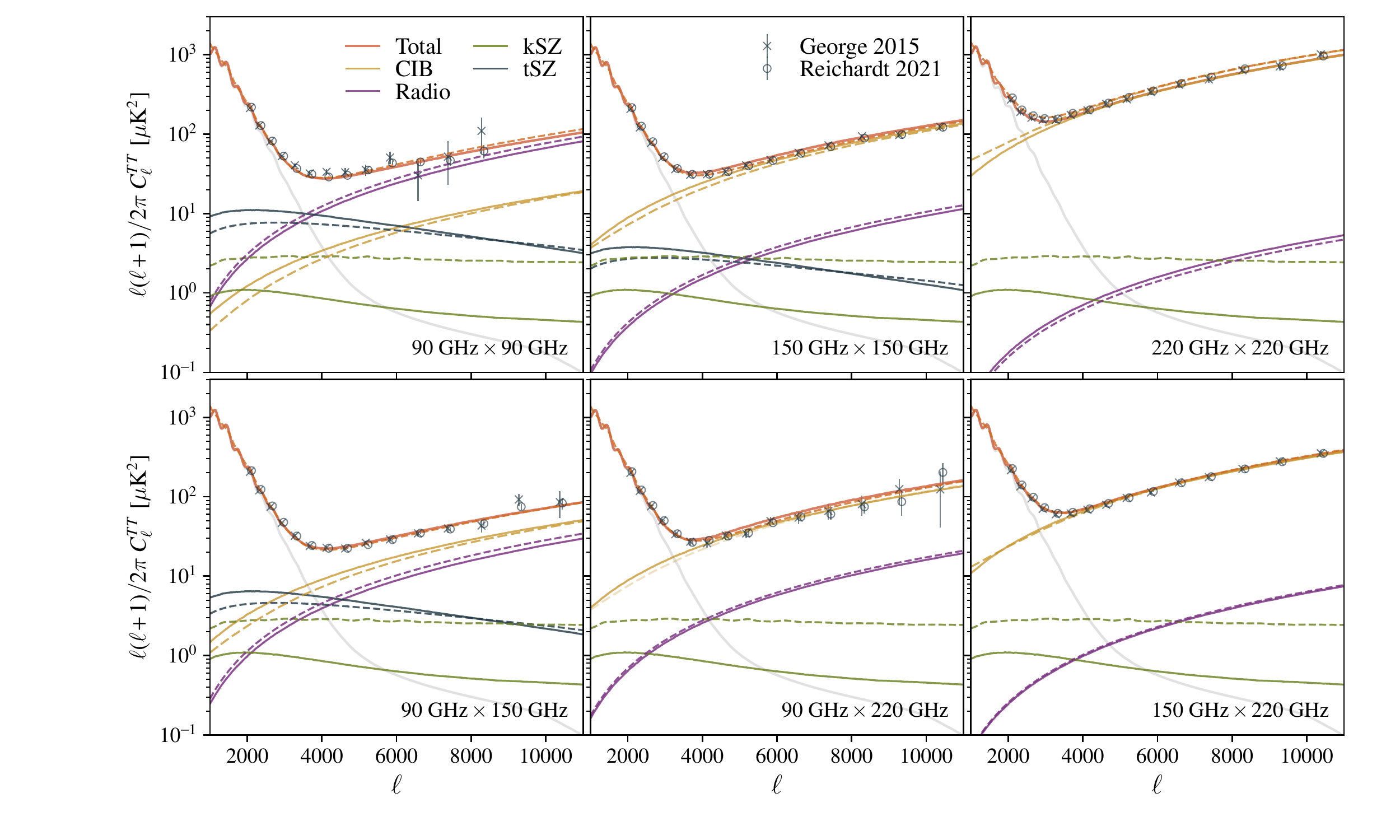}
\caption{The measured auto- and cross-spectra of  mock SPT-SZ 95/150/220 GHz maps (solid orange lines). As a comparison, the data measurements from \protect\cite{george2015} and \protect\citet{reichardt2021} are shown as navy cross/circle points. The measured auto-/cross-spectra of the individual foreground components are also shown (solid coloured lines), as well as the best-fit templates from \protect\citet{reichardt2021} (dashed coloured lines). As noted in the text, the kSZ template used by \protect\citet{george2015} and \protect\citet{reichardt2021} are the sum of the late-time and reionisation kSZ, and therefore their amplitudes are higher relative to the kSZ amplitude measured from the simulation (which only includes the late-time component). }
\label{fig:cls_sptsz_allcross}
\end{center}
\end{figure*}

In this subsection, all the CMB secondary components are combined and added to the lensed CMB map to produce realistic SPT-SZ 95/150/220 GHz frequency maps. The components added are:
\begin{enumerate}[leftmargin=25pt,align=left,labelwidth=\parindent,labelsep=1pt]
\item Lensed CMB map.
\item Lensed Compton-$y$ map using the \bahamas{} 8.0 model, converted into temperature units using Equation \eqref{eq:y2uk}.
\item Lensed CIB map.
\item Lensed kSZ map.
\item Lensed radio sources.
\end{enumerate}

The measured auto- and cross-spectra of the combined maps are shown in Figure \ref{fig:cls_sptsz_allcross}. The total auto- and cross-frequency spectra are found to match well with the measurements from \citet{george2015} and \citet{reichardt2021}. Furthermore, the amplitudes and shapes of the individual CMB secondary components are also found to be consistent at the $\sim10$\% level when compared with the best-fit templates from those studies. It is noted that the shape and amplitude of the kSZ template appears to be different because \citet{george2015} and \citet{reichardt2021} both use the sum of late-time and reionisation kSZ as their template (the latter of which is not incorporated in this work).

\subsection{Galaxy weak lensing}\label{sec:validation_gwl}

\begin{figure}
\begin{center}
\includegraphics[width=1.0\linewidth]{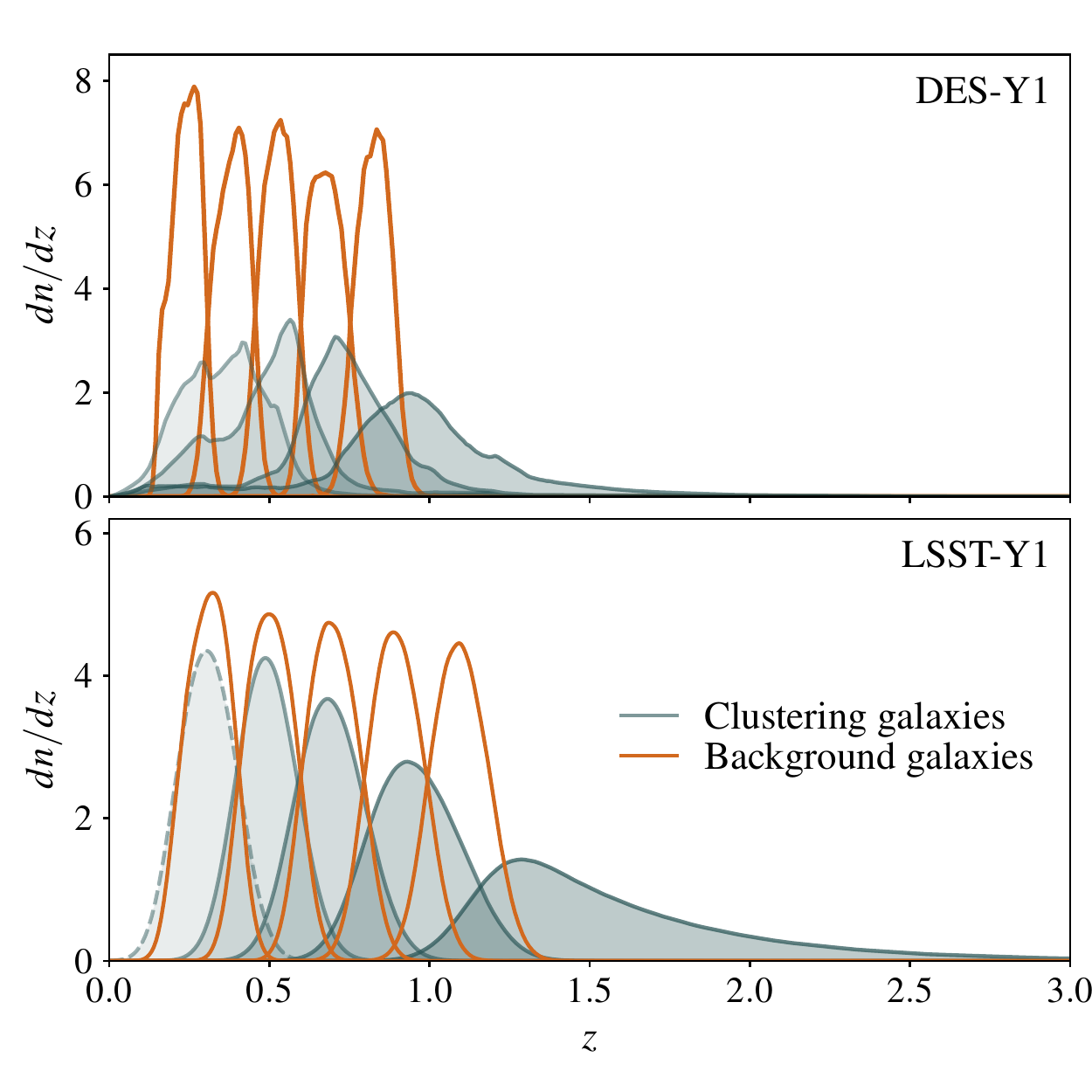}
\caption{ The redshift distributions of DES-Y1 (\textbf{upper}) and LSST-Y1 DESC-SRD galaxy samples (\textbf{lower}). The redshift distributions for the DES-Y1 are extracted from data, whereas the distributions for the LSST-Y1 are derived analytically. The first background galaxy bin for the LSST-Y1 sample is dropped, as the redshift distribution overlaps significantly with the first clustering galaxy bin. }
\label{fig:dndz}
\end{center}
\end{figure}

\begin{figure*}
  \centering
  \begin{minipage}[t]{1.00\textwidth}
    \includegraphics[width=\textwidth]{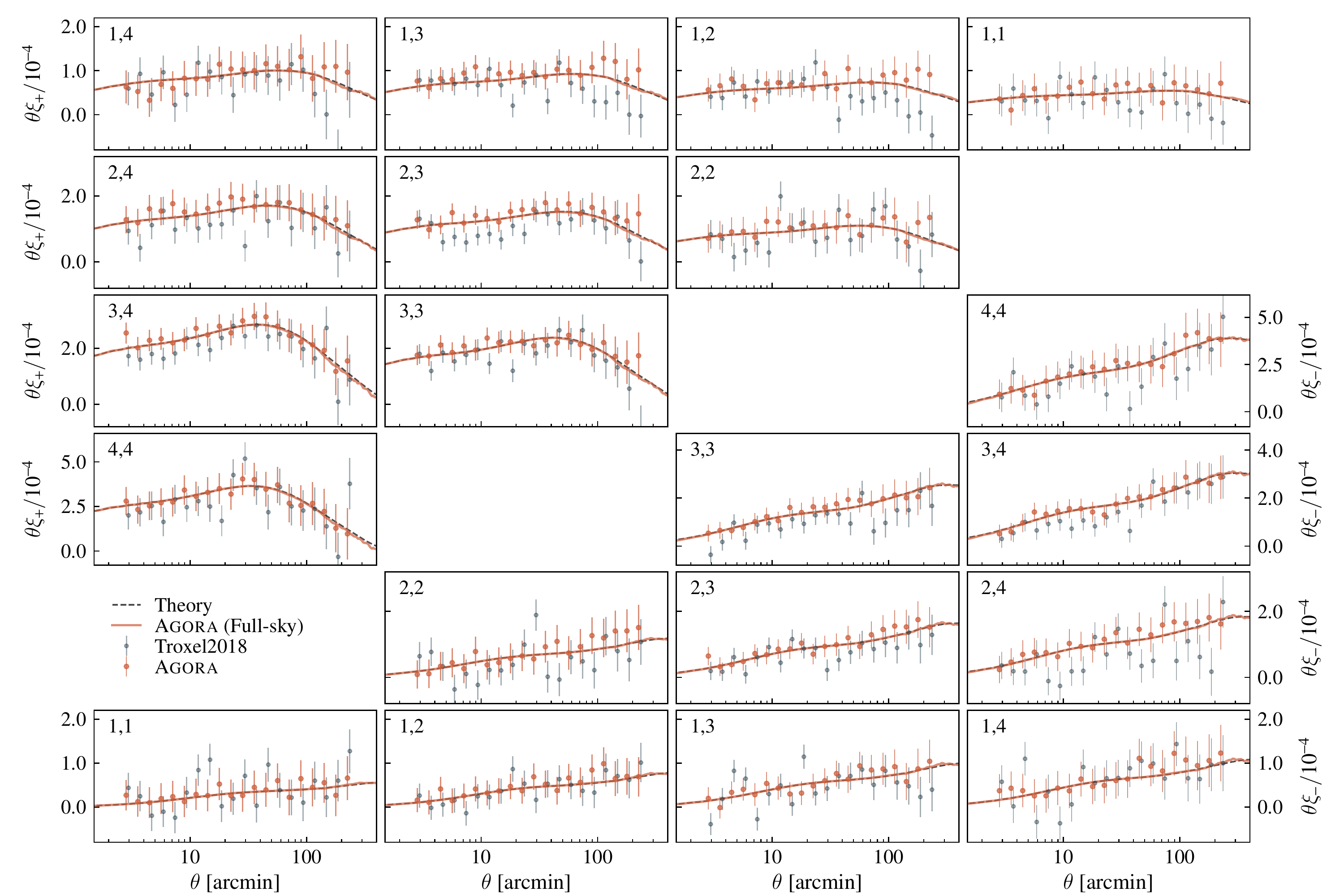}
    \caption{Measurements of $\xi_{\theta,\pm}$ using the simulated DES-Y1 catalogue using the 2-point estimator (with IA added) are shown as orange points, the measured full-sky $C_{\ell}^{\kappa\kappa}$ converted into $\xi_{\theta,\pm}$ are shown as solid orange lines, and the analytical predictions are shown in dashed grey lines. The measurements from DES-Y1 cosmic shear analysis \protect\citep{troxel2018} are shown as grey points.}
    \label{fig:shear_2pt}
  \end{minipage}\par
  \vskip\floatsep
  \begin{minipage}{0.48\textwidth}
    \includegraphics[width=\textwidth]{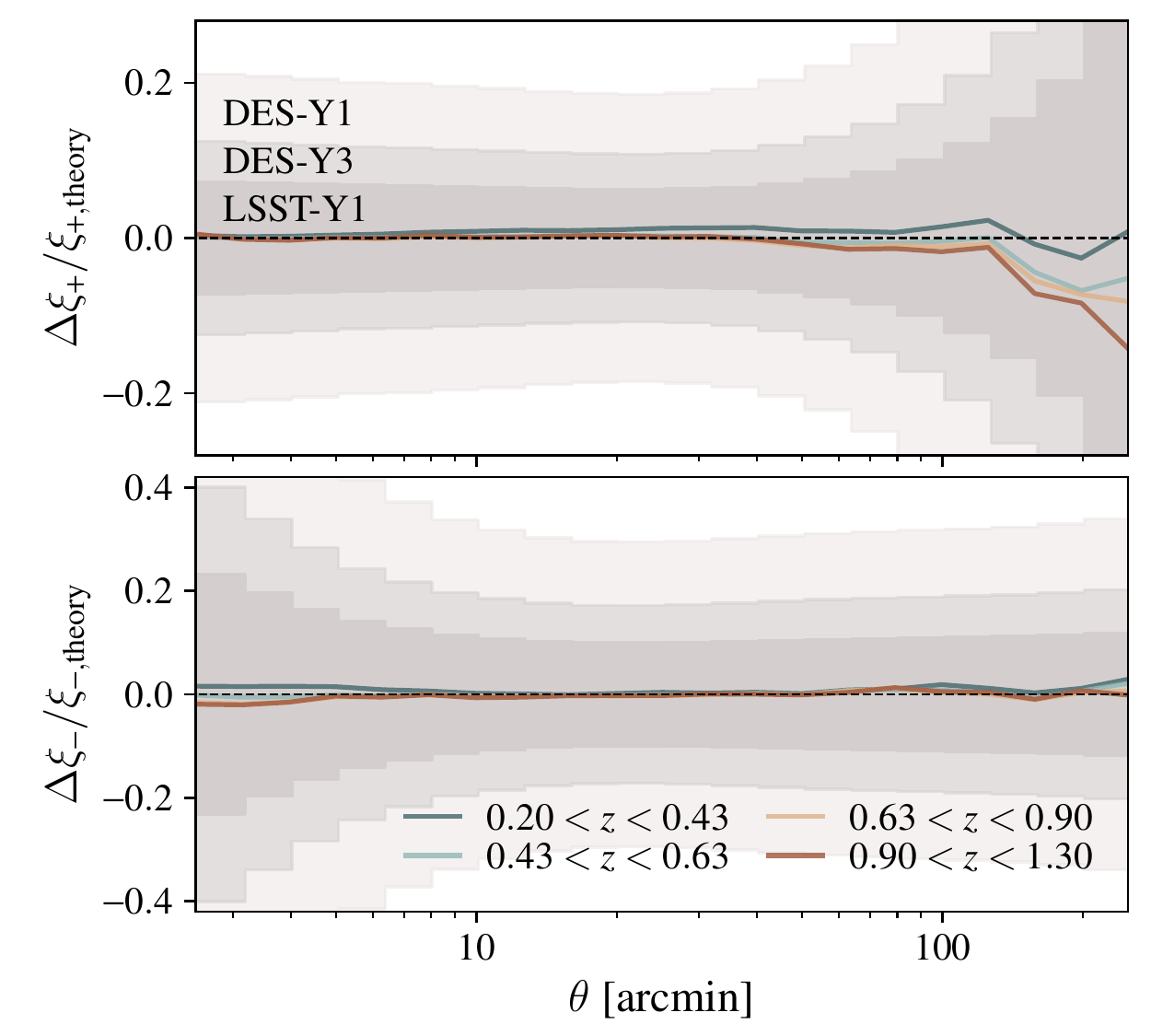}
    \caption{The ratio $\Delta\xi_{\theta,\pm}/\xi_{\theta,\pm}$ where $\Delta\xi_{\theta,\pm}$ is the difference between the measured noiseless shear+IA field and the theoretical model. The three shaded regions represent the DES-Y1, DES-Y3 and LSST-Y1 uncertainties (obtained by scaling the DES-Y3 covariance by ratios of $f_{\rm sky}$).}
    \label{fig:shear_bias}
  \end{minipage}
  \hfill
  \begin{minipage}{0.48\textwidth}
    \includegraphics[width=\textwidth]{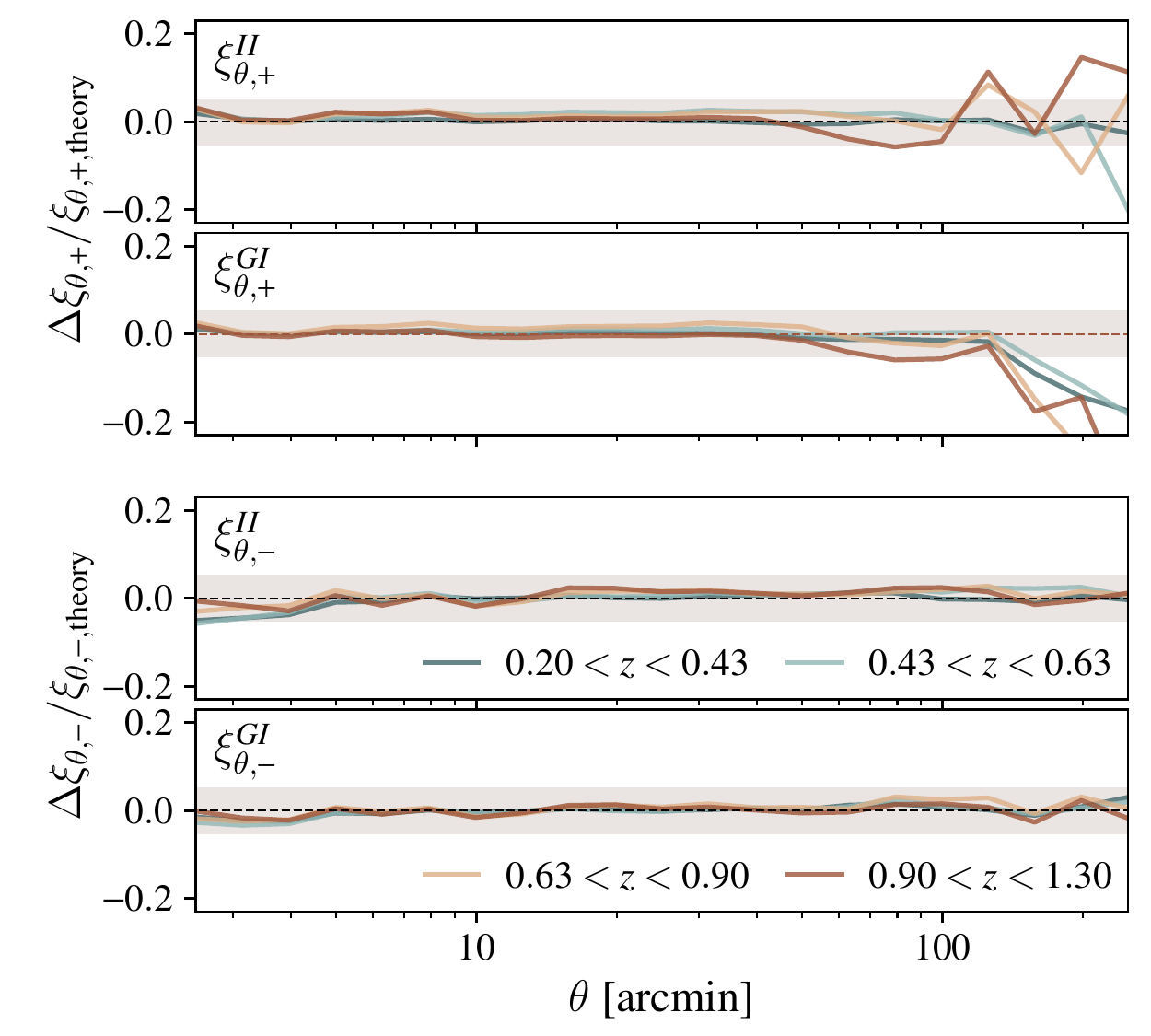}
    \caption{
    Ratios of measured $\xi_{\theta,\pm}^{\rm GI,II}$ to analytical predictions, both assuming the NLA model from Equation \eqref{eq:kappa_IA}.The grey region represents $\pm5\%$ intervals.
    } 
    \label{fig:shear_2pt_IA}
  \end{minipage}
\end{figure*}

A simulated shape catalogue is created in this section using the redshift distributions of the background galaxies from DES-Y1 (shown in the upper panel of Figure \ref{fig:dndz}), and adding shape noise using the random rotation approach described in Section \ref{sec:mock_shearcatalog}. The validity of the shear catalogue is verified by comparing the measured shear two-point correlation $\xi_{\theta,\pm}$ with an analytical model and measurements from \cite{troxel2018}, where the shear two-point correlation function is calculated using \citep{troxel2018}:
\begin{equation}
\xi_{\theta,\pm}^{ij}=\frac{\sum_{ab} W_{a}W_{b}[e_{a,t}^{i}(\vec{\theta})e_{b,t}^{j}(\vec{\theta})\pm e_{a,\times}^{i}(\vec{\theta})e_{b,\times}^{j}(\vec{\theta})]}{\sum_{ab} W_{a}W_{b}}.
\end{equation}
Here, $e_{a,t}$ is the tangential shear component of the ellipticity of galaxy $a$, measured with respect to the line connecting galaxies $a$ and $b$. $W_{a,b}$ denote the weights for the individual galaxies (which are set to one in this work, since systematic effects are ignored), and $i,j$ denote the redshift bin indices. The measurements are shown in Figure \ref{fig:shear_2pt}. The total cosmic shear measurements are shown to be visually consistent with the measurements from \citet{troxel2018}.

Next, the accuracy of the signal component of the lensing maps is examined. Since the goal here is to verify the accuracy of the underlying signal, the measurements are carried out using noiseless full-sky maps. Furthermore, since measuring the position-space two-point correlation function is computationally inefficient for a full-sky map, instead, the auto- and cross-spectra are measured, and converted into position-space correlation function measurements using a Hankel transform:
\begin{align}
\xi^{i,j}_{\theta,\pm}=&\int\frac{\ell {\rm d}\ell}{2\pi}J_{0/4}(\ell\theta)\hspace{0.1cm}C_{\ell}^{\kappa^{i}\kappa^{j}}.
\end{align}
Figure \ref{fig:shear_bias} shows the estimated bias in the signal-only simulation map, quantified as $\Delta(\xi_{\theta,\pm})/\sigma(\theta,\xi_{\theta,\pm})=(\xi^{\rm sim}_{\theta,\pm}-\xi_{\theta,\pm}^{\rm theory})/\sigma(\xi_{\theta,\pm})$, and the uncertainties for DES-Y1, DES-Y3 and LSST-Y1 are considered.\footnote{The covariance matrices for DES-Y1 and DES-Y3 cosmic shear measurements are available at \url{http://desdr-server.ncsa.illinois.edu/despublic/y1a1_files/chains/2pt_NG_mcal_1110.fits} and \url{http://desdr-server.ncsa.illinois.edu/despublic/y3a2_files/datavectors/2pt_NG_final_2ptunblind_02_26_21_wnz_maglim_covupdate.fits} respectively. For DES-Y3, the uncertainties are simply overlaid on top of the DES-Y1 data vectors. Uncertainties for LSST-Y1 are estimated by simply scaling DES-Y3 uncertainties by a factor of 0.58 to account for the difference in $f_{\rm sky}$.} The accuracy of the simulation is shown to be significantly better than the statistical uncertainties of the measurements from   DES-Y1 and DES-Y3. On the other hand, for LSST-Y1, although the measurements of  $\xi_{\theta,-}$ fall within $1\sigma$ of the statistical uncertainties, deviations are noticeable at $\theta>100'$ for $\xi_{\theta,+}$, which may result in a slightly increased $\chi^2$.

The ratios between the measured intrinsic alignment correlation functions in the simulation and analytical predictions are shown in Figure \ref{fig:shear_2pt_IA}. Agreements of order 5\% for $\xi_{\theta,-}$ in the angular range $2.5' <\theta< 250'$ are found for these intrinsic alignment correlations. Agreement is also found for $\xi_{\theta,+}$ up to $\theta=100'$, but the scatter increases beyond $100'$. However, since the amplitudes of the GI and II correlations are no more than 20\% (for the IA model adopted in this work) of the true cosmic shear signal, this level of disparity is unlikely to have a significant impact on the final cosmological results.

Finally, it is emphasized that the simulation catalogues do not include survey systematics such as photometric redshift errors, biases from galaxy selections, shear calibration biases, and baryonic effects on the matter power spectrum, which must be implemented separately to generate more realistic shear catalogues. Since many of these systematic effects are specific to the galaxy sample being analysed, these considerations are left to the users to implement.

\section{Applications}\label{sec:applications}
In the previous section, the power spectra or correlation functions of the individual components were verified to closely match with those measured from observational data and theoretical models. In this section, certain aspects of various analyses are carried out to showcase how the various simulation products can be used in a high-level analysis.

\subsection{Calibration of products}\label{sec:calibration}
So far, all validation tests have been based on raw simulation outputs.  While the simulation products are generally consistent with data, minor differences are expected. To compensate for these minor differences, small calibration factors of no more than 5\% are applied to bring the simulation products closer to observational data. Table \ref{tab:CIB_MC_corrections} summarizes the calibration factors used, and simulation products with these calibration factors applied are used in the sections that follow, unless otherwise specified.

\begin{table}
\begin{center}
\begin{tabular}[t]{lccc}
\toprule
 \multirow{2}{*}{Experiment} & \multirow{2}{*}{Frequency} & \multicolumn{2}{c}{\hspace{-0.2cm}Correction factor}\\
 & & CIB  & Radio  \\
\midrule
SPT-SZ & 90  & 0.95 & 1.00 \\
SPT-SZ & 150 & 0.95 & 1.00\\
SPT-SZ & 220 & 1.03 & 1.00\\
\hline
SPT-3G & 90  & 0.95 & 1.00\\
SPT-3G & 150 & 0.95 & 1.00\\
SPT-3G & 220 & 1.03 & 1.00\\
\hline
{\it Planck} & 100 & 1.05 & 1.00\\
{\it Planck} & 143 & 1.05 & 1.00\\
{\it Planck} & 217 & 0.97 & 1.00\\
{\it Planck} & 353 & 1.00 &  - \\
{\it Planck} & 545 & 1.00 & - \\
{\it Planck} & 857 & 1.00 & -\\
\bottomrule
\end{tabular}
\caption{Table summarizing the calibration factors applied to the CIB/radio maps such that the measured power spectrum match better with observations.}
\label{tab:CIB_MC_corrections}
\end{center}
\end{table}

\subsection{Multi-probe cosmological constraints}\label{sec:6x2pt}
{\it Products used: LSST-Y1-like galaxy density maps, LSST-Y1-like noiseless convergence maps, Simon's Observatory-like CMB lensing map.}\\[0.05cm]

Various studies have combined galaxy clustering, cosmic shear, and galaxy-galaxy lensing to place tight constraints on the cosmological parameters that describe the composition of the Universe and those that characterize the evolution of structure formation in the late-time Universe \citep{desy1_3x2,desy1_6x2,kids1000_3x2,desy3_3x2,miyatake2022,desy3_6x2}. 

Using combinations of observational probes helps break down intricate degeneracies between cosmological and nuisance parameters. A well-known example of this effect is the degeneracy breaking between galaxy bias $b_{\rm g}$ and $\sigma_{8}$ using the combination of galaxy clustering and galaxy-galaxy lensing, which takes advantage of the fact that the two probes rely on different powers of the galaxy bias.

One of the greatest challenges for upcoming experiments is the treatment of survey systematics.  For these surveys to achieve their respective scientific goals,  exquisite understanding and handling of systematic effects will be required. While improvements to the modelling of systematic effects are continuously being made, data will always contain some degree of unknown systematic effects. Identifying the origins of these biases may not always be trivial, especially in scenarios where several effects are intricately related. Therefore, it is often useful to have cross-correlation measurements between observables, which are less prone to systematics, and can therefore be used for consistency checks. For cross-survey correlations, the data will not share the same experimental systematic effects, and therefore the number of possible nuisance effects is significantly reduced.\footnote{Here, it is emphasized that cross-correlations between unrelated surveys nullify survey systematics, but undesired astrophysical and cosmological systematic effects will remain.} 

Future galaxy surveys such as LSST,  {\it Euclid}, and {\it Roman} are expected to benefit greatly from complementary information between surveys, either at the raw pixel level \citep{rhodes2017,chary2019,capak2019} or at the data vector level \citep{eifler2021}. The combination of these surveys would aid in reducing systematic effects and improve the overall figure-of-merit \citep{jain2015}. Furthermore, future CMB surveys will play a role in constraining key systematic effects such as multiplicative shear calibration bias \citep{vallinotto2012,schaan2017} and intrinsic alignment \citep{schaan2017} through cross-correlation measurements, and will therefore provide powerful complementary information. 

For a simulation to be useful in these cosmological analyses, the input signal must behave as expected. Simulation artefacts and biases must be small enough such that when a full cosmological analysis is performed, the final results are consistent with the expected results to well within the statistical uncertainties of the dataset under consideration. This is especially important when testing the impact of additional systematic effects, because distinguishing the source of cosmological parameter shifts in the presence of both simulation artefacts and injected systematics can be challenging.

In this section, an idealized (i.e., signal-only) multi-probe analysis is carried out to ensure that simulation artefacts are sufficiently small, and to show that the input cosmology is correctly recovered when the simulation products are subjected to a full cosmological analysis. To demonstrate this, all the possible probe combinations that could be formed using the galaxy density, galaxy weak lensing and CMB lensing fields are measured, and parameter constraints are calculated using particular combinations of these measurements.

\begin{table}
\begin{center}
\begin{tabular}[t]{cccccc}
\toprule
 & & \hspace{-1.2cm}LSST-Y1 DESC-SRD galaxies\\
\toprule
& Clustering & & & &\\
& $\bar{n}_{{\rm g}, l}$ & $1.78, 1.77, 1.78, 1.78, 1.78$\\
& $b_{\rm g}$ & $1.23,\ 1.36,\ 1.50,\ 1.65,\ 1.80$\\
& $(z_{0},\alpha)$ & (0.26, 0.94)\\
\midrule
& Background & & & &\\
& $\bar{n}_{\rm g, s}$ & $2.25, 3.11, 3.09, 2.61, 2.00$\\
& $\sigma_{\rm e}$ & $0.26, 0.26, 0.26, 0.26, 0.26$\\
& $(z_{0},\alpha)$ & (0.13, 0.78)\\
\bottomrule
\end{tabular}
\caption{Values of linear galaxy bias, galaxy number density (per arcmin$^{2}$), and shape noise assumed for the 5 redshift bins of the LSST-Y1 DESC-SRD clustering and background galaxy samples.}
\label{tab:6x2setup}
\end{center}
\end{table}

Simulated LSST-Y1 galaxy density and noiseless galaxy weak lensing maps are produced for this purpose, using the configurations from the Dark Energy Science Collaboration Scientific Requirement Document (DESC-SRD; \citealt{descsrd}). The values for the galaxy bias are derived from the analytic relation $b_{\rm g}(z)=1.05/D(z)$, and are summarized in Table \ref{tab:6x2setup}. The redshift distributions for the galaxy samples are generated by first computing the parent redshift distributions using:
\begin{equation}\label{eq:descdndz}
\frac{dn}{dz}\propto z^{2}\exp(-z/z_{0})^{\alpha},
\end{equation}
where $(z_{0}, \alpha)=(0.26, 0.94)$ for the clustering galaxies and $(z_{0}, \alpha)=(0.13, 0.78)$ for the background galaxies. These distributions are normalized such that the total number density of galaxies equal 18 and 9.52  galaxies/arcmin$^{2}$ respectively.

The individual redshift bins of clustering galaxies (i.e., $dn_{l}^{j}/dz$) are obtained by selecting galaxies in the redshift range $0.2<z<1.2$ from the first parent distribution and dividing into 5 redshift bins, each with redshift bin width 0.2, and applying a photo-$z$ scatter of $\sigma(z)=0.03(1+z)$. The redshift bins of background galaxies  (i.e., $dn_{\rm s}^{j}/dz$) are obtained by dividing the second parent distribution into 5 redshift bins with equal numbers of galaxies, and applying a photo-$z$ scatter of $\sigma(z)=0.05(1+z)$. These redshift distributions are shown in the lower panel of Figure \ref{fig:dndz}.

The Limber integral:
\begin{equation}
C_{\ell}^{XY}=\int d\chi \frac{1}{\chi^2}W_{ X}(\chi)W_{Y}(\chi)P\left(\frac{\ell+1/2}{\chi},z(\chi)\right),
\end{equation}
is used to model all two-point functions, where $P$ is the nonlinear matter power spectrum and $W_{X},W_{Y}$ are the kernel functions which can be $W_{\kappa_{\rm CMB}}$ or $W_{\delta}$, as defined in Equations \eqref{eq:W_kcmb} and  \eqref{eq:W_gal}, or $W_{\gamma}$ defined as:
\begin{align}
W^{j}_{\gamma}(\chi)&=\frac{3\Omega_{\rm m}H_{0}^{2}}{2c^2}\frac{\chi}{a(\chi)}\int_{\chi}^{\infty}d\chi' \frac{dn^{j}_{\rm s}}{dz}\frac{dz}{d\chi'}\frac{\chi'-\chi}{\chi'}.
\end{align}

The analysis is carried out in harmonic-space as it is easier to incorporate CMB lensing (for which a SO-like CMB lensing map is produced with the lensing noise spectrum computed using the specifications from \citealt{simons}). For the purpose of demonstrating that the recovered cosmological constraints from our simulation are unbiased relative to the input cosmology, additional simplifications are assumed:
\begin{enumerate}[leftmargin=\parindent,align=left,labelwidth=\parindent,labelsep=0pt]
\item A Gaussian covariance matrix is used, ignoring the non-Gaussian and the super-sample covariance contributions since these components have been shown to be sub-dominant \citep{krause2017}. Since the purpose here is to assess the deviations between the recovered and input cosmologies, this is a conservative choice, because the inclusion of extra covariance terms would imply a smaller discrepancy.
\item The cosmological dependence of the so-called $N^{(1)}_{L}$ bias (see e.g., \citealt{planck2015xv}) in the CMB lensing likelihood is ignored, since the CMB lensing auto-spectrum is not the dominant probe in this analysis. 
\end{enumerate}

\subsubsection{Analysis setup}

The analysis choices described in the DESC-SRD are directly adopted in this work. Angular scales in the range $\ell_{\rm max}< k_{\rm max}\chi(\langle z \rangle)-0.5$ are used for galaxy clustering and galaxy-galaxy lensing measurements, where $k_{\rm max}\sim 0.3 h^{-1}{\rm Mpc}$, and multipoles in the range $30<\ell<3000$ are used for cosmic shear measurements. Since CMB lensing is not one of the probes considered in the DESC-SRD, no scale cuts are defined; however, for the purpose of this demonstration, the same angular range as cosmic shear is adopted. Furthermore, only the auto-redshift bin combinations are used for galaxy clustering measurements and only the redshift bin combinations that have the clustering galaxies in front of the background galaxies are used for galaxy-galaxy lensing.\footnote{While this is not explicitly stated in \citealt{descsrd}, the redshift bin combinations included in the forecast can be found in \url{https://github.com/CosmoLike/DESC_SRD}.}

Since the goal is to quantify the bias in the recovered cosmology relative to the projected statistical uncertainties, power spectra measured from noiseless full-sky maps are used as the data vector, and the covariance matrix is computed with noise and assuming a sky area of 12,300$\ {\rm deg}^{2}$ ($f_{\rm sky}=0.3$).

\subsubsection{Results}
\begin{figure*}
\begin{center}
\includegraphics[width=1.0\linewidth]{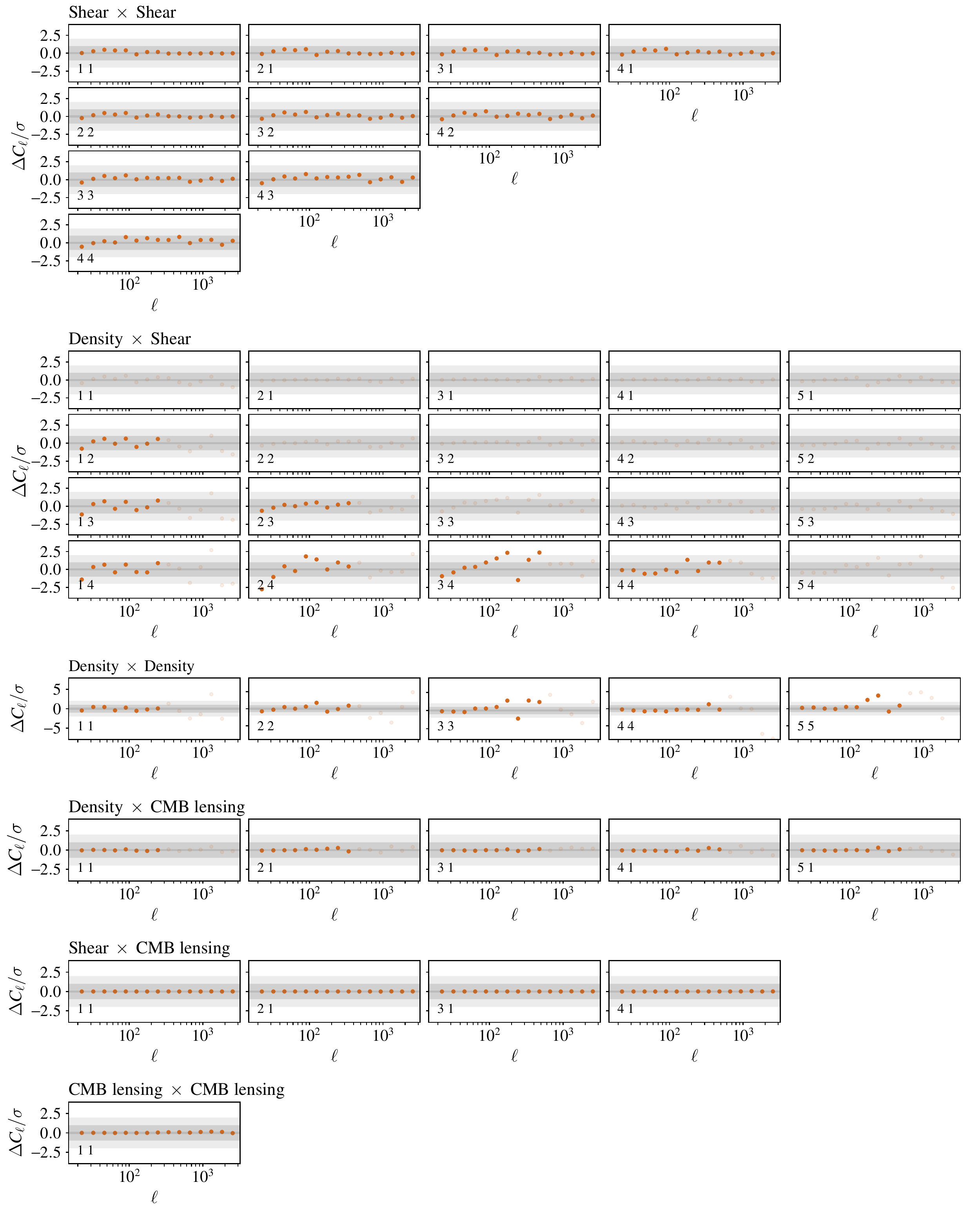}
\caption{Ratios of differences in measured and model spectra to expected statistical uncertainties for cosmic shear, galaxy-galaxy lensing, galaxy clustering, galaxy-CMB lensing, shear-CMB lensing, and CMB lensing spectra. The $1\sigma/2\sigma$ statistical errors are shown as light/dark grey bands. 
}
\label{fig:6x2_cldiffs}
\end{center}
\end{figure*}

\begin{figure*}
\centering
\includegraphics[width=1.0\linewidth]{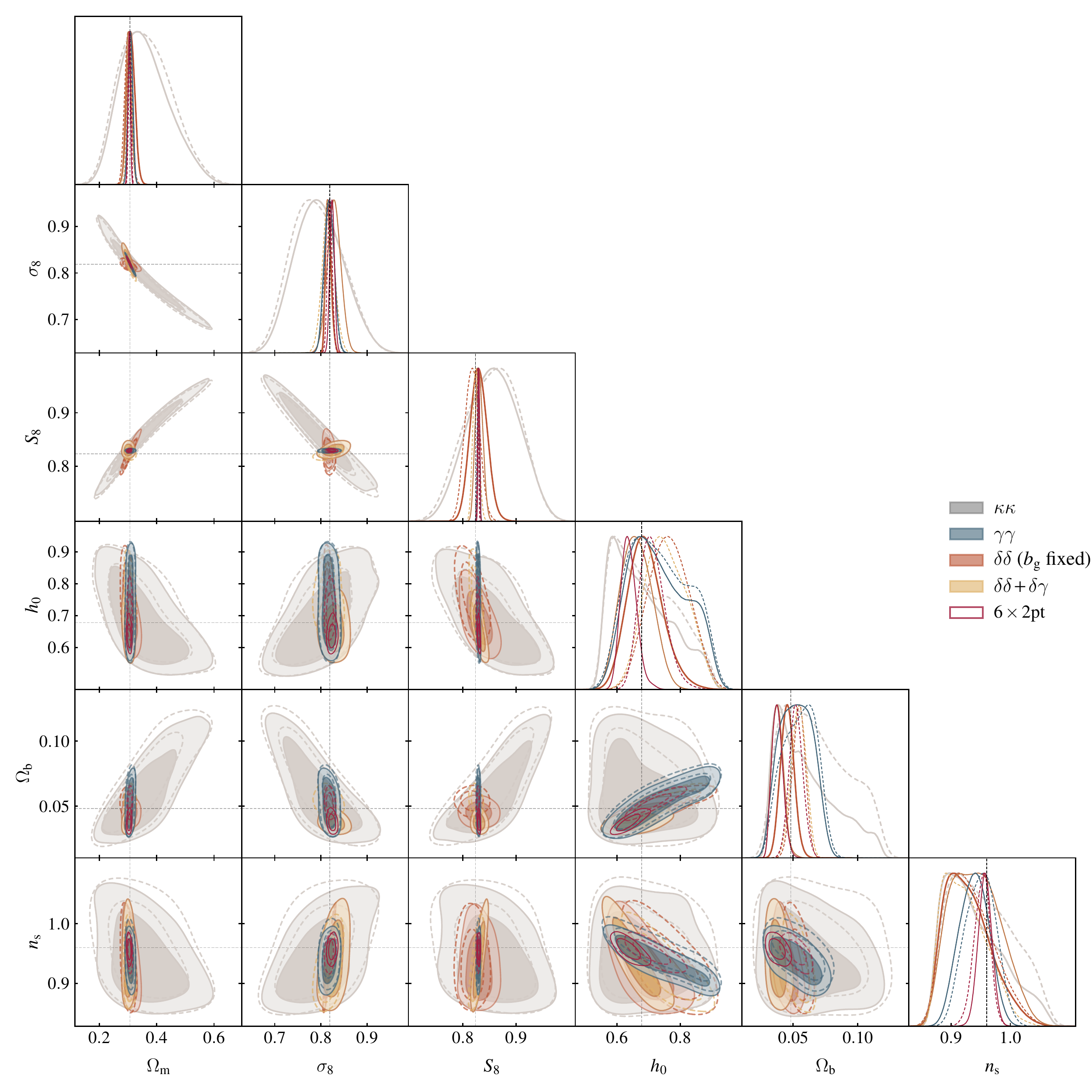}\llap{\raisebox{10.25cm}{\includegraphics[height=7.25cm]{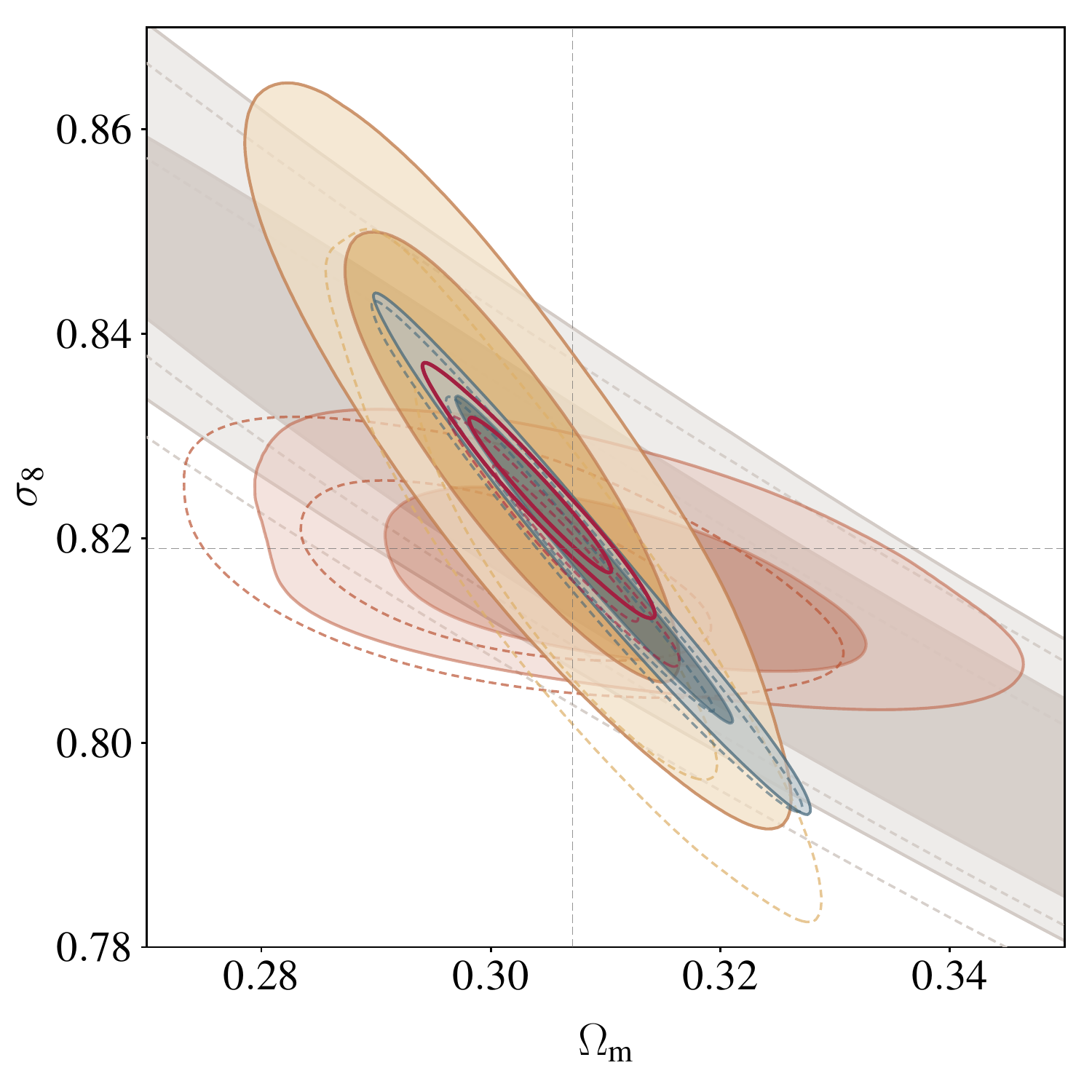}}}
\caption{
 Cosmological constraints obtained from CMB lensing (grey), cosmic shear (blue), galaxy clustering with fixed bias (red), galaxy clustering + galaxy-galaxy lensing (gold) and the combination of all possible two-point functions (crimson). The recovered cosmology is consistent with the input cosmology to within $\sim1.5\sigma$ in the $\Omega_{\rm m}$-$\sigma_{8}$ plane. Also shown as open dashed lines are the corresponding constraints obtained using simulated data vectors.}
\label{fig:constraints_6x2pt}
\end{figure*}

Figure \ref{fig:6x2_cldiffs} compares the differences in the measured and model spectra with the expected statistical uncertainties. For cosmic-shear, CMB lensing auto-spectrum, and the two cross-spectra that involve CMB lensing, the measured noiseless spectra lie well within the statistical uncertainties. However, for galaxy clustering and galaxy-galaxy lensing, the scatter in the noiseless data vector is larger than the expected scatter. This is mainly due to the high number density assumed for the clustering galaxy sample, necessitating a highly accurate and precise simulation, which is numerically challenging to produce. As a result, additional contributions to the total $\chi^2$ coming from the simulation itself are inevitable. Nonetheless, the constraints on cosmological parameters from the individual probes $\cnn{}$ (with fixed galaxy bias), $\gamma\gamma$, $\ckk{}$ as well as the combination of $\cnn{}+\cggl{}$ (2$\times$2pt) and the full combination (6$\times$2pt) are shown in Figure \ref{fig:constraints_6x2pt}, and are all  shown to be consistent with the input cosmology to within  $1.5\sigma$ in the $\Omega_{\rm m}$-$\sigma_{8}$ plane.

\subsection{Testing Compton-$y$ component separation }\label{sec:app_compsep}
\subsubsection{Planck MILCA $y$-map }
{\it Simulation products used: lensed Planck 100/143/217/353/545/857 GHz maps and Galactic foreground maps from PySM3.}\\[0.2cm]
In this section, simulated {\it Planck} frequency maps with Galactic dust and synchrotron added\footnote{Here, the \texttt{d1}, \texttt{s1} models from \textsc{PySM3} \citep{thorne2017} are used for the dust and synchrotron components, respectively. The dust templates are scaled by factors of 1.0, 1.1, 1.20, 1.14, 1.05, 0.69 at 100, 143, 217, 353, 545, 857 GHz channels respectively, such that the mean level at the Galactic centre matches the data.} are passed through the Compton-$y$ map making procedure described in \citet{planck2015xxii}. More specifically, the MILCA Compton-$y$ map is produced using the MILCA weights, filters, and full focal plane simulation (FFP8) noise realisations.\footnote{The noise realisations and MILCA weights, as well as the weight propagation code are available on the {\it Planck} legacy archive.} The same reconstruction procedure used for data is applied directly to the simulated frequency maps, with and without noise, and the measured auto- and cross-spectra are compared to those measured from data.

\begin{figure}
\begin{center}
\includegraphics[width=1.00\linewidth]{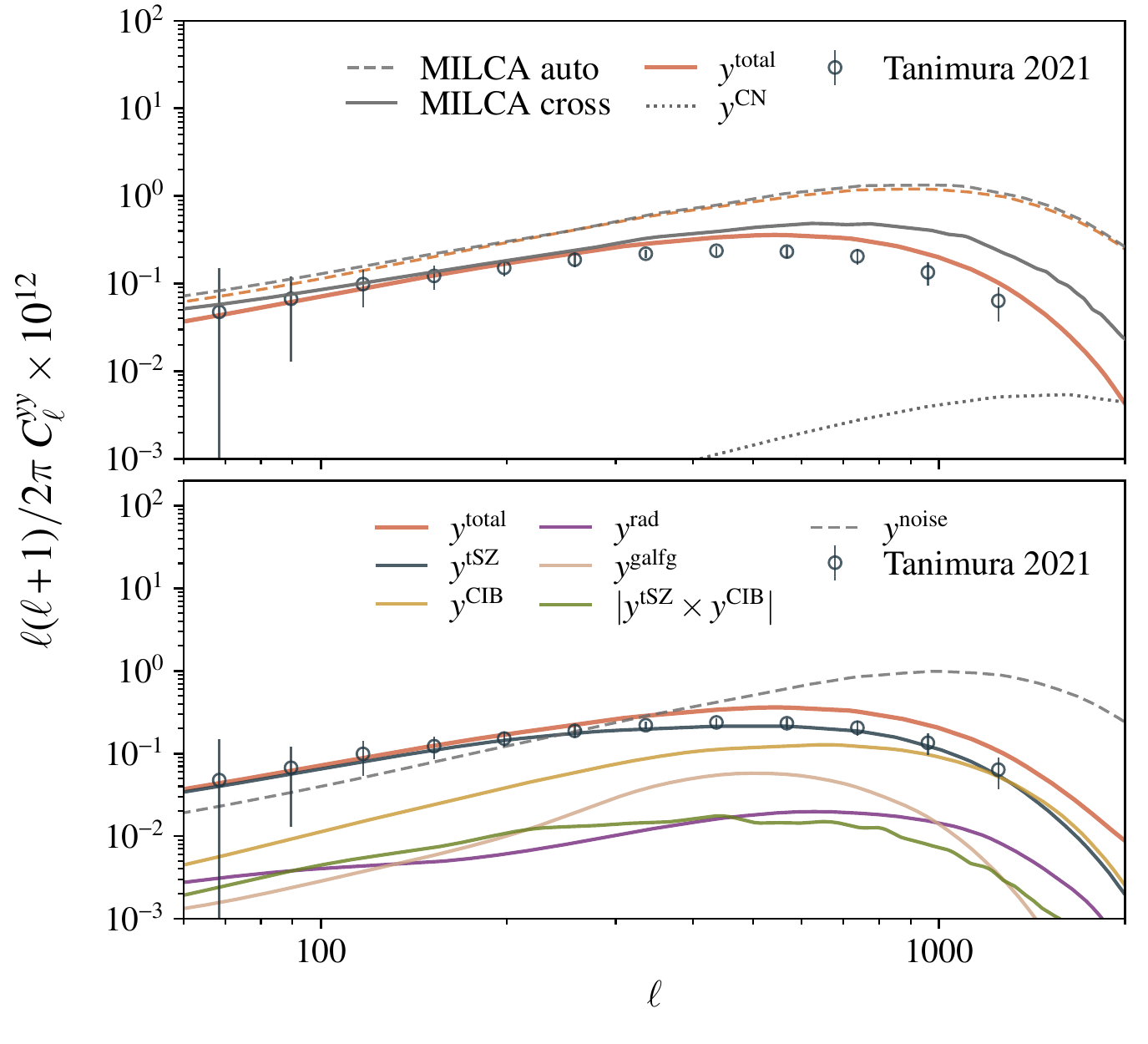}
\caption{ {\bf Upper:} Comparison of the auto and half-depth cross-spectrum of the data MILCA Compton-$y$ map and those from the simulated MILCA Compton-$y$ map. While the auto-correlations (dashed orange and dashed grey lines) are in good agreement, the half-depth cross-spectrum for the simulated MILCA Compton-$y$ (for which noiseless maps are used instead) is lower than the measurement from data at $\ell>400$. {\bf Lower:} The no-noise total Compton-$y$ spectrum (solid orange line in the upper panel) broken up into various foreground components obtained by passing the individual component maps through the MILCA reconstruction pipeline: tSZ (blue line), CIB (gold line), radio (purple line), galactic foregrounds (beige line), and tSZ-CIB correlation (green line). 
}
\label{fig:mdpl2_milca_yyspec}
\end{center}
\end{figure}

The results are shown in Figure \ref{fig:mdpl2_milca_yyspec}. The total $C_{\ell}^{yy}$ auto-spectrum measured from the simulation is found to be in good agreement with the power spectrum measured from the data MILCA map. For half-depth cross-spectrum, a slight deficit in power beyond $\ell>500$ is found for the simulated MILCA map. It has been confirmed that the difference is not due to the correlated noise contribution,\footnote{The correlated noise power spectrum is taken from Table 3 of \citet{bolliet2018}. } which is not accounted for in the simulated MILCA Compton-$y$ map. Investigation on this point is left for future work.

Individual foreground components are passed through the Compton-$y$ map making procedure separately to investigate how each of these leak into the resulting Compton-$y$ map. These contributions are denoted as $y^{X}$ where $X\in\{$tSZ, kSZ, CIB, CMB, radio, galactic foreground, noise$\}$.

The power spectrum of the total Compton-$y$ map, as well as the individual foreground components are shown in the lower panel of Figure \ref{fig:mdpl2_milca_yyspec}. At small scales, the CIB is the leading foreground component that contributes the most to the total power, which is consistent with findings from other studies (such as those from \citealt{planck2015xxii,shirasaki2019b}). At large scales, tSZ dominates the signal, but there are small contributions from CIB and Galactic foregrounds. The recovered tSZ signal is also consistent with the inferred measurements from \citet{tanimura2021}.

\subsubsection{Biases in SPT-SZ + Planck Compton-$y$ map and impact on cross-correlation measurements}

{\it Simulation products used: mock Planck 100/143/217/353 GHz maps, mock SPT-SZ 95/150/220 GHz maps, and mock LSST-Y1 galaxy weak lensing maps.}\\[0.2cm]
\begin{figure}
\begin{center}
\includegraphics[width=1.0\linewidth]{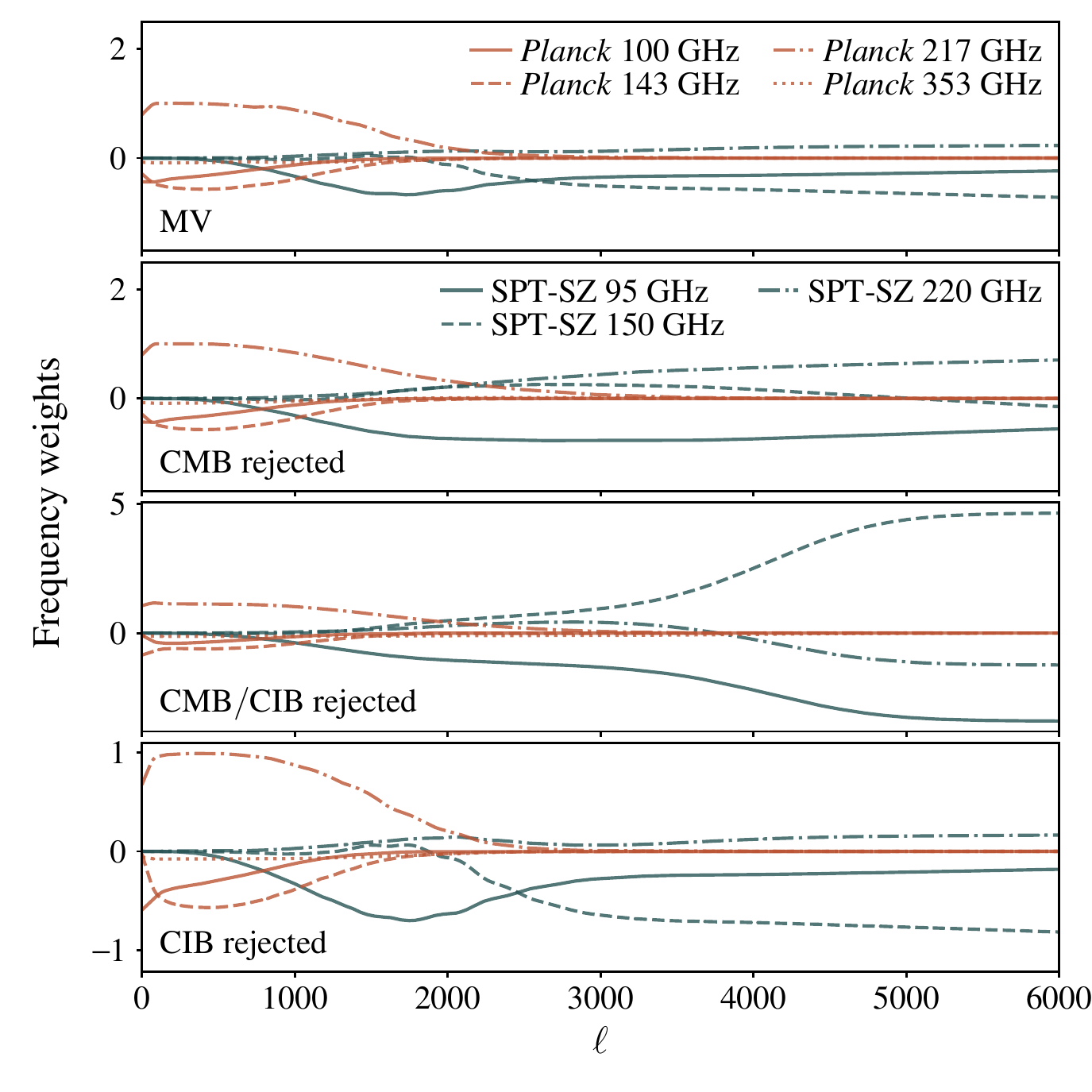}
\caption{Weights for the four variants of the Compton-$y$ map from \citet{bleem2021}, which are directly applied to the simulated frequency maps. The weights for the SPT channels (navy lines) and \planck{} channels (orange lines)  are shown to illustrate the contributions of the two experiments at different $\ell$. }
\label{fig:ymap_weights}
\end{center}
\end{figure}

\begin{figure}
\begin{center}
\includegraphics[width=1.0\linewidth]{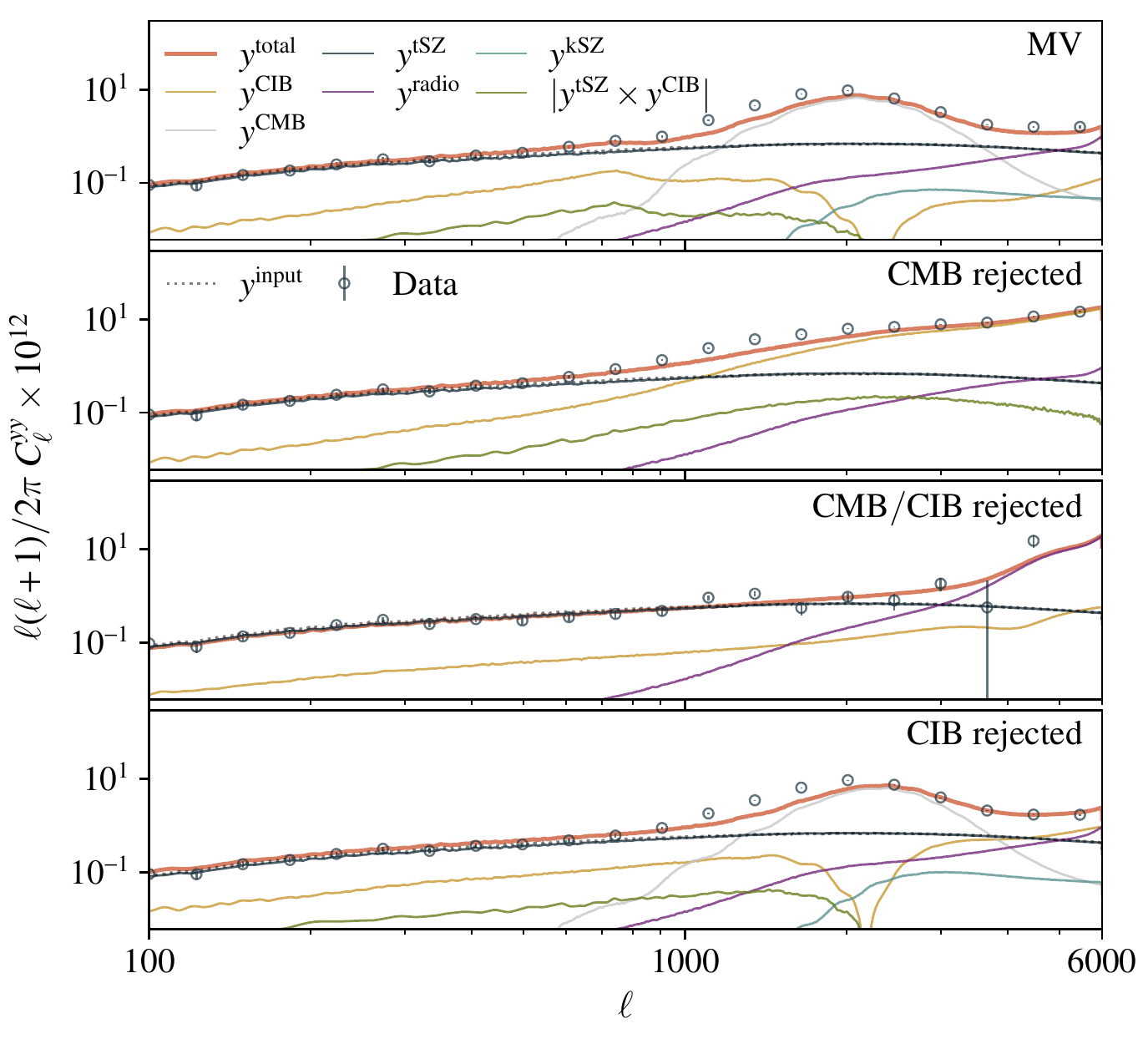}
\caption{Figure showing the recovered Compton-$y$ power spectra obtained by applying the frequency weights from \protect\citet{bleem2021} to the simulated SPT-SZ and \planck{} frequency maps (orange solid line). Power spectra of the other components in the Compton-$y$ map are obtained by applying the same weights to the individual components separately (coloured lines). As a comparison, the measured spectra from the data (navy points) and the input tSZ auto-spectrum (black dotted lines) are also shown. }
\label{fig:powerspec_ymap}
\end{center}
\end{figure}

In this section, frequency weights derived in \citet{bleem2021}, and shown in Figure \ref{fig:ymap_weights}, are applied to simulated SPT-SZ 95/150/220 GHz and \planck{} 100/143/217/353 GHz frequency maps to generate four variants of the Compton-$y$ map, and the power spectra of those maps are compared to those from data. The results are shown in Figure \ref{fig:powerspec_ymap}: overall, the simulation and data spectra are consistent in the low- and high-$\ell$ regimes, but a mild deficit of power is found at $\ell \sim 1000$ in all the power spectra measured from the simulation, which is especially pronounced for the minimum variance reconstruction. 
 
Next, the frequency weights are applied separately to each extragalactic component (CMB, CIB, tSZ, kSZ, radio). The tSZ component  of the reconstructed Compton-$y$ maps (i.e., $y^{\rm tSZ}$) are shown to match well with the input tSZ signal for all the variants, as expected. On the other hand The amplitudes of the other extragalactic components vary depending on the frequency weights used:
 \begin{enumerate}[leftmargin=\parindent,align=left,labelwidth=\parindent,labelsep=0pt]
\item[-] Minimum variance reconstruction: the CIB, CMB and radio sources are the dominant residuals at low-$\ell$, intermediate-$\ell$ and high-$\ell$ regimes respectively. It is noted that the minimum variance map from  \citet{bleem2021} is not a pure minimum variance map but simultaneously attempts to minimize contributions from the CIB using a template subtraction approach (which is equivalent to the partially constrained ILC method proposed by \citealt{abylkairov2021}), and thus the CIB residual amplitude is lower than a true minimum variance map.\vspace{0.2cm}
\item[-] CMB-rejected reconstruction: the CMB residual is nulled in this case as expected, as well as the kSZ residual. However, as a consequence, the CIB residual amplitude is boosted, and dominates the auto-spectrum at all scales beyond $\ell=1000$. \vspace{0.2cm}
\item[-]  CMB/CIB-rejected reconstruction: while the CIB amplitude is reduced, it is not fully removed. This is due to the difference in the CIB SED assumed in this work and in \citet{bleem2021}, as well as the residual Poisson CIB component leaking into the reconstruction, which is not explicitly deprojected in \citet{bleem2021}. At $\ell>3000$, the radio sources become the dominant contamination in the map. \vspace{0.2cm}
\item[-]  CIB-rejected reconstruction: despite the design of this Compton-$y$ map variant to reduce CIB leakage, it is found that the amplitude of the CIB residual is higher in this map compared to the minimum variance map. This is primarily due to the effectiveness of the template-based nulling used in the minimum variance reconstruction, compared to a SED based approach.
\end{enumerate}

\begin{figure}
\begin{center}
\includegraphics[width=1.0\linewidth]{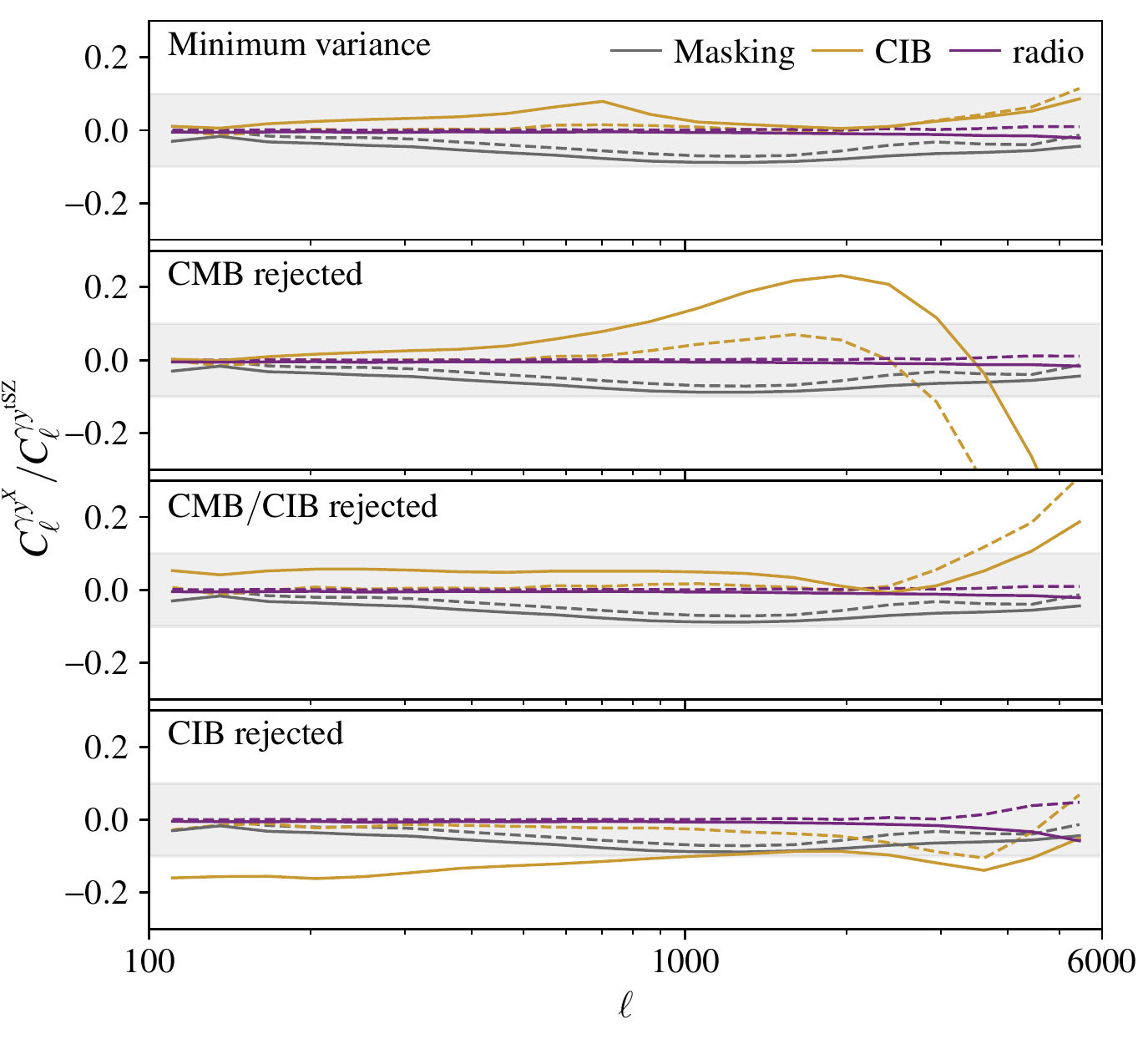}
\caption{Fractional biases to the shear-Compton-$y$ correlation due to CIB (gold lines), radio sources (purple lines), and masking (grey lines) for the four variants of the simulated Compton-$y$ map generated using frequency weights from \protect\cite{bleem2021}. The solid lines correspond to biases in the correlations measured using the fifth redshift bin of the background galaxies from LSST-Y1 as described (see Section \ref{sec:method_gwl}), and the dashed lines represent measurements for the first redshift bin. The grey bands represent an offset of  $\pm10\%$. }
\label{fig:yxg_bias}
\end{center}
\end{figure}

{\it \noindent Biases in tSZ-galaxy weak lensing measurements}\\[0.25cm]
The tSZ component of the reconstructed Compton-$y$ map ($y^{\rm tSZ}$) and the residuals from CIB and radio sources ($y^{\rm CIB}$, $y^{\rm rad}$) are used to investigate the biases in cross-correlation measurements between the Compton-$y$ maps and LSST-Y1 galaxy weak lensing maps (described in Section \ref{sec:6x2pt}). The results for all the Compton-$y$ map variants are shown in Figure \ref{fig:yxg_bias}. Cross-correlations between galaxy lensing maps and the minimum variance Compton-$y$ map produce results that are mildly biased in comparison with the true tSZ signal. The dominant bias comes from the CIB, followed by the masking bias (caused by the point source mask removing regions of high Compton-$y$ values, due to tSZ-radio and tSZ-CIB correlations). Radio sources also have a small impact ($\lesssim 3\%$) on measurements at small scales. The bias from radio sources is also found to be small for the other Compton-$y$ map variants. In contrast, the shape of the CIB bias is found to vary significantly depending on the frequency weights used.

Compton-$y$ maps, as demonstrated by these tests, are intricate maps containing residuals from various astrophysical sources, and the amplitudes of these foreground components are highly dependent on the frequency weights used to construct the Compton-$y$ map. The foreground component maps and frequency channel maps presented in this work can be used to investigate the amplitudes of residuals in a given Compton-$y$ map.

\subsection{Delensing}\label{sec:delensing}

{\it Simulation products used: noiseless ray traced CMB lensing map and lensed CMB maps.}\\[0.2cm]

According to the standard cosmological model, our Universe went through a period of accelerated expansion known as inflation, during which a stochastic background of gravitational waves was created. This leaves a distinct degree-scale "$B$-mode" pattern in the CMB's polarization. One of the primary scientific goals of current and future CMB experiments is to precisely measure these $B$-modes (or, more commonly, the tensor-to-scalar ratio $r$) to constrain the energy scale of inflation in our early Universe.

Ongoing experiments, such as BICEP/{\it Keck} \citep{BK18}, are rapidly approaching a regime in which $B$-mode measurements are no longer limited by instrumental noise, but rather by $B$-modes generated from deflected $E$-modes, also known as lensing $B$-modes. To recover the true primordial $B$-mode signal, these lensing $B$-modes must be removed through a process known as delensing \citep{cmbs4}. Delensing will be a crucial step in obtaining the best possible constraints on $r$, but will also be useful for sharpening the acoustic peaks of the $TT/EE/TE$ spectrum to obtain better constraints on quantities such as $N_{\rm eff}$ \citep{green2017,hotinli2022}.

\begin{figure*}
\begin{center}
\includegraphics[width=1.0\linewidth]{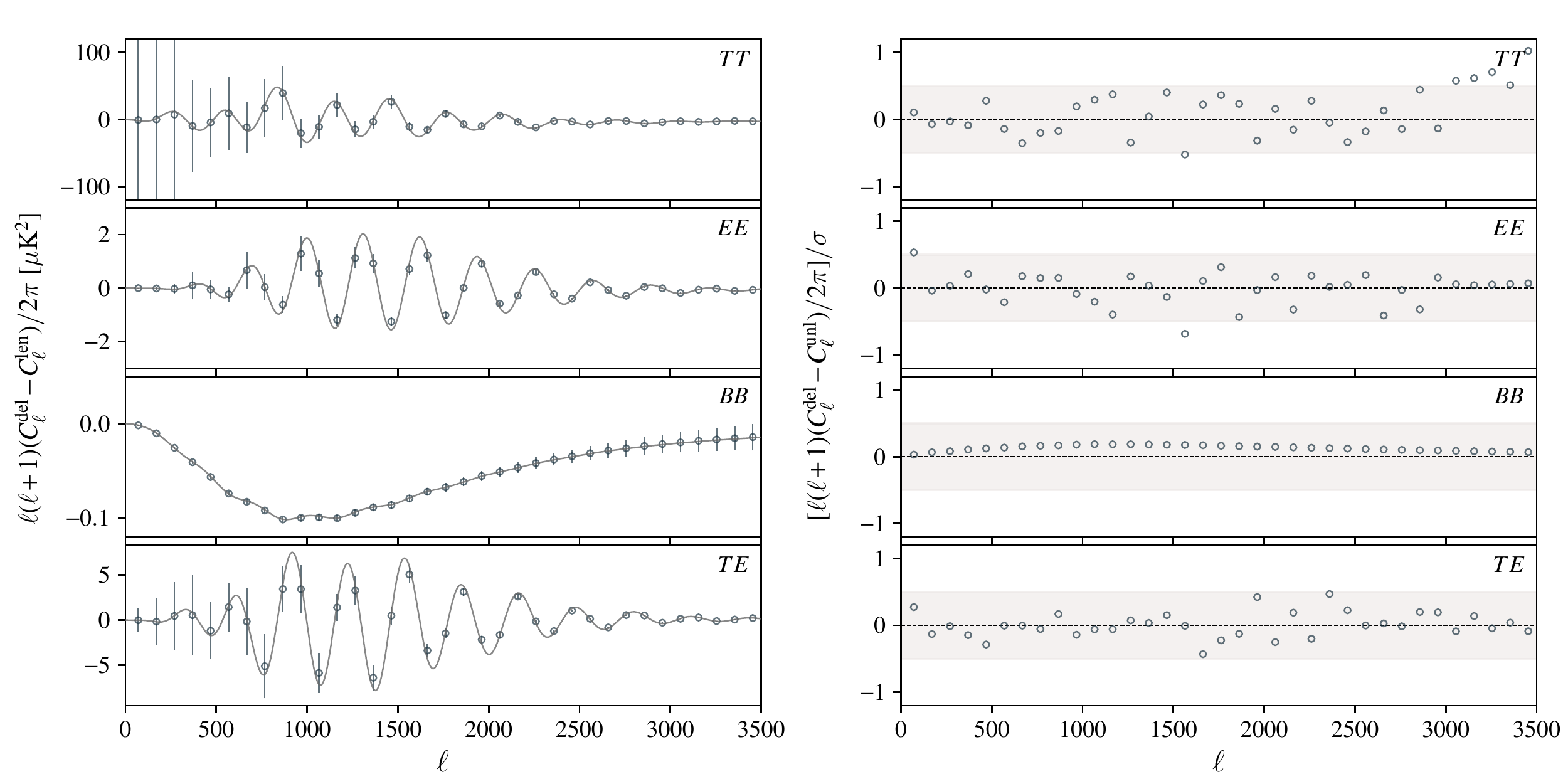}
\caption{ 
Comparison of the difference between the measured delensed $T\!T$/$E\!E$/$T\!E$ spectra and lensed spectra and theoretical predictions. The left panels show the absolute differences with the grey line showing the theoretical predictions and the error bars are obtained by taking the variance over 20 realizations with different primary CMB realization but the same CMB lensing realization. The panels on the right show the comparison of the differences between delensed and unlensed spectra, and the expected statistical uncertainties. The grey bands represent margins of $\pm0.5\sigma$. 
}
\label{fig:fully_delensed}
\end{center}
\end{figure*}

The goal of this section is to demonstrate that the simulation's CMB lensing field has sufficient accuracy and resolution to be used for delensing purposes. We begin by noting that the lensed and unlensed fields are related to each other by the following relations \citep{carron2017}:\ 
\begin{align}
X^{\rm len}=X^{\rm unl}(\hat{n}+\vec{\alpha}(\hat{n})),\\
X^{\rm unl}=X^{\rm len}(\hat{n}+\vec{\beta}(\hat{n})),
\end{align}
where $X\in T/Q/U$ and $\vec{\alpha}=\vec{\nabla}\phi$. The goal is  then to find $\vec{\beta}$, which maps back a lensed map to an unlensed one. A common choice is to use ${\beta}=\vec{\nabla}\phi^{\rm inv}=-\vec{\nabla}\phi$. However, this is known to leave residuals proportional to $(\vec{\alpha}\cdot\vec{\nabla})\vec{\alpha}$, where $\vec{\alpha}$ is the deflection angle \citep{green2017}. While the significance of this bias is small for experiments like \planck{}, it will become increasingly important for future surveys  \citep{carron2017}. To improve on this, a reverse deflection $\vec{\beta}$ could be determined by taking the relationship \citep{carron2017,diegopalazuelos2020}:
\begin{equation}
\hat{n}+\vec\beta(\hat{n})+\vec{\alpha}(\hat{n}+\vec\beta(\hat{n}))=\hat{n},
\end{equation}
from which $\vec{\beta}$ can be solved for using the Newton-Raphson iterative method:
\begin{equation}
\vec{\beta}_{i+1}(\hat{n})=\vec{\beta}_{i}(\hat{n})-M^{-1}(\hat{n}+\vec{\beta}_{i}(\hat{n}))\left[\vec{\beta}_{i}(\hat{n}) + \vec{\alpha}(\hat{n}+\vec{\beta}_{i}(\hat{n})) \right],
\end{equation}
where $M^{-1}$ is the inverse magnification matrix, 
\begin{equation}
M^{-1}=\begin{pmatrix}
1-\kappa-\gamma_{1} & \gamma_{2} \\
\gamma_{2} & 1-\kappa+\gamma_{1} 
\end{pmatrix},
\end{equation}
which can be computed at each pixel using the outputs from ray tracing, taking the starting point to be  $\vec{\beta}_{0}(\hat{n})=0$. This relation is iteratively solved, and the solution after 5 steps is used as $\vec{\beta}=\vec{\nabla}\phi^{\rm inv}$. Once this is determined, $\lenspix$ is used to undeflect the lensed CMB map. The delensed  $TT$/$EE$/$BB$/$TE$ spectra are shown in Figure \ref{fig:fully_delensed}. The differences between the delensed and unlensed spectra are compared with the statistical uncertainties expected for SPT-3G at full depth (see Table \ref{tab:ilc_expt_setup}) to quantify how accurately the unlensed fields are recovered. The residuals are shown to be within $0.5\sigma$ in the multipole range $30<\ell<3000$.

\subsection{Biases in CMB lensing due to foregrounds}\label{sec:lensingbias}
{\it Simulation products used: lensed SPT-3G 95/150/220 GHz maps, LSST-Y1 galaxy density maps}\\[0.2cm]
As experiments push toward lower noise, lensing map reconstruction will rely more heavily on the CMB polarization field, as it provides a more direct probe of the gravitational lensing effect \citep{hu2002}. However, for ongoing and near-term experiments with $n_{\rm lev}\gtrsim5 \muk{}$-arcmin, temperature-based lensing reconstruction will still have a non-negligible contribution to the total signal-to-noise ratio, and thus cannot be completely dismissed \citep{simons}.

Previous studies have shown that temperature-based lensing reconstruction could be severely contaminated by CMB secondary effects if no treatments are made. Numerous studies have identified and quantified the biases that could be present in the auto-spectrum and cross-correlations with large-scale structure \citep{vanengelen2012,osborne2014,coulton2018,baxter2019,omori2022}. These types of biases are of concern for experiments like SO, due to the relative predicted noise levels of the temperature and polarization maps. Various approaches for minimizing such biases have been proposed, including the bias hardening method \citep{namikawa2013}, the gradient cleaning method \citep{madhavacheril2018}, and the shear-only lensing reconstruction method \citep{schaan2018}. Furthermore, various studies have explored the optimal combination of methods to extract the highest signal-to-noise ratio with minimal foreground bias \citep{darwish2021,sailer2021}. In this section, a demonstration of how the simulation products can be used to quantify the amplitudes of foreground biases in the reconstructed lensing map is given.

\subsubsection{Input maps}
\begin{figure}
\begin{center}
\includegraphics[width=1.0\linewidth]{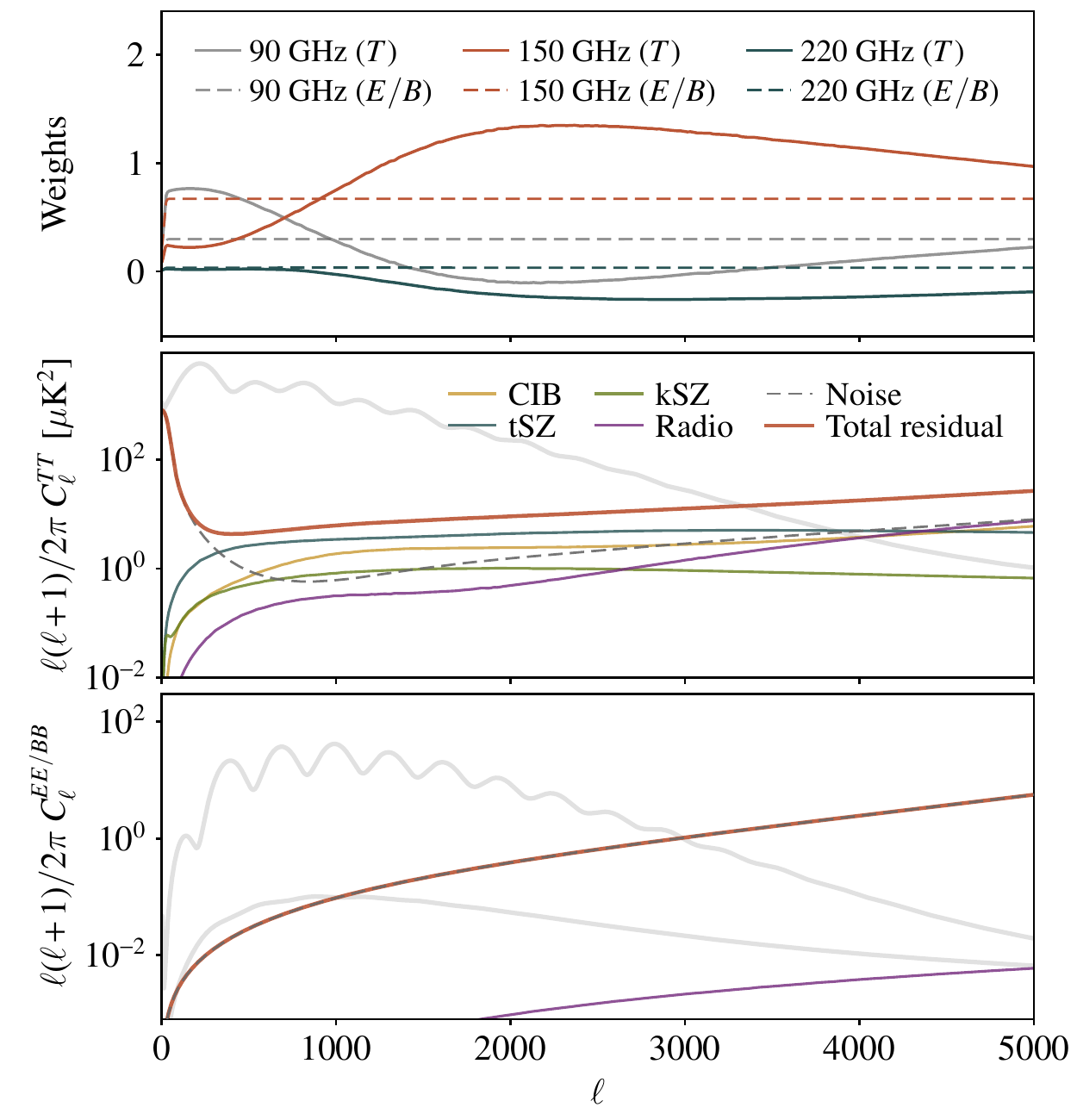}
\caption{ {\bf Upper:} Frequency weights for the 95/150/220\ GHz channels to construct the minimum variance SPT-3G $T$/$Q$/$U$ maps. {\bf Middle:} Amplitude of noise and the various extragalactic components in the minimum variance temperature map. {\bf Lower:} Similar to the middle panel, but for polarization. For polarization, the total residual is dominated by the noise, and hence the amplitude of the noise spectrum overlaps perfectly with the total residual spectrum.}
\label{fig:ilc_weights_res}
\end{center}
\end{figure}

For this exercise, the projected SPT-3G full depth noise levels from \citet{bender2018} are used  (summarized in Table \ref{tab:ilc_expt_setup}) and the noise power spectra for 95, 150, and 220 GHz channels are computed using:
\begin{equation}
N_{\ell}=\left(n_{\rm lev}\right)^{2}\exp\left(\ell(\ell+1)\frac{\theta^{2}_{\rm FWHM}}{8\ {\rm Ln}\ 2}\right)\left(1+\left(\frac{\ell_{\rm knee}}{\ell}\right)^{\alpha}\right),
\end{equation}
where $n_{\rm lev}$ is the white noise level in $\mu {\rm K}$-arcmin, $\theta_{\rm FWHM}$ is the beam FWHM and $\ell_{\rm knee}$ is the atmospheric noise transition point. Gaussian noise realizations generated from these noise spectra are added to their corresponding input frequency maps (discussed in Section \ref{sec:seconadary_modeling}). Next, the frequency weights that return the minimum variance temperature map are computed based on the total residual power spectrum (i.e., the sum of all the non-CMB components and noise) after masking point sources (both IR and radio) detected above 6 mJy and $\sim$24,000 positions centred at the local maxima of the Compton-$y$ map (which roughly corresponds to $5\sigma$ detection in the SPT-SZ data). The frequency weights, and the resulting residual power in the minimum variance temperature/polarization maps are shown in Figure \ref{fig:ilc_weights_res}.

Using the measured auto-spectrum of the minimum variance total foreground map, 150 Gaussian realizations are generated. These Gaussian realizations are used to compute the mean-field bias, lensing response function, and noise biases in the CMB lensing auto-spectrum, which are described in the following section.

\subsubsection{Lensing reconstruction}
The curved-sky quadratic estimator \citep{okamoto2003,planck2015xv} is used to estimate the lensing potential:
\begin{align}\label{eq:QuadraticEstimator}
\bar{\phi}_{LM}&[\bar{X},\bar{Z}]=\nonumber\\
&\frac{(-1)^{M}}{2}\sum_{\ell_{1}m_{1}\ell_{2}m_{2}}\begin{pmatrix}
\ell_{1} & \ell_{2} & L\\
m_{1} & m_{2} & -M
\end{pmatrix}
\times W_{\ell_{1}\ell_{2}L}^{\phi}\bar{X}_{\ell_{1}m_{1} }\bar{Z}_{\ell_{2}m_{2}}
\end{align}
where $\bar{X}_{\ell,m},\bar{Z}_{\ell,m}\in [\bar{T}_{\ell m}, \bar{E}_{\ell m}, \bar{B}_{\ell m }]$ and the over-bars imply that the maps have been inverse-noise filtered (i.e. $\bar{X}_{\ell m}=F_{\ell}X_{\ell m}=X_{\ell m}/(C_{\ell}^{XX}+N_{\ell}^{XX})$), the term in the bracket is the Wigner-$3j$ symbol and  $W^{\phi}_{\ell_{1}\ell_{2}L}$ is the lensing weight functions, which is different for each estimator $TT$,$EE$,$TE$,$TB$, and $EB$. The readers are referred to \cite{okamoto2003} for the full derivation and expressions for the weight functions. For temperature maps, multipoles in the range $300<\ell<3000$ are used, whereas multipoles in the range $300<\ell<5000$ are used for polarization maps.

Once $\bar{\phi}_{LM}$ is obtained, the spurious lensing-like signal induced by the mask (known as the mean-field $\phi^{\rm MF}$) is subtracted, a response function $\mathcal{R}_{L}$ is deconvolved to correct for the filtering applied to the input $T/Q/U$ maps, and the lensing potential map is converted into a convergence map: 
\begin{align}\label{eq:biased_kappatt}
\hat{\kappa}_{LM}=\mathcal{R}_{L}^{-1}\bar{\kappa}_{LM}=\frac{1}{2}\mathcal{R}_{L}^{-1}L(L+1)(\bar{\phi}_{LM}-{\phi}_{LM}^{\rm MF}).
\end{align}
Similar to the approach used in \cite{omori2017,omori2022}, the response function is computed by cross-correlating the reconstructed lensing maps ${\bar\kappa}^{\rm out}$ and input ${\bar \kappa}^{\rm in}$ maps, and normalizing by the auto-spectrum of the input maps:
\begin{equation}
\mathcal{R}_{L}^{\rm MC}=\frac{\langle C_{L}(\bar{\kappa}_{LM}^{\rm out} \bar{\kappa}_{LM}^{\rm in})\rangle }{ \langle C_{L}(\bar{\kappa}_{LM}^{\rm in} \bar{\kappa}_{LM}^{\rm in})\rangle  },
\end{equation}
where the average is taken over 150 realizations.
Lensing maps are reconstructed using all the estimator $TT$,$EE$,$TE$,$TB$,$EB$ and a minimum variance combination is formed by taking:\footnote{Here we simply follow the definition of ``minimum variance" from \cite{planck2015xv}, although \cite{maniyar2021} notes that this approach neglects correlations between certain modes and hence does not yield the true minimum variance (also coined as the ``global minimum variance").} 
\begin{equation}
\hat\kappa_{LM}^{\rm MV} = \frac{\sum_{\alpha} \kappa_{LM}^{\alpha}\mathcal{R}_{L}^{{\rm MC},\alpha} }{ \sum_{\alpha} \mathcal{R}_{L}^{{\rm MC},\alpha} }\hspace{1.25cm} \alpha\in[TT,EE,TE,TB,EB].
\end{equation}
\\
\begin{figure}
\begin{center}
\includegraphics[width=1.0\linewidth]{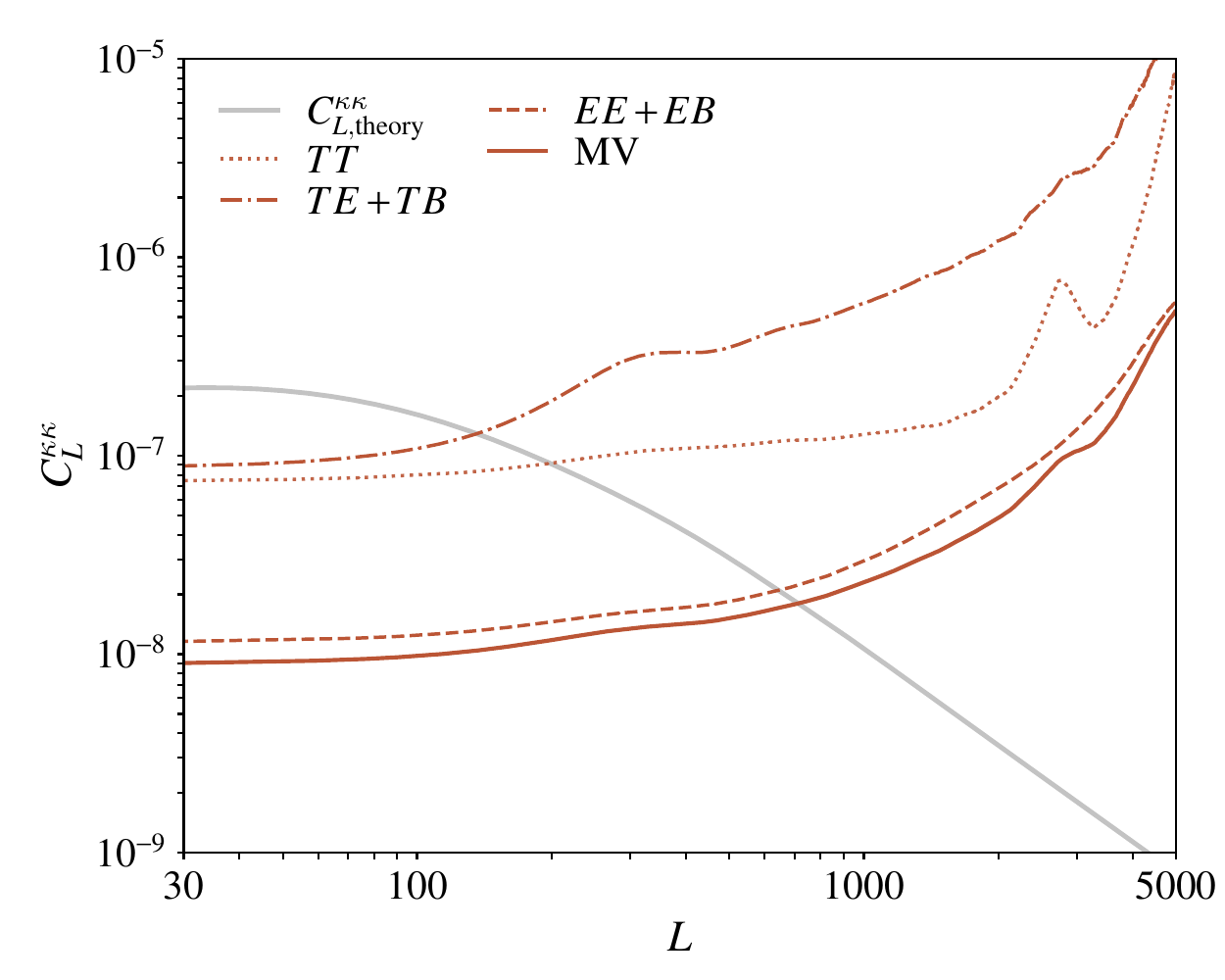}
\caption{CMB lensing noise levels for the temperature-only reconstruction ($TT$; dotted orange line), polarization-only reconstruction ($EE$+$EB$; orange dashed line), temperature-polarization-cross reconstruction ($TE$+$TB$; dot dashed orange line) and the minimum variance combination (MV; solid orange line), compared with the input CMB lensing spectrum (light grey line).  }
\label{fig:lensing_noise}
\end{center}
\end{figure}
The noise levels for the temperature-only ($TT$), polarization-only ($EE$+$EB$), and the minimum variance reconstructions are shown in Figure \ref{fig:lensing_noise}.

\begin{table}
\begin{center}
\begin{tabular}[t]{ccccccc}
\toprule
$\nu$ [GHz] & $n_{\rm lev}^{T}\ (n_{\rm lev}^{P})$ [$\mu$K-arcmin] & $\ell_{\rm knee}^{T}$ & $\alpha_{\rm knee}^{T}$  & $\ell_{\rm knee}^{P}$ & $\alpha_{\rm knee}^{P}$ \\
\midrule
$95$  & 3.0 (4.2) & 1200 & 3  & 300 & 1 \\
$150$ & 2.0 (2.8) & 2200 & 4  & 300 & 1 \\
$220$ & 9.0 (12.4) & 2300 & 4  & 300 & 1 \\
\bottomrule
\end{tabular}
\caption{Assumed experimental set up in this exercise. The noise levels are taken from \protect\citet{bender2018}, the values for $\ell_{\rm knee}$ are taken from \protect\citet{aylor2019}, and $\alpha_{\rm knee}$ are assumed values.  }
\label{tab:ilc_expt_setup}
\end{center}
\end{table}

\subsubsection{Foreground biases in the lensing auto-spectrum}\label{sec:biases_clkk}
The raw measured power spectrum of a reconstructed CMB lensing map contains noise bias terms resulting from the disconnected portion of the 4-point function, which must be subtracted off even in the absence of foregrounds. These are the so-called the $N^{(0)}_{L}$ and $N_{L}^{(1)}$ biases, and they can be computed using two sets of simulated lensed CMB skies: one set consisting of 150 primary CMB realizations lensed by 150 independent lensing potential realizations, and a second set consisting of 150 different primary CMB realizations lensed by the same lensing potential as the first set \citep{story2015,omori2017,omori2022}. More specifically, the ``realisation-dependent" $N_{L}^{(0)}$ bias is computed by replacing one of the input maps with a CMB map that is treated as ``data" (In this case, a CMB map that has been deflected using the lensing map obtained from ray tracing). This will be referred to as $N_{L}^{(0),{\rm RD}}$.

\begin{figure*}
\begin{center}
\includegraphics[width=1.0\linewidth]{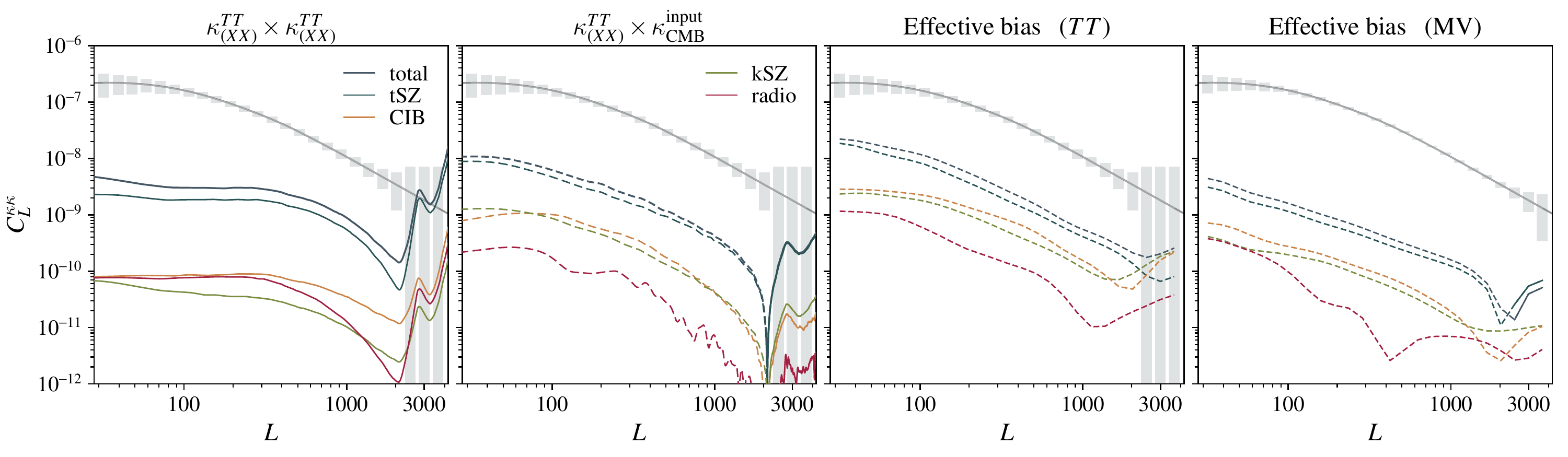}
\caption{Biases in the CMB lensing auto-spectrum. The biases are split into trispectum terms (first panel from the left) and bispectrum terms (second panel). For the bispectrum terms, the dashed lines imply negative correlations. The total effective bias (described in the text) from the various components when using the $TT$ estimator alone is shown in the third panel, and an equivalent plot for the MV combination is shown in the fourth panel.  }
\label{fig:lensing_biases_auto}
\end{center}
\end{figure*}

After subtracting the $N_{L}^{(0),{\rm RD}}$ and $N_{L}^{(1)}$ noise bias terms from the raw reconstructed lensing spectrum, what remains is the foreground contaminated CMB lensing power spectrum.  One approach to estimate the foreground biases in the CMB lensing power spectrum is to run the quadratic estimator on the individual foreground maps using the same filtering function $F_{\ell}$ and $\ell_{\rm min}/\ell_{\rm max}$. 
The left most panel of Figure \ref{fig:lensing_biases_auto} shows the measured auto-spectra of such maps, and the second panel shows the cross-spectra obtained by correlating the reconstructed maps with the noiseless input convergence map. 
Subtracting these foreground biases from the contaminated lensing spectrum, however, does not recover the input lensing spectrum since these terms themselves have their own $N_{L}^{(0)}$ and $N^{(1)}_{L}$-like terms. Instead, the ``effective" lensing biases can be evaluated by taking the difference between the lensing auto-spectrum computed from CMB maps with non-Gaussian and Gaussian foregrounds:
\begin{equation}\label{eq:bias4_bias2}
{\rm Bias}=
C_{L}^{\kappa\kappa}({\rm CMB}^{\rm NG}+{\rm FG}^{\rm NG}) - C_{L}^{\kappa\kappa}({\rm CMB}^{\rm NG}+{\rm FG}^{\rm G}). 
\end{equation}
Since the non-Gaussian foreground maps are correlated with the underlying lensing field, the first term effectively captures both the trispectrum and the bispectrum-type biases, and the second term subtracts the underlying reconstructed lensing field (which may contain its own biases due to simulation artefacts even in the absence of foregrounds). Estimates of biases obtained this way tend to have large scatter due to the difference in the amplitudes of the raw reconstructed lensing spectra and foreground biases. To reduce this scatter, the same procedure is repeated 5 times using different CMB realizations (but lensed with the same lensing field obtained from ray tracing), and their averages are taken. The results are shown in the third panel of Figure \ref{fig:lensing_biases_auto}.

The biases are generally negative at large scales, and are primarily driven by the bispectrum-type biases. Somewhat surprisingly, the prominent rise seen at high-$\ell$ in the raw foreground trispectrum are not seen in the effective biases. This implies that the procedure of subtracting $N_{L}^{(0),{\rm RD}}$ and $N_{L}^{(1)}$ from the raw CMB lensing spectrum reduces the amplitude of these biases.

Next, the lensing map constructed using the $TT$ estimator is combined with lensing maps constructed using the $EE$, $TE$, $TB$, and $EB$ estimators to form a minimum variance lensing map, and the amplitudes of the foreground biases are re-evaluated. Since non-$TT$ estimators are less affected by CMB secondary effects (except for radio sources at very high-$\ell$), foreground biases are reduced by forming a minimum variance combination. In the fourth panel of Figure \ref{fig:lensing_biases_auto}, the evaluated biases for the minimum variance lensing map are shown, and it can be seen that the amplitude of the total bias is of order ${\rm bias}/\sigma\sim0.3$.

\subsubsection{Foreground biases in cross-correlations}
\begin{figure*}
\begin{center}
\includegraphics[width=1.0\linewidth]{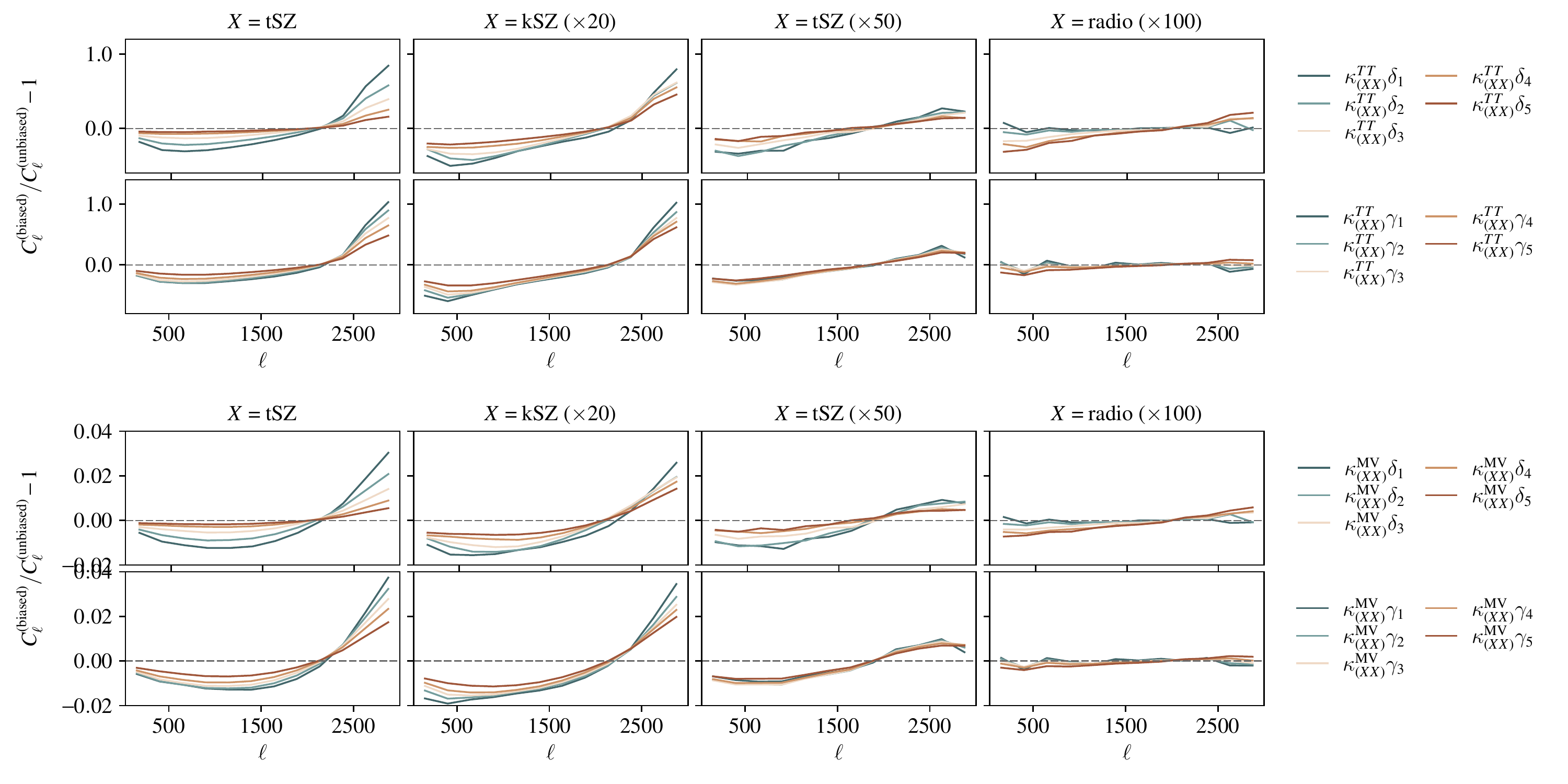}
\caption{ 
Biases in the cross-correlation measurements between CMB lensing and galaxy density (first and third rows) and galaxy lensing (second and fourth rows) due to tSZ, kSZ, CIB, and radio sources (respectively for each column). The first and second rows are for biases in temperature-based CMB lensing maps, whereas the third and fourth rows are for the minimum variance combination. The different coloured lines in each panel correspond to different redshift bins of the galaxy density and weak lensing (blue corresponds to lower redshifts and orange corresponds to higher redshifts). Only the single-foreground terms (i.e.$\kappa_{(XX)}$) are considered, and the terms that involve two secondary components (i.e.$\kappa_{(XY)}$) have been ignored. Note that for kSZ, CIB, and radio sources, the biases are multiplied by factors of 20, 50, and 100, respectively.}
\label{fig:lensing_biases_cross}
\end{center}
\end{figure*}

Biases in cross-correlations are estimated by correlating LSST-Y1 galaxy density/weak lensing maps (described in Sections \ref{sec:method_gal}, \ref{sec:method_gwl} and \ref{sec:6x2pt}) with lensing-like maps generated by passing tSZ, kSZ, CIB, radio source maps through the CMB lensing reconstruction pipeline (described Section \ref{sec:biases_clkk}). The fractional bias is computed by taking the ratio:
\begin{equation}
{\rm Bias}=C_{\ell}^{\alpha\kappa_{X}}/C_{\ell}^{\alpha\kappa_{\rm CMB}^{\rm input}},
\end{equation}
where $\alpha\in \delta_{\rm g},\kappa_{\rm g}$ and  $X\in$\ tSZ, kSZ, CIB, radio sources. These biases are quantified for both temperature-based and the minimum variance lensing map, and assuming that none of the foregrounds considered here impact the polarization-based lensing map. The results are shown in Figure \ref{fig:lensing_biases_cross}. When a temperature-based lensing map is used, the measured cross-correlations are heavily contaminated by the tSZ effect, and the kSZ effect also biases the measurements by a few percent. On the other hand, biases due to  the CIB and radio sources are not dominant, which is likely due to the fact that CIB is very close to a Gaussian field (and hence does not get picked up by the lensing estimator), and that radio sources occupy halos of a wide range of masses, and hence very little correlation is detected. For the minimum variance lensing map, the amplitudes of the biases are significantly suppressed, even for tSZ, for which the bias is  at most 3\%, and the biases from the other foregrounds are significantly smaller.

\subsection{Forecasting multi-tracer delensing}\label{sec:app_multitracer}
{\it Simulation products used: SPT-3G (95/150/220 GHz) maps, Planck CIB 545 GHz map and biased density maps (mock LSST-Y1 lens sample)}\\

This section provides a demonstration of how the simulation products can be used to estimate the delensing efficiency and partially delensed spectra ($C_{\ell}^{TT,{\rm del}}$, $C_{\ell}^{EE,{\rm del}}$, $C_{\ell}^{TE,{\rm del}}$ and $C_{\ell}^{BB,{\rm del}}$) when a given tracer is used for delensing. The various astrophysical components implemented in Section \ref{sec:seconadary_modeling} are used as tracers of the lensing potential, and the measured correlations between the tracers and the input lensing potential, as well as the correlation between the tracers, are also used to optimally combine the different tracers. Both ``internal" (temperature-based, polarization-based, and minimum variance CMB lensing maps) and ``external" tracers  (CIB at 545 GHz from \planck{} and LSST-Y1 galaxy density maps) are used.

\subsubsection{Internal lensing tracers}
\begin{figure}
\begin{center}
\includegraphics[width=1.0\linewidth]{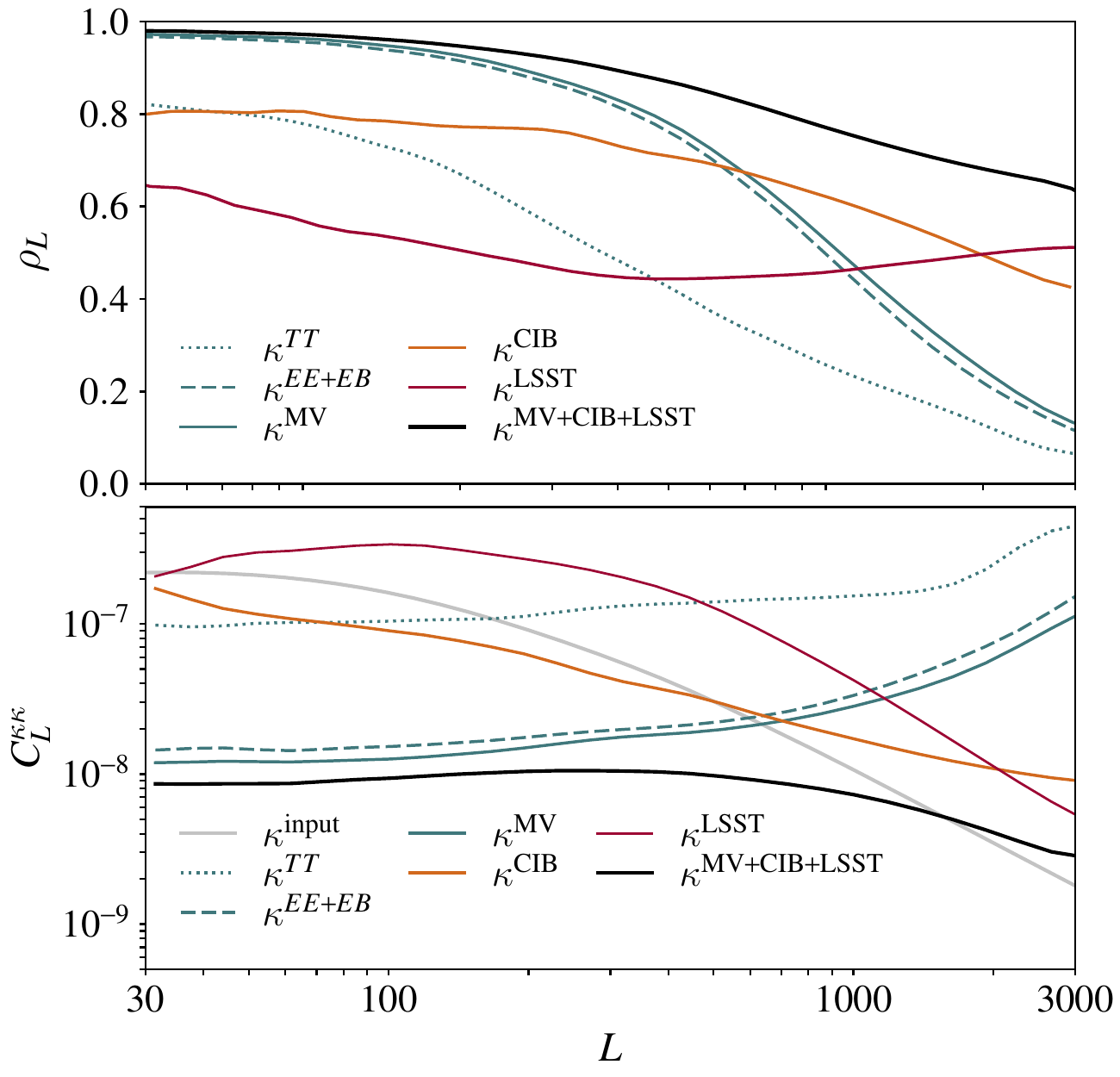}
\caption{ {\bf Upper:} The measured correlation factors between the input convergence and CMB lensing from temperature-only reconstruction (dotted teal line),  polarization-only reconstruction (dashed teal line), the minimum variance combination (solid teal line), \planck{} 545 GHz CIB map (orange line) and LSST-Y1 clustering galaxy sample (purple line), and all the tracers combined (black line). {\bf Lower:} Comparison between the CMB lensing signal (light grey line), and  the effective lensing noise level computed using $N_{L}=C_{L}^{\kappa\kappa}(1/\rho_{L}^2-1)$ where $\rho_{L}$ from the upper panel are used.}
\label{fig:delensing_rho}
\end{center}
\end{figure}

\begin{figure}
\begin{center}
\includegraphics[width=1.0\linewidth]{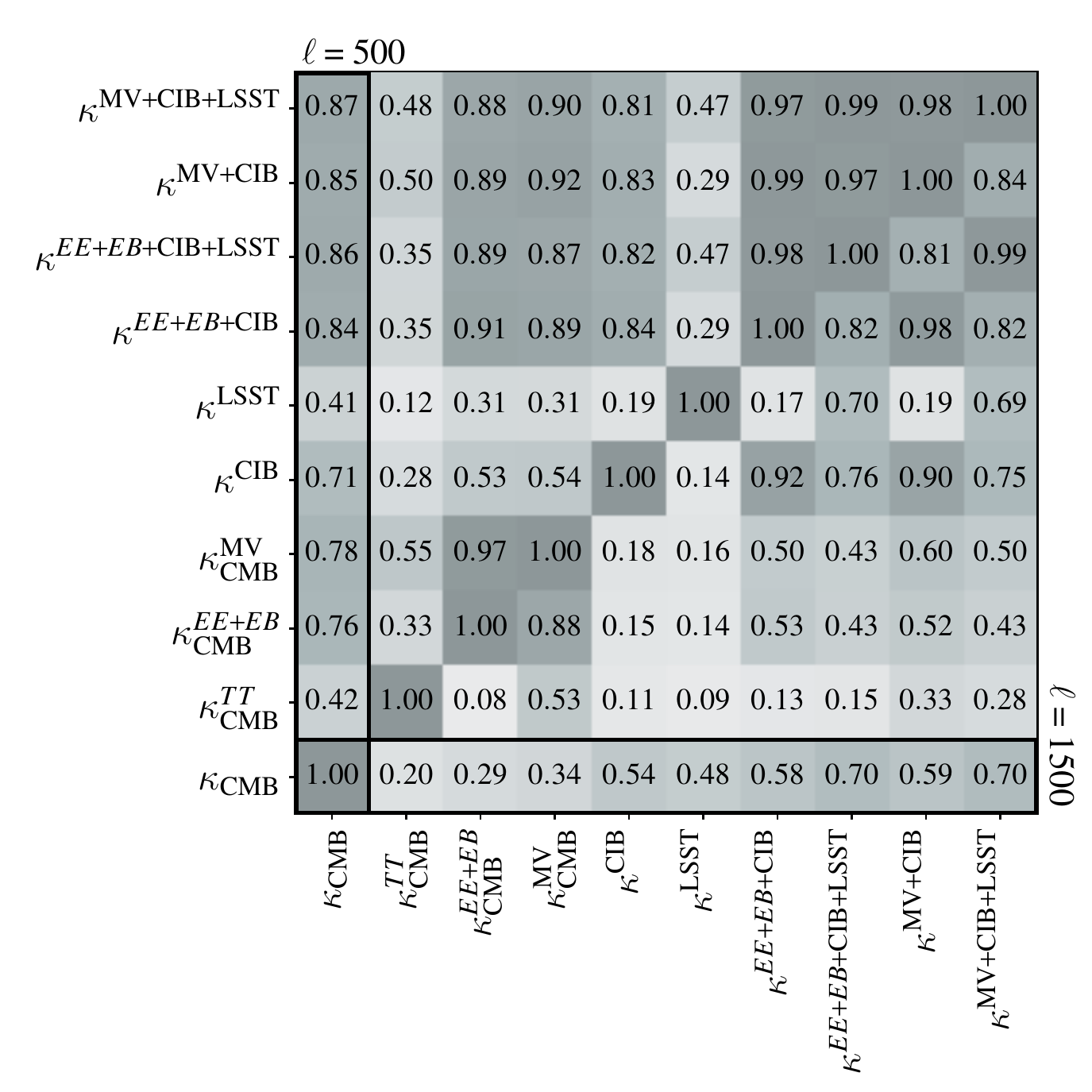}
\caption{Correlation coefficients between various tracers of the lensing potential. The upper triangle corresponds to the correlation values measured at $\ell=500$ and the lower triangle correspond to the correlation values measured at $\ell=1500$, with darker shades corresponding to higher correlation values (denoted in each cell). The first row and column correspond to the correlation with the input lensing field.}
\label{fig:corrcoeff}
\end{center}
\end{figure}

Starting with the temperature-only ($TT$) reconstruction, the correlation factor with the input lensing signal $\rho_{\ell}$ is found to vary between $0.04-0.86$, in the multipole range $30<L<3000$. The fractional $B$-mode delensing efficiency, defined as:
\begin{equation}
\eta^{\alpha}_{\rm del}=\left\langle\frac{C^{BB}_{\ell,{\rm len}}-C^{BB}_{\ell,{\rm del}} }{C^{BB}_{\ell,{\rm len}}}\right\rangle_{\ell<1500},
\end{equation}
is also computed (where $\alpha$ denotes the tracer used for delensing), and is found to be $\eta^{\kappa_{TT}}_{\rm del}=0.29$.  Lensing reconstruction from polarization-only and the minimum variance combination, on the other hand, have higher correlation factors ranging between $0.10 - 0.98$ and with $\eta^{\kappa_{EE}+\kappa_{EB}}_{\rm del},\eta^{\kappa_{\rm MV}}_{\rm del}=0.62,0.64$ respectively. The similarity between these two values suggests that the temperature-based lensing map adds little value and that it may be preferable to use only the polarization-based lensing maps to avoid potential biases caused by foreground contamination in the temperature-based lensing maps. The correlation factors for these internal lensing estimates are high at large scales but decreases rapidly beyond $L=500$. This is due to the shape of the CMB lensing noise spectrum (shown in the lower panel of Figure \ref{fig:delensing_rho}): while at low-$L$, the measurements are signal dominated, beyond $L>1000$ the noise quickly takes over.  Since, the lensing $L<500$ modes  contribute the most to the $B$-mode spectra in the $\ell<1500$ range \citep{simard2015}, internal lensing estimates are very effective for delensing. These results are shown in Figures \ref{fig:delensing_rho} and \ref{fig:corrcoeff}.

\subsubsection{External tracers (LSST galaxies and CIB )}
Using LSST-Y1 galaxy density maps, the correlation factor is found to be in the range $0.40-0.78$ as shown in Figure \ref{fig:delensing_rho}, with a delensing efficiency of $\eta^{\rm LSST}_{\rm del}=0.25$. While the correlation amplitude is lower at large scales compared to internal tracers, the correlation coefficient remains relatively constant even at $L>500$, indicating that galaxy density maps could provide complementary information. Although galaxy weak lensing is also expected to be a high signal-to-noise probe, it is sensitive to structures at lower redshifts than the galaxy density maps, making it a less effective tracer for delensing.

CIB is also an efficient tracer for delensing purposes, since the redshift kernel extends over a broad range of redshifts $0<z<4$ (see Figure \ref{fig:cib_dIdz}) and the peak of the kernel lies near the peak of the CMB lensing kernel ($z\sim2$). In addition, high signal-to-noise ratio maps of the CIB, such as the GNILC maps \cite{planck2015xlviii} and the maps from \cite{lenz2019} are publicly available. For CIB, the correlation factor is found to be in the range 0.40-0.84, and the delensing efficiency is found to be  $\eta^{\rm CIB}_{\rm del}=0.52$. 

There are, caveats to using the CIB as a lensing potential tracer because it is composed of high redshift galaxies whose properties are not well understood. There are very few physical models capable of simultaneously describing CIB observables such as the auto- and cross-power spectra, as well as their statistical properties such as the luminosity function, number counts, and dust temperature evolution. Furthermore, because light from the CIB is emitted from dust surrounding star forming galaxies, the emission properties are very similar to those from Galactic dust, making it challenging to disentangle the two \citep{planck2015xlviii}. These issues can be investigated further using the simulation products of this work.

\subsubsection{Combined tracer}
\begin{figure*}
\begin{center}
\includegraphics[width=1.0\linewidth]{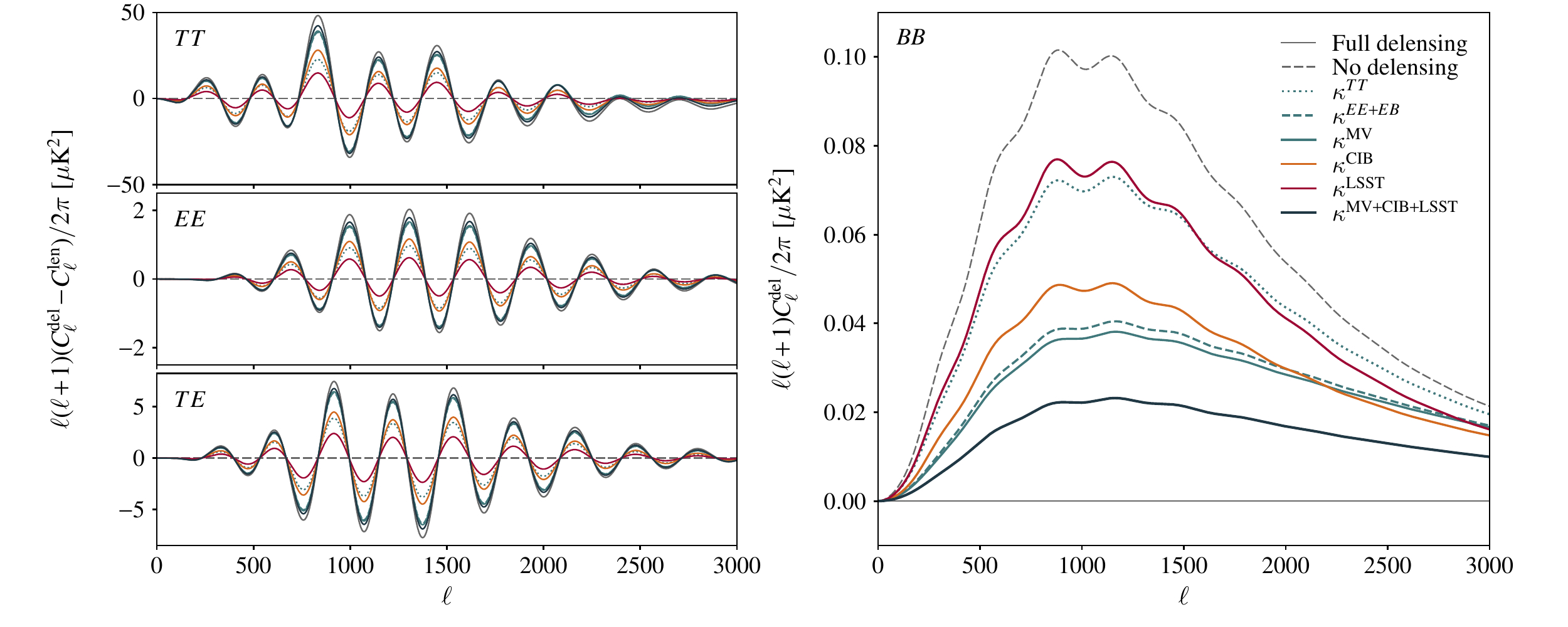}
\caption{Analytically computed delensed $TT$, $EE$, $TE$, and $BB$ spectra using the correlation factors measured from the simulation when using $\kappa^{TT}$ (dotted teal line), $\kappa^{EE+EB}$ (dashed teal line), $\kappa^{\rm MV}$ (solid teal line), $\kappa^{\rm CIB}$ (orange line), $\kappa^{\rm LSST}$ (purple line) and, $\kappa^{\rm MV+CIB+LSST}$ (navy line) as tracers. The case of perfect delensing is shown as solid grey line. }
\label{fig:multitracer_delensedspec}
\end{center}
\end{figure*}

With the individual internal and external tracers in hand, an optimally combined tracer can be constructed using the approach described in \citet{planck2018viii}. For this, optimal weights for the individual tracers are first calculated using:
\begin{equation}
w_{L}^{i}=\sum_{j}\rho_{L}^{\kappa I_{j}}(\rho_{L}^{I_{j}I_{i}})^{-1},
\end{equation}
where
\begin{align}
\rho^{I_{i}I_{j}}&=\frac{C_{L}^{I_{i}I_{j}} }{\sqrt{C_{L}^{I_{i}I_{i}}C_{L}^{I_{j}I_{j}}} },\\
\rho^{\kappa I_{j}}&=\frac{C_{L}^{\kappa I_{j}}}{\sqrt{C_{L}^{\kappa\kappa}C_{L}^{I_{j}I_{j}}}},
\end{align}
are the correlation factors between the tracers, and between tracers and the input CMB lensing map. Here, $\kappa$ is the noiseless input convergence field, and $I_{j}$,$I_{j}$ are all the other tracers (with noise), including reconstructed CMB lensing. The maximum-a-posteriori map is then formed by using:
\begin{equation}
{\hat \kappa}^{\rm MAP}_{LM}=\sum_{i} w_{L}^{i}I^{i}_{LM}\sqrt{\frac{C_{L}^{\kappa\kappa,{\rm fid}}}{C_{L}^{I_{i} I_{i}}}}.
\end{equation}
Using this optimally combined tracer, the correlation coefficient is found to be in the range $0.63-0.99$ and the delensing efficiency is found to be $\eta^{\rm MV+CIB+LSST}_{\rm del }=0.78$. Comparisons of the delensed spectra using different combinations of tracers are shown in Figure \ref{fig:multitracer_delensedspec}. 

The delensing efficiencies and delensed spectra presented in this section were calculated using the forecasted experimental noise levels of SPT-3G at full depth. Further improvements in these quantities can be expected using future surveys with lower instrumental noise and greater sky coverage, such as CMB-S4.

\subsection{Extracting $\rho_{\rm SFR}$ from the CIB}\label{sec:app_csfr}
{\it Products used: \planck{} 353/545/857 GHz CIB maps.}\\[0.1cm]
The evolution of the mean cosmic star formation rate density is one of the key astrophysical quantities that can be extracted from the CIB auto- and cross-spectra. Numerous studies have attempted to constrain the cosmic SFRD by studying the CIB this way (see, e.g., \citealt{maniyar2018,mccarthy2021}). In this work, the true SFRD is based on the results from \um{} and is hence known (shown in Figure \ref{fig:um_cosmicsfr}). The CIB maps in the simulation can be used to investigate whether a given CIB model is sufficiently flexible to be able to recover the input signal. 

\subsubsection{Model}
For this demonstration, the CIB halo model from \cite{maniyar2020} is considered. In this model, the CIB power spectrum is written as the sum of the 1-halo, 2-halo and a shot-noise term:
 \begin{equation}
 C_{\ell}^{\rm tot}=C_{\ell}^{\rm 1h}+C_{\ell}^{\rm 2h}+C_{\ell}^{\rm shot}.
 \end{equation}
 The 1-halo term is written as:
 \begin{align}
 &C_{\ell,\nu,\nu'}^{\rm 1h}=\int\int\frac{d\chi}{dz}\left(\frac{a}{\chi}\right)^{2}\biggl[\frac{dj_{\nu,{\rm c} }}{d {\rm log}M_{\rm h}} \frac{dj_{\nu',{\rm sub}}}{d {\rm log}M_{\rm h}} u(k ,M_{\rm h},z)\nonumber\\
 & + \frac{dj_{\nu',{\rm c} }}{d {\rm log}M_{\rm h}} \frac{dj_{\nu,{\rm sub}}}{d {\rm log}M_{\rm h}} u(k ,M_{\rm h},z)\nonumber\\
 & + \frac{dj_{\nu,{\rm sub} }}{d {\rm log}M_{\rm h}} \frac{dj_{\nu',{\rm sub}}}{d {\rm log}M_{\rm h}} u^{2}(k ,M_{\rm h},z)\biggl]
\left(\frac{{\rm d}^{2}N}{{\rm d}{\rm log}M_{\rm h}{\rm d}V} \right)^{-1} dz d{\rm log}M_{\rm h},
\end{align}
where $dj_{\nu,{\rm c/sub}}/d\log M_{\rm h}$ are the specific emissivity of the central and satellite sub-haloes, $u$ is the Fourier transform of the Navarro-Frenk-White (NFW) density profile and $d^2N/d\log M_{\rm h} dV$ is the halo mass function. The 2-halo term is written as:
 \begin{align}
 &C_{\ell,\nu,\nu'}^{\rm 2h}=\int\int\int\frac{d\chi}{dz}\left(\frac{a}{\chi}\right)^{2}\biggl[\frac{dj_{\nu,{\rm c} }}{d {\rm log}M_{\rm h}}+ \frac{dj_{\nu,{\rm sub}}}{d {\rm log}M_{\rm h}} u(k ,M_{\rm h},z)\biggl]\nonumber\\
 & + \biggl[\frac{dj_{\nu',{\rm c} }}{d {\rm log}M'_{\rm h}}+ \frac{dj_{\nu',{\rm sub}}}{d {\rm log}M'_{\rm h}} u(k ,M_{\rm h},z)\biggl]\nonumber\\
 & \times b_{\rm h}(M_{\rm h},z)b_{\rm h}(M'_{\rm  h},z)P_{\rm lin}(k,z)d{\rm log}M_{\rm h}d{\rm log}M'_{\rm h}dz,
\end{align}
where $b_{\rm h}$ is the halo bias and $P_{\rm lin}$ is the linear matter power spectrum. Emissivity per logarithmic mass for the central and satellite galaxies are defined as:
\begin{equation}
\frac{dj_{\nu,{\rm c} }}{d{\rm log} M_{\rm h}}(M_{\rm h},z)=\frac{d^{2}N}{d{\rm log} M_{\rm h} dV}\chi^{2}(1+z)\frac{{\rm SFR}_{\rm c}}{K}\hspace{0.5em} S_{\nu}^{\rm eff},
\end{equation}
\begin{align}
\frac{dj_{\nu,{\rm sub} }}{d{\rm log} M_{\rm h}}(M_{\rm h},z)=&\frac{d^{2}N}{d{\rm log} M_{\rm h} dV}\chi^{2}(1+z)\nonumber\\
&\int\frac{dN}{d{\rm log}m_{\rm sub}}(m_{\rm sub}|M_{\rm h})\frac{{\rm SFR}_{\rm sub}}{K}\hspace{0.5em} S_{\nu}^{\rm eff}d\log m_{\rm sub},
\end{align}
where ${\rm SFR_{c/sub}}$ are the star formation rates for the central and subhalos, $K=1\times10^{-10}{\rm M}_{\sun}{\rm yr}^{-1}{\rm L}_{\odot}^{-1}$ is the Kennicutt's constant and $S_{\nu}^{\rm eff}$ is the effective SED of infrared galaxies.\footnote{ $S_{\nu}^{\rm eff}$ provided with the modelling package found at \url{https://github.com/abhimaniyar/halomodel_cib_tsz_cibxtsz} are directly adopted. } 

The key characteristic that distinguishes  the \citet{maniyar2020} halo model from others (such as \citealt{shang2012}) is the implementation of SFR. In this model, the ratio of SFR and the baryonic accretion rate (BAR) is modelled as :
 \begin{equation}
 \frac{\rm SFR}{\rm BAR}(M_{\rm h},z)=\eta=\eta_{\rm max}\exp\left(-\frac{(\log{M_{\rm h}-{\rm log}M_{\rm max}})^2}{2\sigma_{\rm M_{\rm h}}(z)}\right),
 \end{equation}
where BAR is given by:
 \begin{align}
 {\rm BAR}(M_{\rm h},z)&=46.1 {\rm M}_{\sun}{\rm yr}^{-1}\left(\frac{M_{\rm h}}{10^12 {\rm M}_{\sun}}\right)^{1.1}\nonumber\\ &\times(1+1.11z)\sqrt{\Omega_{\rm m}(1+z)^3+\Omega_{\Lambda})},
 \end{align}
and the halo mass dispersion $\sigma_{M_{\rm h}}$ is given by:
\begin{equation}
\sigma_{M_{\rm h}}(z)=\sigma_{M_{\rm h0}}-\tau\times{\rm max}(0,z_{\rm c}-z).
\end{equation}
Finally the SFR for central and satellite galaxies are assigned separately where:
\begin{equation}
{\rm SFR}_{\rm c}(M_{\rm h},z)=\eta(M_{\rm h},z)\times {\rm BAR}(M_{\rm h},z),
\end{equation}
and
\begin{equation}
{\rm SFR}_{\rm s}(M_{\rm h},z)= {\rm SFR}_{\rm c}(M_{\rm h},z) \times\frac{m_{\rm sub}}{M_{\rm h}}.
\end{equation}
The model CIB power spectra $C_{\ell}^{\rm tot}$ are then compared with the measured CIB power spectra from the simulation, and an MCMC is used to determine the best-fit values for ${\rm log}M_{\rm max}$, $\eta_{\rm max}$, $\sigma_{M_{\rm h0}}$, and $\tau$, as well the  shot noise amplitudes for the three frequency channels $A_{\rm shot}^{353}, A_{\rm shot}^{545}, A_{\rm shot}^{857}$.

To extract $\rho_{\rm SFR}$, the best-fit values for  ${\rm log}M_{\rm max}$, $\eta_{\rm max}$, $\sigma_{M_{\rm h0}}$, $\tau$ are used to compute the star formation rate densities for the central and satellite galaxies:
\begin{align} 
{\rm SFRD}_{\rm c}(z)&=\int\frac{d^{2}N}{d{\rm log}M_{\rm h}dV}{\rm SFR}_{\rm c}(M_{\rm h},z)d\log(M_{\rm h})\\
{\rm SFRD}_{\rm s}(z)&=\int\frac{d^{2}N}{d{\rm log}M_{\rm h}dV}\nonumber\\
&\times\biggl[\int\frac{dN}{d\log M_{\rm sub}}{\rm SFR}_{\rm s}(m_{\rm sub},z)d\log(M_{\rm sub}) \biggl] d\log(M_{\rm h}),
\end{align}
and the sum of the two term are taken to compute the total cosmic star formation density:
\begin{equation}
{\rm SFRD}(z)={\rm SFRD}_{\rm c}(z)+{\rm SFRD}_{\rm s}(z).
\end{equation}

\begin{figure}
\begin{center}
\includegraphics[width=1.0\linewidth, bb=20 20 560 560]{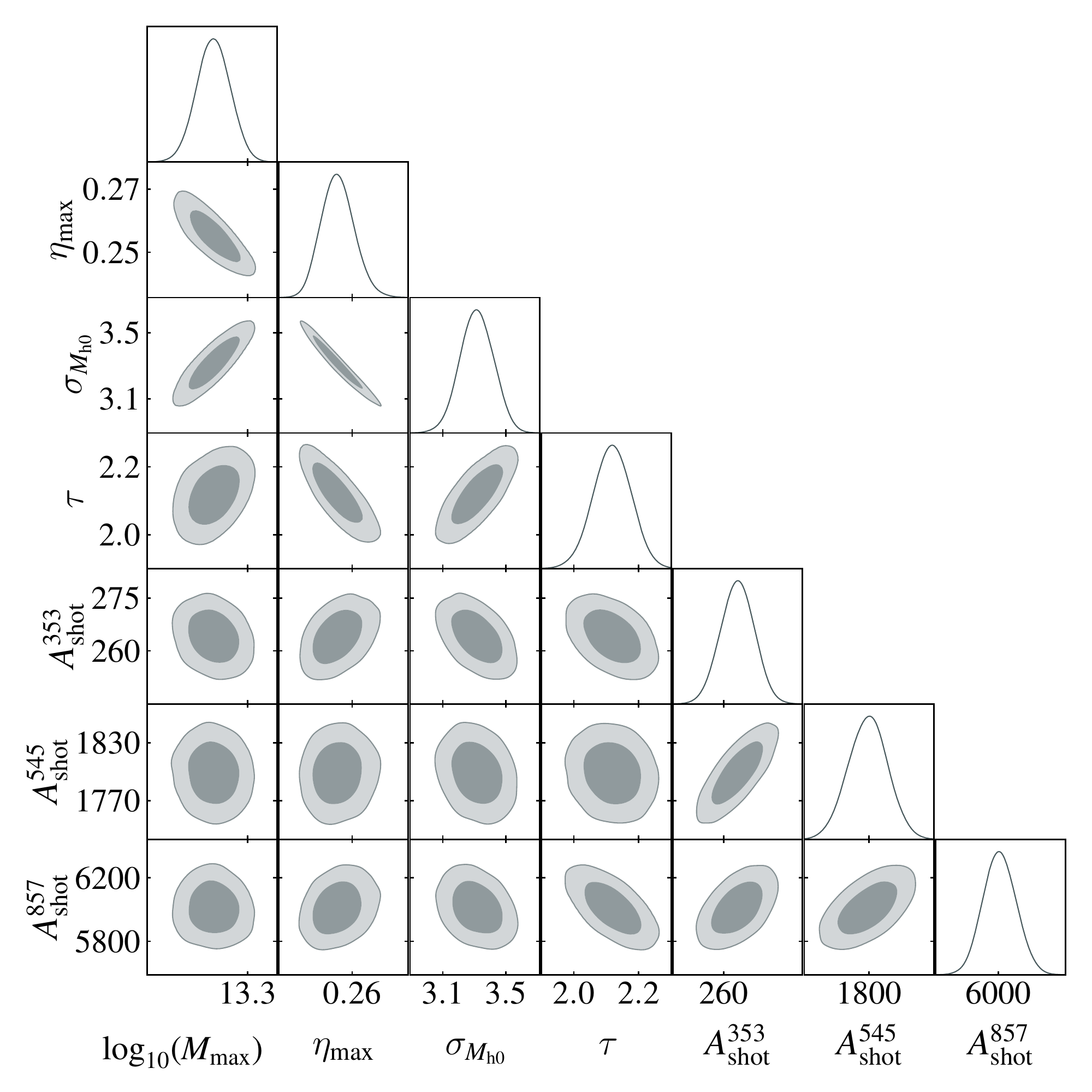}\llap{\makebox[\wd1][l]{\raisebox{0.25cm}{\includegraphics[bb=169 494 600 -600]{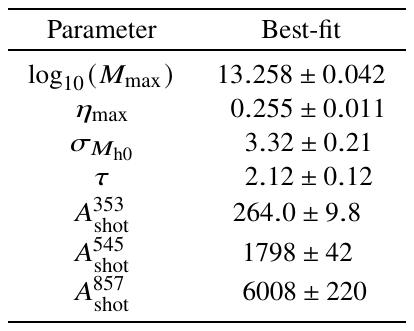}}}}
\caption{Constraints obtained for the  \protect\cite{maniyar2020} CIB halo model parameters obtained by comparing the simulated CIB spectra at 353/545/857 GHz.   }
\label{fig:cib_hod_maniyar_bestfit}
\end{center}
\end{figure}

\subsubsection{Results}

\begin{figure}
\begin{center}
\includegraphics[width=1.0\linewidth]{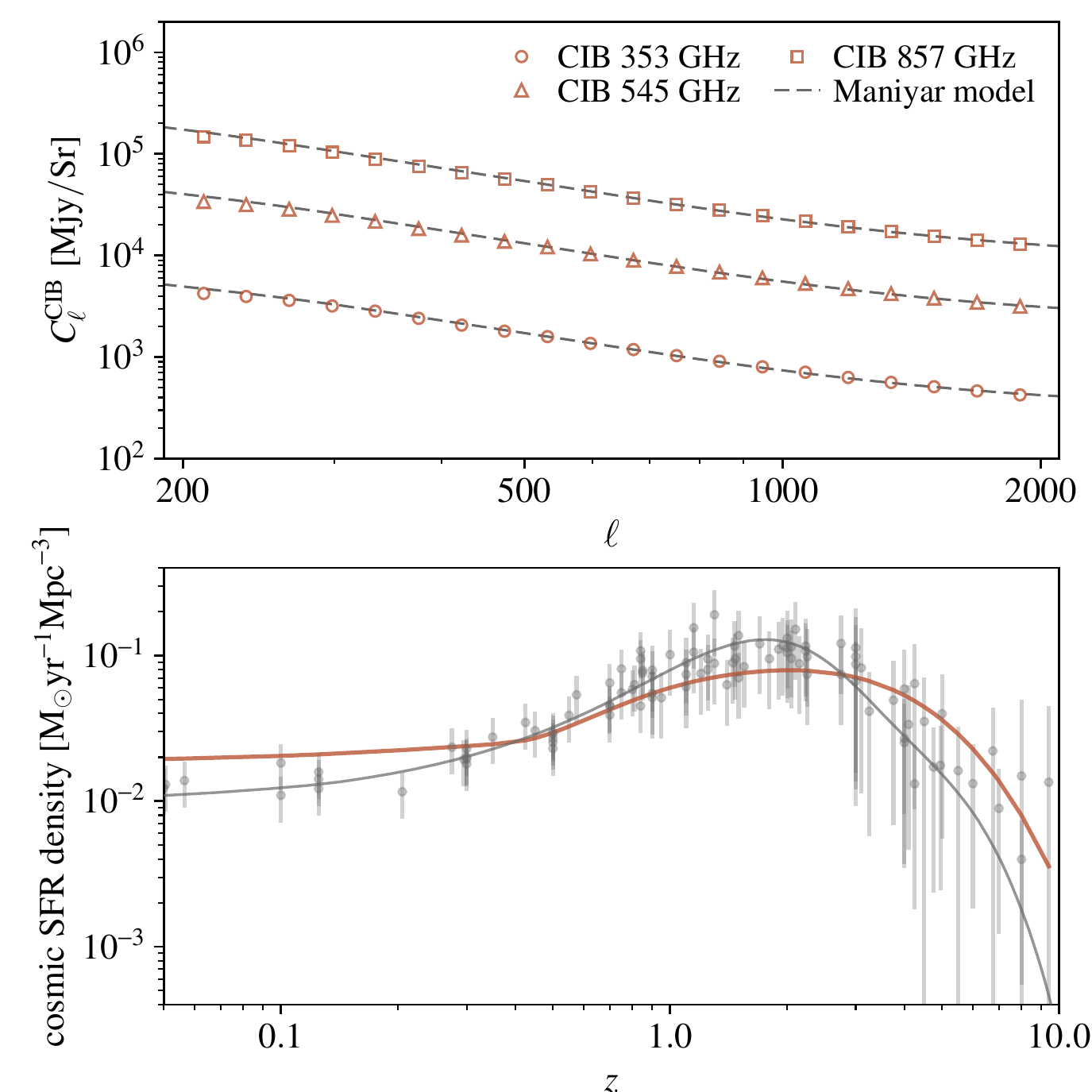}
\caption{ {\bf Upper:} Measured CIB auto-spectra from the simulation at 353/545/857 GHz (orange points), and the best-fit model from the \protect\citet{maniyar2020} halo model (dashed grey  lines). {\bf Lower:} The derived cosmic star formation rate density from the best-fit halo model parameters in the simulation (orange line), the input cosmic star formation rate from \um{} (grey line), and the various data sets that were used to constrain the \um{} model described in Figure \ref{fig:um_cosmicsfr} (grey points). }
\label{fig:cib_hod_maniyar_sfrd}
\end{center}
\end{figure}

Figure \ref{fig:cib_hod_maniyar_bestfit} shows the constraints on the parameters of the CIB halo model obtained by comparing the model with CIB spectra measured from the simulation. The CIB auto-spectra at 353, 545, 857 GHz computed using the best-fit parameters are shown in the upper panel of Figure \ref{fig:cib_hod_maniyar_sfrd}, and the best-fit SFRD obtained are shown in the lower panel of Figure \ref{fig:cib_hod_maniyar_sfrd}: overall, a flatter cosmic SFRD trend is obtained across the redshift range of  $0<z<10$, but the characteristics (such as the SFRD peak at $z\sim2$) are roughly recovered. 

\section{Discussion}\label{sec:discussion}

While as many observables as possible were included in this work, some key observables were left out. As described in Section \ref{sec:method_gal}, one was the inclusion of realistic galaxies, which could be implemented based on either the local density field \citep{chuang2015,wechsler2021} or a halo-based approach, as done in many other studies. Galaxies were specifically excluded from this work because their implementation would necessitate extensive validation, which is beyond the scope of this paper.

While the CMB secondary effects that commonly appear in analyses have been implemented, there are also a plethora of astrophysical effects that would be interesting for future surveys, that could be implemented in the simulation. Moreover, there are other astrophysical and cosmological observables that are of great interest to implement, such as gravitational waves. These additional observables will be continuously added to this simulation suite.

Finally, a parallel and complementary work by 
Sato-Polito et al. (in prep.) which implements various emissions lines (e.g., 21cm, CO, CII and Lyman-$\alpha$) in the same lightcone, using a flexible framework should be mentioned. In that work, the same simulation products from the MDPL2 simulation and \textsc{UniverseMachine} catalogues are used. Furthermore, the same physical modelling to derive infrared luminosity from the SFR and $M_{*}$ is adopted, which can be used as an intermediate step to calibrate line luminosities from independent observations and is demonstrated for the case of CO emission. The maps are then rotated consistently using the same rotation matrices as in this work to ensure that the simulation products are correlated. Having such a simulation will be important for studying various biases in cross-correlation measurements between emission lines and CMB secondary effects, and will be a useful tool for analysing data from upcoming intensity mapping experiments.

\section{Summary}\label{sec:summary}

In this work, synthetic maps of the CMB secondary effects (CMB lensing, tSZ, kSZ, CIB radio sources), galaxy shear catalogues (DES-Y1 and LSST-Y1), and galaxy density maps (LSST-Y1) were created using the particles and halo catalogues of the MDPL2 $N$-body simulation.

Descriptions of how each of the components is derived from the halo catalogues and dark matter particle density maps are described in Section \ref{sec:seconadary_modeling}. The weak lensing maps (both CMB and galaxy) were produced by projecting all the particles in the simulation onto spherical shells and running a ray tracing algorithm through the lightcone. The tSZ and kSZ maps were generated by pasting gas profiles extracted from the \bahamas{} hydrodynamical simulation onto halos in the lightcone and also using maps of the velocity fields (for kSZ). The CIB maps were generated based on the astrophysical quantities provided with the \um{} catalogues, and using a data-calibrated prescription to translate SFR and $M_{*}$ to IR brightness. Radio sources were also based on the \um{} catalogue, with an implement based on the outputs of the code \trinity{}, which predicts the number of active black holes. The population of halos with active black holes assigned was then abundance matched with the 5GHz radio luminosity function from \cite{tucci2021}, and a frequency scaling was used to convert those fluxes from 5GHz to 95/150/220 GHz. Finally, maps of galaxy density and galaxy weak lensing were generated based on the density shells and ray traced weak lensing shells, respectively. The maps were integrated up to $z=3$ for tSZ/kSZ, and $z=8.6$ for the matter density, galaxy shear, CIB, and CMB lensing fields (a Gaussian realization was added to extend the CMB lensing component to $z=1089$).

The measured power spectra, correlation functions and number counts from the simulation products are 
presented in Section \ref{sec:secondary_validation}. Maps and catalogues were shown to be consistent with observational data, and that these simulation products are realistic with sufficient precision to be used in real data analysis.

In Section \ref{sec:applications}, some example usages of the simulation products are demonstrated. In particular, it was shown that the synthetic maps can be used for multi-probe analysis consisting of galaxy clustering, galaxy lensing and CMB lensing (Section \ref{sec:6x2pt}),  validating component separation pipelines (Section \ref{sec:app_compsep}), evaluating biases in temperature based lensing reconstruction (Section \ref{sec:biases_clkk}), forecasting  multi-tracer delensing efficiencies (Section \ref{sec:app_multitracer}), as well as testing models of CIB (Section \ref{sec:app_csfr}). 

Although the simulation products of this work cannot be used for, for example, covariance estimation, it is anticipated that the maps will be useful for many other aspects, such as for end-to-end pipeline validation and estimating the biases in auto- and cross-correlation measurements. These high-resolution simulation maps are useful not only for ongoing and future CMB experiments but also for LSS surveys, such as {\it Euclid}, {\it Roman}, and LSST, as well as for analyses that measure cross-correlations between two surveys, as simulations geared towards such studies are currently limited. Much like an ``Agora," it is hoped that this simulation will serve as an ``open place of assembly" where members of the CMB and LSS communities (as well as line intensity mapping and others in the future) can discuss and analyse correlations between various observables.

\section*{Acknowledgements}

The author would like to thank: Dhayaa Anbajagane, Eric Baxter, Chihway Chang, Tom Crawford, Judit Prat, and Lucas Secco for reviewing and providing detailed comments on the draft, Peter Behroozi for making the \um{} catalogues available, Albert Chuang and Gustavo Yepes for helping out with the transferring of the raw MDPL2 particles, Srinivasan Raghunathan for providing an ILC code, and Marcelo Alvarez and Risa Wechsler for helpful discussions.

The author is supported by DOE grant DE-SC0021949. The author gratefully acknowledges the computing resources provided on Crossover, a high-performance computing cluster operated by the Laboratory Computing Resource Center at Argonne National Laboratory. Some of the computing for this project was performed on the Sherlock cluster. The author would like to thank Stanford University and the Stanford Research Computing Center for providing computational resources and support that contributed to these research results. This research used resources of the National Energy Research Scientific Computing Center (NERSC), a U.S. Department of Energy Office of Science User Facility located at Lawrence Berkeley National Laboratory, operated under Contract No. DE-AC02-05CH11231 using NERSC award HEP-ERCAP0021264. The author acknowledges the University of Chicago’s Research Computing Center for their support of this work.

The CosmoSim database used in this paper is a service by the Leibniz-Institute for Astrophysics Potsdam (AIP). The MultiDark database was developed in cooperation with the Spanish MultiDark Consolider Project CSD2009-00064.

The author gratefully acknowledges the Gauss Centre for Supercomputing e.V. (\url{www.gauss-centre.eu}) and the Partnership for Advanced Supercomputing in Europe (PRACE, \url{www.prace-ri.eu}) for funding the MultiDark simulation project by providing computing time on the GCS Supercomputer SuperMUC at Leibniz Supercomputing Centre (LRZ, \url{www.lrz.de}). The Bolshoi simulations have been performed within the Bolshoi project of the University of California High-Performance AstroComputing Center (UC-HiPACC) and were run at the NASA Ames Research Center.

\bibliographystyle{mnras}
\bibliography{bibl}

\appendix

\section{Simulation products}\label{sec:simproducts}

Listed below are the simulation products that will be made public with the initial release upon publication:

\begin{enumerate}[leftmargin=\parindent,align=left,labelwidth=\parindent,labelsep=1pt]
\item Lensed CMB: lensed CMB maps ($T/Q/U$) using the ray traced lensing field, the Gaussianized realizations of ray traced lensing field, as well as purely Gaussian lensing field.  
\item CMB lensing: ray traced CMB lensing map that covers the entire redshift range of $0<z<1089$.
\item CIB maps: lensed CIB maps for SPT-SZ/SPT-3G channels (95,150,220 GHz) and {\it Planck} (100-857 GHz). Two variants of the maps will be provided: one in units of $\muk{}$ and the other in Jy/Sr. 
\item tSZ maps: lensed tSZ maps (as Compton-$y$) for the \bahamas{} 7.6, 7.8, 8.0 models.  
\item kSZ maps: lensed kSZ maps for the \bahamas{} 7.6, 7.8, 8.0 models (as $\Delta T$, in units of $\mu {\rm K}$).
\item Radio catalogue: a catalogue of lensed radio sources at SPT-SZ frequencies (95/150/220 GHz).
\item Mock LSST-Y1 DESC-SRD galaxy density maps (5 bins).
\item Mock LSST-Y1 DESC-SRD shear maps (5 bins).
\item Rotation matrix to transform coordinates.
\item Relevant \camb{} files.
\end{enumerate}
Additionally, some derived products presented in Section \ref{sec:applications}  (with the calibration factors described in Section \ref{sec:calibration} applied) are also to be provided:
\begin{enumerate}[leftmargin=\parindent,align=left,labelwidth=\parindent,labelsep=1pt]
\item Mock \planck{} frequency maps (100/143/217/353/545/857 GHz).
\item Mock SPT-SZ frequency maps (95/150/220 GHz).
\item Mock SPT-3G frequency maps (95/150/220 GHz).
\item Mock MILCA Compton-$y$ map.  
\item Mock SPT-SZ Compton-$y$ maps.
\item Multi-tracer delensing templates.
\end{enumerate}
Additional products will be made available in future releases. Other simulation products may be available upon reasonable request.

\section{Unit conversion and colour correction}\label{sec:uc_cc}
The simulated skies are generated by convolving each map with the spectral transmission function $\tau(\nu)$ of a certain experiment. Maps are converted to temperature units using the unit conversion formulation given in \cite{planck2013ix}. For radio sources and CIB, we use:
\begin{equation} 
U({\rm Mjy/Sr}\ {\rm to}\ {\rm K}_{\rm CMB}) = \left[ \frac{\int d\nu \tau(\nu)b'_{\nu}}{\int d\nu \tau(\nu)(\nu_{\rm c}/\nu) } \times 10^{20} \right]^{-1}
\end{equation}
where $\nu_{\rm c}$ is the reference frequency of an experiment (i.e. 95/150/220 GHz for SPT-SZ and 100/143/217/353/545/857 GHz for ${\it Planck}$) and 
\begin{align}
b'_{\nu}&=\frac{\partial B_{\nu}(T,\nu)}{\partial T}|_{T=2.7255\ {\rm K}}\nonumber \\
&=\left[\frac{2h\nu^{3}}{c^{2}(\exp(h\nu/kT)-1)}\right].
\end{align}
We note that the choice of $\nu_{\rm c}$ {\it does} 
affect the values of unit conversion and colour correction. In this study, the same frequency values adopted by the various analyses are used. For SPT-SZ $\nu_{c}=$ 97.9, 154.1, and 219.6 GHz from \citealt{george2015}, and for {\it Planck} we simply use 100, 143, 217, 353, 545, 857 GHz.
For CIB, we additionally calculate the colour correction factor
\begin{equation}
C=\frac{\int(I_{\nu}/I_{\nu_{0}})\tau(\nu)d\nu}{\int (\nu_{0}/\nu)\tau(\nu)d\nu}.
\end{equation}
To convert Compton-$y$ into temperature, the following equation is used \citep{planck2013ix}:
\begin{equation}\label{eq:y2uk}
U(y_{\rm SZ}\ {\rm to}\ {\rm K}_{\rm CMB}) = \left[ \frac{\int d\nu \tau(\nu)b'_{\nu}}{\int \tau(\nu)b'_{\nu}  T_{\rm CMB} \left( \left( \frac{h\nu}{kT}\right)\frac{{\rm exp}[h\nu/kT_{\rm CMB}]+1}{{\rm exp}[h\nu/kT_{\rm CMB}]-1}-4\right)}  \right]^{-1}.
\end{equation}

\section{Correlated Gaussian realizations}\label{sec:appendix_corrgauss}
The method described in \cite{giannantonio2008} is used to generate correlated Gaussian maps used in Section \ref{sec:implementation_primarycmb}. For the simplistic case of generating two correlated maps, we take the auto- and cross-spectra $C_{\ell}^{XX}, C_{\ell}^{XY}, C_{\ell}^{YY}$:
\begin{align}
a_{\ell m}^{1}&=\eta_{1}T_{11}\nonumber\\
              &=\eta_{1}\sqrt{C_{\ell}^{11}}\\
a_{\ell m}^{2}&=\eta_{1}T_{12}+\eta_{2}T_{22}\nonumber\\
&=\eta_{1}C_{\ell}^{12}/ \sqrt{C_{\ell}^{11}}+\eta_{2}\sqrt{C_{\ell}^{22}-(C_{\ell}^{12})^2/C_{\ell}^{11} },
\end{align}
where $\eta$ are complex numbers with unit variance and zero mean, and  $T_{ij}$ are coefficients that can be written as a function of power spectra $C_{\ell}^{ij}$. It can be shown that the coefficients of a given map can be generated from the previous step, and therefore correlated maps can be generated successively using the relation:
\begin{equation}
T_{ij}=
\begin{cases}
		 \sqrt{C^{ji}-\sum_{k=1}^{j-1}T_{ik}^{2}}	, & \text{if $i=j$;}\\[10pt]
            \frac{C^{ji}-\sum_{k=1}^{j-1}T_{ik}T_{jk} }{T_{jj}}, & \text{if $i>j$.}
		 \end{cases}
\end{equation}

\label{lastpage}
\end{document}